\documentclass{article}
\usepackage{amsmath,amsfonts,amssymb,setspace,comment,url}
\usepackage[a4paper]{geometry}
\usepackage{cite}
\usepackage{xcolor}
\usepackage{hyperref}
\newcommand{\be}{\begin{equation}}
\newcommand{\ee}{\end{equation}}

\newcommand{\EG}[1]{\mathrm{E}_{#1(#1)}}

\newcommand{\SU}[1]{\mathrm{SU}( #1 )}
\newcommand{\SL}[1]{\mathrm{SL}( #1 )}
\newcommand{\GL}[1]{\mathrm{GL}( #1 )}
\newcommand{\SO}[1]{\mathrm{SO}( #1 )}

\newcommand{\UO}{\mathrm{U}(1)}

\newcommand{\GH}{\mathrm{G}_{\mathrm{half}}}

\newcommand{\hK}{\hat{K}}
\newcommand{\K}{K}
\newcommand{\J}{J}
\newcommand{\hJ}{\hat{J}}

\newcommand{\bomega}{\bar{\omega}}
\newcommand{\homega}{\omega}

\newcommand{\hsigma}{\hat{\sigma}}
\newcommand{\comega}{\hat{\omega}}
\newcommand{\hphi}{\hat{\phi}}
\newcommand{\tsigma}{\star_2 \sigma}

\newcommand{\by}{\bar{y}}

\newcommand{\obf}[1]{\overline{\mathbf{#1}}}
\newcommand{\mbf}[1]{\mathbf{#1}}
\newcommand{\gL}{\mathcal{L}}

\newcommand{\gM}{\mathcal{M}}

\newcommand{\z}{z}

\numberwithin{equation}{section}

\newcommand\Tstrut{\rule{0pt}{3ex}}         
\newcommand\Bstrut{\rule[-1.3ex]{0pt}{0pt}}   

\newcommand{\KKD}{\tilde{D}}
\newcommand{\tD}{\tilde{D}}
\newcommand{\tvol}{vol_{\tilde{S}^2}}
\newcommand{\tTh}{\tilde{\theta}}

\begin{document}

\begin{titlepage}
\vfill

\begin{center}
   \baselineskip=16pt
   	{\Large \bf Supersymmetric AdS$_{7}$ and AdS$_6$ vacua and their \\consistent truncations with vector multiplets}
   	\vskip 2cm
   	{\large \bf Emanuel Malek$^a$\footnote{\tt Emanuel.Malek@aei.mpg.de}, Henning Samtleben$^b$\footnote{\tt Henning.Samtleben@ens-lyon.fr}, Valent\'{i} Vall Camell$^{c}$\footnote{\tt Valenti.Vall-Camell@itp.uni-hannover.de}}
   	\vskip .6cm
   	{\it $^a$ Max-Planck-Institut f\"{u}r Gravitationsphysik (Albert-Einstein-Institut), \\
   	Am M\"{u}hlenberg 1, 14476 Potsdam, Germany \\ \ \\
    \it $^b$ Univ Lyon, Ens de Lyon, Univ Claude Bernard, CNRS,\\
    Laboratoire de Physique, F-69342 Lyon, France \\ \ \\
    \it $^c$ Institut f\"ur Theoretische Physik and Riemann Center for Geometry and Physics,\\
Leibniz Universit\"at Hannover, Appelstrasse 2, D-30167 Hannover, Germany \\ \ \\}
   	\vskip 1cm
\end{center}
\vfill

\begin{abstract}
	Using exceptional field theory we construct supersymmetric warped AdS$_7$ vacua of massive IIA and AdS$_6$ vacua of IIB, as well as their consistent truncations including vector multiplets. We show there are no consistent truncations of massive IIA supergravity around its supersymmetric AdS$_7$ vacua with vector multiplets when the Roman's mass is non-vanishing. For AdS$_6$ vacua of IIB supergravity, we find that in addition to the consistent truncation to pure $\mathrm{F}(4)$ gauged SUGRA, the only other half-maximal truncations that are consistent result in $\mathrm{F}(4)$ gauged SUGRA coupled to one or two Abelian vector multiplets, to three non-Abelian vector multiplets, leading to an $\mathrm{ISO}(3)$ gauged SUGRA, or to three non-Abelian plus one Abelian vector multiplet, leading to an $\mathrm{ISO}(3) \times \UO$ gauged SUGRA. These consistent truncations with vector multiplets exist when the two holomorphic functions that define the AdS$_6$ vacua satisfy certain differential conditions which we derive. We use these to deduce that no globally regular AdS$_6$ solutions admit a consistent truncation to $\mathrm{F}(4)$ gauged SUGRA with two vector multiplets, and show that the Abelian T-dual of the Brandhuber-Oz vacuum allows a consistent truncation to $\mathrm{F}(4)$ gauged SUGRA with a single vector multiplet.
\end{abstract}
\vskip 0cm

\vfill

\end{titlepage}

\tableofcontents

\newpage

\section{Introduction}
Supersymmetric AdS vacua of 10-/11-dimensional SUGRA play an important role in our modern understanding of theoretical physics. For example, they have led to many important insights into superconformal field theories via the AdS/CFT correspondence. For many holographic applications, it is useful to have a consistent truncation of 10-/11-dimensional SUGRA around a supersymmetric AdS vacuum. Such a consistent truncation allows us to uplift solutions of a lower-dimensional (usually gauged) SUGRA to solutions of 10-/11-dimensional SUGRA. This makes them a powerful tool in studying deformations of the AdS vacua, for example those breaking supersymmetry. Moreover, since the AdS radius of supersymmetric AdS vacua is typically of the same scale as the compactification radius, lower-dimensional SUGRA theories do not arise by integrating out the Kaluza-Klein tower of the compactification. Thus, consistent truncations are the only way to study AdS vacua via lower-dimensional supergravities.

However, constructing consistent truncations is a notoriously difficult task which has until recently largely eluded a systematic approach. For some purposes, it may even be enough to know that a consistent truncation of 10-/11-dimensional SUGRA exists, even without having the explicit truncation Ans\"{a}tze. Yet, to date there is no classification of what consistent truncations exist around a given supersymmetric AdS vacuum, although it is conjectured that for every warped supersymmetric AdS$_D$ vacuum of 10-/11-dimensional SUGRA, there exists a ``minimal'' consistent truncation to $D$-dimensional gauged SUGRA keeping only the gravitational supermultiplet \cite{Gauntlett:2007ma}, which has been proven in some cases.

Powerful tools in constructing consistent truncations have recently come from exceptional field theory (ExFT) \cite{Berman:2010is,Berman:2011cg,Berman:2012vc,Hohm:2013pua} and exceptional generalised geometry (EGG) \cite{Pacheco:2008ps,Coimbra:2011ky,Coimbra:2012af}, which reformulate 10-/11-dimensional SUGRA in a way which unifies the metric and flux degrees of freedom. In this framework, consistent truncations preserving all supersymmetries arise as ``generalised Scherk-Schwarz'' truncations \cite{Aldazabal:2011nj,Geissbuhler:2011mx,Grana:2012rr,Berman:2012uy}, generalising consistent truncations on group manifolds \cite{Scherk:1979zr} to the more general setting of ``generalised (Leibniz) parallelisable spaces'' \cite{Lee:2014mla}, which includes certain homogeneous spaces. This has led to a proof of the consistency of the maximally supersymmetric $S^5$ truncation of IIB supergravity \cite{Lee:2014mla,Hohm:2014qga,Baguet:2015sma}, and to new consistent truncations giving rise to compact and dyonic gaugings \cite{Malek:2015hma,Baguet:2015iou,Ciceri:2016dmd,Cassani:2016ncu,Inverso:2016eet,Inverso:2017lrz,Malek:2017cle}. Moreover, all currently known maximally supersymmetric consistent truncations, including the truncations of 11-dimensional SUGRA on $S^4$ and $S^7$ \cite{Lee:2014mla,Hohm:2014qga}, first found in \cite{Nastase:1999cb,Nastase:1999kf,deWit:1986oxb}, and the truncation of massive IIA on $S^6$ \cite{Ciceri:2016dmd,Cassani:2016ncu}, first constructed in \cite{Guarino:2015jca,Guarino:2015vca}, are nicely captured by the framework of generalised Scherk-Schwarz truncations.

Recently, \cite{Malek:2016bpu,Malek:2017njj} has shown how use this framework to define consistent truncations breaking half of the supersymmetry. Such half-maximal truncations of type II/11-dimensional SUGRA then lead to a half-maximal gauged SUGRA in lower dimensions. Furthermore, \cite{Malek:2017njj} proved the half-maximal case of the conjecture of \cite{Gauntlett:2007ma}, i.e. that every half-maximally supersymmetric warped AdS$_D$ vacuum of 10-/11-dimensional SUGRA admits a consistent truncation to half-maximal $D$-dimensional gauged SUGRA keeping only the gravitational supermultiplet.

Moreover, ExFT and EGG lead to a new geometric description of supersymmetric AdS vacua of 10-/11-dimensional SUGRA where the compactification manifold is characterised by ``generalised holonomy'', or a (weakly) integrable generalised $G$-structure, \cite{Coimbra:2014uxa,Coimbra:2015nha,Ashmore:2016qvs,Coimbra:2016ydd,Malek:2017njj,Coimbra:2017fqv} in analogy to supersymmetric Minkowski vacua without fluxes arising from special holonomy compactifications \cite{Candelas:1985en}. Moreover, as showed in \cite{Malek:2017njj}, once the generalised $G$-structure underyling the supersymmetric AdS vacuum is constructed, the ``minimal'' consistent truncation can be obtained immediately. Therefore, this framework is ideally suited to studying supersymmetric AdS vacua and their consistent truncations, which we will undertake in this paper.

In this work, we will focus on supersymmetric AdS$_{7}$ solutions of massive IIA SUGRA and supersymmetric AdS$_{6}$ solutions of IIB. Building on previous work \cite{Lozano:2012au,Apruzzi:2014qva,Kim:2015hya}, families of infinitely many such vacua have recently been constructed in the literature \cite{Apruzzi:2013yva,DHoker:2016ujz,DHoker:2016ysh,DHoker:2017mds,DHoker:2017zwj}, where the AdS$_7$ solutions are characterised by a cubic function on an interval \cite{Cremonesi:2015bld} and the AdS$_6$ solutions by two holomorphic functions on a Riemann surface\footnote{An alternative characterisation of the AdS$_6$ vacua in terms of a real harmonic function is given in \cite{Apruzzi:2018cvq}.}. These AdS vacua admit a ``universal'' consistent truncation to pure 7-dimensional $\SU{2}$ gauged SUGRA \cite{Townsend:1983kk} and 6-dimensional $\mathrm{F}(4)$ gauged SUGRA \cite{Romans:1985tw}, which takes the same form for any of the cubic functions / holomorphic functions defining the AdS vacua \cite{Passias:2015gya,Hong:2018amk,Malek:2018zcz}. In a recent paper \cite{Malek:2018zcz}, we showed that in ExFT these infinite families of AdS solutions are described by the same universal generalised half-maximal structure and used this to explain the universal form of the AdS$_7$ consistent truncations and derive the consistent truncation around the AdS$_6$ vacua.

It has remained an interesting open problem to find any consistent truncations around the AdS$_{6,7}$ vacua keeping more modes than just the gravitional supermultiplet. Supersymmetry implies that any extra modes kept will have to form vector multiplets of the 6- and 7-dimensional gauged SUGRA obtained after truncation. Here we will use the framework of ExFT, and specifically the tools developed in \cite{Malek:2016bpu,Malek:2017njj}, to address this problem: we will classify all possible consistent truncations with vector multiplets around the supersymmetric AdS$_{6,7}$ vacua that are compatible with the Ansatz proposed in \cite{Malek:2016bpu,Malek:2017njj}. Assuming the Ansatz of \cite{Malek:2016bpu,Malek:2017njj} to be the most general Ansatz for consistent truncations with vector multiplets, our results give a full classification of the consistent truncations around supersymmetric AdS$_{6,7}$ vacua. We find that
\begin{itemize}
	\item there are no consistent truncations with vector multiplets around the supersymmetric AdS$_{7}$ vacua of massive IIA SUGRA when the Roman's mass is non-vanishing,
	\item supersymmetric AdS$_{6}$ vacua of IIB SUGRA admit consistent truncations with vector multiplets when the holomorphic functions characterising them admit certain differential conditions which we  give explicitly. We construct the non-linear consistent truncation Ans\"{a}tze that give rise to less than four vector multiplets.
\end{itemize}

Our paper is organised as follows. First, we give a summary of our results in \ref{s:ResultsSummary}. In section \ref{s:ExFTReview}, we give a brief introduction to the relevant aspects of ExFT, while in section \ref{s:AdSReview}, we review the techniques developed in \cite{Malek:2016bpu,Malek:2017njj} to describe supersymmetric AdS vacua of 10-/11-dimensional SUGRA and their minimal consistent truncations, in which only the gravitational supermultiplet is kept. In section \ref{s:VecTruncation}, we review how to define consistent truncations with matter multiplets as described in \cite{Malek:2016bpu,Malek:2017njj}. Next, we show how to compute the generalised metric from the half-maximal structure underlying the AdS vacua in section \ref{s:GenMetric}. In sections \ref{s:AdS7} and \ref{s:AdS6}, we show how one can easily construct the supersymmetric AdS$_7$ vacua of massive IIA SUGRA and AdS$_6$ vacua of IIB SUGRA, respectively, using half-maximal structures of ExFT, before deriving their minimal consistent truncations in \ref{s:ConsTruncation}. Finally, in section \ref{s:VecTruncationAdS7} we show that there are no consistent truncation with vector multiplets around the supersymmetric AdS$_7$ vacua of massive IIA and in section \ref{s:VecTruncationAdS6} we classify all possible consistent truncations with vector multiplets around the supersymmetric AdS$_6$ vacua of IIB SUGRA. These consistent truncations require the holomorphic functions characterising the AdS$_6$ vacua to satisfy certain differential constraints which we derive. We also explicitly construct the non-linear consistent truncation Ans\"{a}tze yielding less than four vector multiplets. We conclude with a discussion and outlook in section \ref{s:Conclusions}.

\subsection{Summary of results} \label{s:ResultsSummary}
We summarise here our results. In sections \ref{s:AdS7} and \ref{s:AdS6} we construct and classify all supersymmetric AdS$_7$ vacua in mIIA theory and AdS$_6$ vacua in IIB, respectively. As we review in section \ref{s:RevConsTruncation}, for each of them one can construct a consistent truncation to a minimal half-supersymmetric gauged supergravity with a gravitational supermultiplet, which we explicitly construct in section \ref{s:ConsTruncation}.  Finally, in sections \ref{s:VecTruncationAdS7} and \ref{s:VecTruncationAdS6} we analyse the possibility of having consistent truncations with matter multiplets around these vacua, using the methods of \cite{Malek:2016bpu,Malek:2017njj}. From the latter, it follows that we can have at most three (four) vector multiplets in consistent truncations around supersymmetric AdS$_7$ (AdS$_6$) vacua and, for the truncation to be consistent, the compactification space has to satisfy certain conditions. These extra conditions imply that there are no consistent truncations with vector multiplets around AdS$_7$ vacua for non-vanishing Roman's mass. For the AdS$_6$ case, we find that only a small subset of 6-dimensional half-maximal gauged SUGRAs admitting supersymmetric AdS$_6$ vacua \cite{Karndumri:2016ruc} can arise as a consistent truncation of IIB SUGRA, and we derive explicit differential constraints on the compactification space for the consistent truncations to exist. More concretely, our findings for each of the cases are the following:

\subsubsection*{AdS$_7$ in mIIA}
In section \ref{s:AdS7} we construct and classify all geometries in mIIA theory consisting of the warped product
\begin{equation}
\text{AdS}_7\times I\times S^2\,,
\end{equation}
with $I$ an interval, that preserve supersymmetry, the minimal amount being 16 supercharges in seven dimensions. We encounter that they can be classified in terms of a function $t(z)$ on the interval $I$ satisfying
\begin{equation}
\dddot{t}=-\frac{m}{2}\,, \qquad\text{and}\qquad t(z)\ge 0\,,
\end{equation}
where equality in the last condition holds on the endpoints of $I$ and ensures that the total internal space has no boundaries. The parameter $m$ is the Roman's mass of mIIA. We study all possibilities of having consistent truncations with vector multiplets around these vacua and find that the only possibility is to keep a single vector multiplet in the truncation and only if $m = 0$. This consistent truncation is just a consistent subsector of the maximally supersymmetric consistent truncation around the AdS$_7 \times S^4$ solution of 11-dimensional supergravity dimensionally reduced to IIA supergravity.

\subsubsection*{AdS$_6$ in IIB}
Similarly, in section \ref{s:AdS6} we construct and classify all geometries in IIB theory consisting of the warped product
\begin{equation}
\text{AdS}_6\times \Sigma\times S^2\,,
\end{equation}
where $\Sigma$ is a Riemann surface (with boundaries), that preserve 16 supercharges. We find that they can be classified in terms of two holomorphic functions $f^\alpha$, $\alpha = 1, 2$, on the Riemann surface. These functions have to satisfy the condition 
\begin{equation}
i\, \partial f^\alpha \bar{\partial}\bar{f}_\alpha \geq 0\,, \qquad r \geq 0
\end{equation}
where equality holds on the boundary of $\Sigma$, ensuring that the total internal space has no boundaries. The function $r$ is a real function of the Riemann surface defined  up to an integration constant through the differential equation
\begin{equation}
dr=-p_\alpha\,d k^\alpha\,,
\end{equation}
where $p^\alpha$ and $k^\alpha$ are the real/imaginary parts $f^\alpha = - p^\alpha + i\, k^\alpha$. We also study which consistent truncations with vector multiplets around these vacua exist, and our results are summarised in table \ref{t:AdS6-vect-mult}. We explicitly construct the consistent truncations containing one, two and three vector multiplets.

\begin{table}[h!]
\begin{center}
\begin{tabular}{|c|c|c|c|}
\hline
$N$ & $\SU{2}_R$ rep & Consistent truncations with vect. mult. & Gauging \Tstrut\Bstrut \\
\hline\hline
1 & $\mbf{1}$ & Only if $\exists\, g \in \UO$ s.t. $\partial(g\,\partial f^\alpha)\in \text{real functions on }\Sigma$ & $\SU{2} \times \UO$ \Tstrut\Bstrut\\
\hline
2 & $\mbf{1} \oplus \mbf{1}$ & NO (due to global issues) & $\SU{2} \times \UO^2$ \Tstrut\Bstrut\\
\hline
3 & $\mbf{1} \oplus \mbf{1} \oplus \mbf{1}$ & NO & N/A \Tstrut\Bstrut\\
\hline 
3 & $\mbf{3}$ & Only if  $r\, d\pi^\alpha=p^\alpha \pi^\beta\wedge\pi_\beta$ & $\mathrm{ISO}(3)$ \Tstrut\Bstrut\\
\hline 
4 & $\mbf{1} \oplus \mbf{1} \oplus \mbf{1} \oplus \mbf{1}$ & NO & N/A \Tstrut\Bstrut \\
\hline
\rule{0pt}{4.3ex} 4 & $\mbf{3} \oplus \mbf{1}$ & Only if $\exists\, \textbf{3}$ and $\exists\, \textbf{1}$ with $g=\pm\dfrac{1}{2}\left(\dfrac{p_\alpha\bar{\partial}\bar{f}^\alpha}{p_\beta\partial f^\beta}\right)$ & $\mathrm{ISO}(3) \times \UO$ \Tstrut\Bstrut\\[8pt]
\hline
\end{tabular}
\end{center}
\caption{Possible consistent truncations with $N$ vector multiplets around supersymmetric AdS$_6 \times S^2$ vacua in IIB and the resulting gauging of the gauged SUGRA. Consistency requires that $N \leq 4$ and that the vector multiplets form representations of $\SU{2}_R$, the $R$-symmetry of the AdS$_6$ vacua. The one-forms $\pi^\alpha$ are explicitly defined in terms of the background functions $f^\alpha$, see equations \eqref{eq:piSingleParam} and \eqref{eq:piSingleParam-g}.  \label{t:AdS6-vect-mult}}

\end{table}

\section{Review of exceptional field theories} \label{s:ExFTReview}

In this section, we review the structure of the relevant exceptional field theories. Exceptional field theories (ExFTs) are the manifestly duality covariant formulations of maximal higher-dimensional supergravity theories \cite{Hohm:2013pua,Hohm:2013vpa,Hohm:2013uia}. For our purposes, we will need the ExFTs built on the groups ${\rm E}_{5(5)}\equiv {\rm SO}(5,5)$ \cite{Abzalov:2015ega}, and ${\rm E}_{4(4)}\equiv {\rm SL}(5)$ \cite{Musaev:2015ces}, respectively.

The reformulation of the higher-dimensional supergravities is based on the split of their coordinates into $D$ {\em external} coordinates $x^\mu$ and the remaining  {\em internal} coordinates $y^i$ , with the latter embedded into a set of generalised internal coordinates $Y^M$ transforming in a representation ${R}_1$ of the duality group ${\rm E}_{d(d)}$, with $d=11-D$\,. Internal diffeomorphisms and tensor gauge transformations of the higher-dimensional supergravity  combine into a single symmetry structure of {\em generalised diffeomorphisms} in the coordinates $Y^M$ \cite{Coimbra:2011ky,Berman:2012vc}. Parametrised by a gauge parameter $\xi^M$ in ${R}_1$, the generalised Lie derivative of a generalised vector field $V^M$ in ${R}_1$ reads
\begin{equation}
{\cal L}_\xi V^M =
\xi^N\partial_N V^M-\partial_N \xi^M\, V^N + Y^{MN}{}_{KL}{}\,\partial_N \xi^K\,V^L
\,.
\label{eq:gen_diff}
\end{equation}
The constant ${\rm E}_{d(d)}$-invariant tensor $Y^{MN}{}_{KL}$ encodes the deviation from standard diffeomorphisms. Its presence implies that the transformations \eqref{eq:gen_diff} close into an algebra only after imposing the {\em section constraints}
\begin{equation}
Y^{MN}{}_{KL}{}\,\partial_M \otimes \partial_N = 0
\,,
\label{eq:section}
\end{equation}
where the internal derivatives act on any pair of fields or gauge parameters. Solutions of the section constraints restrict the internal coordinate dependence of all fields to linear subspaces of $R_1$ upon which one recovers the standard supergravity theories.
The action (\ref{eq:gen_diff}) can be rewritten as
\begin{equation}
{\cal L}_\xi V^M =
\left( \xi^N\partial_N 
+ \frac1{9-d}\, \partial_N \xi^N
-a_d\,(t^\alpha)^K{}_L\,\partial_K \xi^L\, t_\alpha \cdot \right) V^M
\,,
\label{eq:gen_diff_gen}
\end{equation}
with constant $a_d$, and $t_\alpha$ labelling the generators of ${\rm E}_{d(d)}$\,. From this formula one also reads off the action of generalised Lie derivatives on different representations. Modulo the section constraints \eqref{eq:section}, the transformations \eqref{eq:gen_diff} close into an algebra defining the E-bracket
\begin{equation}
[\xi_1, \xi_2]^M \equiv
2\,\xi_{[1}^N \partial_N \xi_{2]}^M -
Y^{MN}{}_{KL}\, \xi_{[1}^K \partial_N \xi_{2]}^L
\,.
\label{eq:Ebracket}
\end{equation}
The presence of the tensor $Y^{MN}{}_{KL}$ implies the existence of trivial gauge parameters and non-associativity of the algebra. Generalised diffeomorphisms are realised as local symmetries of ExFT (i.e.\ with parameters $\xi^M$ depending on internal and external coordinates) by introducing covariant external derivatives ${\cal D}_\mu \equiv \partial_\mu - {\cal L}_{{\cal A}_\mu}$ with the ExFT vector fields ${\cal A}_\mu{}$ in the $R_1$ representation. Non-associativity of the algebra \eqref{eq:Ebracket} implies that the standard Yang-Mills field strength based on \eqref{eq:Ebracket} is not a tensor w.r.t.\ the generalised Lie derivative \eqref{eq:gen_diff}. Rather it has to be completed by a coupling to the two-forms ${\cal B}_{\mu\nu}$ of the theory following the structure of the tensor hierarchy~\cite{deWit:2005hv}
\begin{equation}
{\cal F}_{\mu\nu} = 2\, \partial_{[\mu} {\cal A}_{\nu]} - \left[ {\cal A}_\mu,\, {\cal A}_{\nu} \right] + d {\cal B}_{\mu\nu} \,,
\label{eq:ExFTFieldStrength}
\end{equation}
Here, the bracket $\left[ {\cal A}_\mu,\, {\cal A}_{\nu} \right]$ refers to \eqref{eq:Ebracket} while the $d$ operator in the last term denotes a covariant differential operator from the $R_2$ representation of two-forms into $R_1$. Explicitly, it takes the form
\begin{equation}
(d {\cal B}_{\mu\nu})^{M} = Y^{MN}{}_{KL} \,\partial_N  {\cal B}_{\mu\nu}{}^{KL}
\,,
\end{equation}
with the two-forms ${\cal B}_{\mu\nu}{}^{KL}$ living in (a sub-representation of) the symmetric tensor product $R_2 \subset (R_1 \otimes R_1)_{\rm sym}$\,. Continuing the tensor hierarchy gives rise to the couplings of three-forms ${\cal C}_{\mu\nu\rho}\subset R_3$, four-forms ${\cal D}_{\mu\nu\rho\sigma}\subset R_4$, etc., with the lowest non-abelian field strengths given by
\begin{equation}
\begin{split}
{\cal H}_{\mu\nu\rho} &= 3\, {\cal D}_{[\mu} {\cal B}_{\nu\rho]} - 3 \partial_{[\mu} {\cal A}_{\nu} \wedge {\cal A}_{\rho]} + {\cal A}_{[\mu} \wedge \left[ {\cal A}_{\nu} ,\, {\cal A}_{\rho]} \right] + d {\cal C}_{\mu\nu\rho} \,, \\
{\cal J}_{\mu\nu\rho\sigma} &= 
4\, {\cal D}_{[\mu} {\cal C}_{\nu\rho\sigma]}+
2\,{\cal F}_{[\mu\nu} \wedge {\cal B}_{\rho\sigma]} - d {\cal B}_{[\mu\nu} \wedge  {\cal B}_{\rho\sigma]} 
-\frac43\, {\cal A}_{[\mu} \wedge \left( {\cal A}_\nu \wedge \partial_{\rho} {\cal A}_{\sigma]} \right)
\\
&\quad +\frac13\, {\cal A}_{[\mu} \wedge \left( {\cal A}_\nu \wedge [{\cal A}_{\rho}, {\cal A}_{\sigma]} ]\right)
+ d {\cal D}_{\mu\nu\rho\sigma}
\,.
\end{split} \label{eq:ExFTFieldStrengths}
\end{equation}
Again, the $d$ operator denotes the covariant internal differential operators mapping $R_p \longrightarrow R_{p-1}$, while the wedge $\wedge$ represents algebraic maps
\begin{equation}
(R_1 \otimes R_1)_{\rm sym} \longrightarrow R_2\;,\qquad
R_1 \otimes R_2  \longrightarrow R_3\;,
\label{eq:wedges}
\end{equation}
etc.. 
Just as $p$-form field strengths are tensor with respect to the Lie derivative, the field strengths \eqref{eq:ExFTFieldStrength}, \eqref{eq:ExFTFieldStrengths}, are tensors with respect to the generalised Lie derivative. 

Let us now make these structures explicit for the theories we will be using in the following. For $d=4$, the ${\rm E}_{4(4)}={\rm SL}(5)$ ExFT is based on coordinates $Y^{ab}=Y^{[ab]}$, in the $R_1={\bf 10}$ representation of ${\rm SL}(5)$, with $a, b =1, \dots, 5$ labelling the fundamental representation. The  $Y$-tensor in (\ref{eq:gen_diff}) is given by $Y^{ef,gh}{}_{ab,cd} = 6\,\delta_{abcd}^{efgh}$, and induces a tensor hierarchy of $p$-forms living in representations $R_p$ as
\begin{equation}
{\cal A}_\mu{}^{ab} : {\bf 10} \,,\quad
{\cal B}_{\mu\nu\,a} : \obf{5} \,,\quad
{\cal C}_{\mu\nu\rho}{}^{a} : {\bf 5}\,, \quad
{\cal D}_{\mu\nu\rho\sigma\,ab} : \obf{10} \,. \label{eq:7DTH}
\end{equation}
The relevant $\wedge$ products \eqref{eq:wedges} and the $d$ operators in \eqref{eq:ExFTFieldStrength}, \eqref{eq:ExFTFieldStrengths} are explicitly given by
\begin{equation}
 \begin{split}
  ({\cal A}_1 \wedge {\cal A}_2)_a &= \frac14 \epsilon_{abcde}\, {\cal A}_1{}^{bc}  {\cal A}_2{}^{de}\,, \qquad ({\cal A} \wedge {\cal B})^a =  {\cal A}^{ab}  {\cal B}_{b} \,,
 \end{split} \label{eq:7Dwedge}
\end{equation}
\begin{equation}
(d{\cal B})^{ab} = \frac12 \epsilon^{abcde}\,\partial_{cd} {\cal B}_{e} \,,\qquad (d{\cal C})_{a} = \partial_{ba} {\cal C}^b \,, \qquad (d{\cal D})^a = \frac12 \epsilon^{abcde} \partial_{bc} {\cal D}_{de} \,. \label{eq:7Dd}
\end{equation}
For what follows, it will be similarly useful to define $\wedge: R_1 \otimes R_3 \longrightarrow R_4$ and $\wedge: R_2 \otimes R_3 \longrightarrow \mbf{1}$ as
\begin{equation}
 \begin{split}
  \left( {\cal A} \wedge {\cal C} \right)_{ab} &= \frac14 \epsilon_{abcde}\, {\cal A}^{cd} {\cal C}^e \,, \qquad {\cal B} \wedge {\cal C} = {\cal B}_a {\cal C}^a \,. \label{eq:7Dwedge2}
 \end{split}
\end{equation}
Moreover, the theory features 14 scalar fields, parameterising the coset space $\SL{5}/\SO{5}$, which are most conveniently described by a group-valued generalised metric ${\cal M}_{ab}$\,. The ExFT dynamics comes from an action \cite{Musaev:2015ces}, giving rise to standard second order field equations. In particular, the 4-form field strength is dual to the 3-form field strength via the first order equation
\begin{equation}
{\cal J}_{\mu\nu\rho\sigma}{}^a = \frac1{3!} \,\sqrt{|{\cal G}|} \epsilon_{\mu\nu\rho\sigma\kappa\lambda\tau} \,
\gM^{ab} \,{\cal H}^{\kappa\lambda\tau}{}_b \,, 
\label{eq:D7DualityRelations}
\end{equation}
with the scalar matrix $\gM^{ab}$\,, and where $|{\cal G}|$ is the determinant of the external metric, ${\cal G}_{\mu\nu}$, of the ExFT, which is used to raise/lower the external indices on the field strengths.

For $d=5$, the ${\rm E}_{5(5)}={\rm SO}(5,5)$ ExFT is based on coordinates $Y^{M}$, in the $R_1={\bf 16}$ spinor representation of $\SO{5,5}$, with $M =1, \dots, 16$. The $Y$-tensor in \eqref{eq:gen_diff} is given by $Y^{PQ} {}_{MN}=\frac12\, (\gamma^I)_{MN} (\gamma_I)^{PQ}$ in terms of the $\SO{5,5}$ gamma matrices, with the index $I=1, \dots 10$, labelling the vector representation, raised and lowered by the constant $\SO{5,5}$ invariant metric $\eta_{IJ}$\,.
It induces a tensor hierarchy of $p$-forms living in representations $R_p$ as
\begin{equation}
{\cal A}_\mu{}^{M} : {\bf 16} \,,\quad
{\cal B}_{\mu\nu}{}^I : {\bf 10} \,,\quad
{\cal C}_{\mu\nu\rho\,M} : \obf{16} \,,\quad
{\cal D}_{\mu\nu\rho\sigma}{}^{[IJ]} : {\bf 45} \,. \label{eq:6DTH}
\end{equation}
Strictly speaking, the theory also carries additional covariantly constrained 4-forms ${\cal D}_{\mu\nu\rho\sigma\,M}$, but for our purposes we will only consider equations in which all four-forms drop out. The relevant $\wedge$ products \eqref{eq:wedges} and the $d$ operators in \eqref{eq:ExFTFieldStrength}, \eqref{eq:ExFTFieldStrengths} are explicitly given by
\begin{equation}
 \begin{split}
  ({\cal A} \wedge {\cal A})^I &= \frac12 (\gamma^I)_{MN}\, {\cal A}^{M}  {\cal A}^{N}\,, \qquad ({\cal A} \wedge {\cal B})_M = \frac12 (\gamma_I)_{MN} \,{\cal A}^{N}  {\cal B}^I \,,
 \end{split} \label{eq:6Dwedge1}
\end{equation}
\begin{equation}
(d{\cal B})^{M} = (\gamma_I)^{MN}\,\partial_{N} {\cal B}{}^I \,,\qquad (d{\cal C})_{I} = \frac12 (\gamma_{I})^{MN} \partial_{M} {\cal C}_{N} \,. \label{eq:6Dd}
\end{equation}
Once again, it will be useful to also define $\wedge: R_2 \otimes R_2 \longrightarrow \mbf{1}$ as
\begin{equation}
 {\cal B}_1 \wedge {\cal B}_2 = \eta_{IJ}\, {\cal B}_1{}^I {\cal B}_2{}^J \,. \label{eq:6Dwedge2}
\end{equation}
Moreover, the theory features 25 scalar fields, parameterising the coset space $\SO{5,5}/(\SO{5} \times \SO{5})$, which are most conveniently described by a group-valued generalised metric ${\cal M}_{MN}$ in the spinor representation, or by a group-valued generalised metric ${\cal M}_{IJ}$ in the vector representation.
The ExFT dynamics comes from a pseudo-action \cite{Abzalov:2015ega}, which has to be supplemented by first order duality and self-duality equations among the $p$-form field strengths
\begin{equation}
\begin{split}
{\cal J}_{\mu\nu\rho\sigma\,M} &= \frac1{2} \sqrt{|{\cal G}|} \epsilon_{\mu\nu\rho\sigma\kappa\lambda} \gM_{MN}\, {\cal F}^{\kappa\lambda}{}^N \,,
\\
{\cal H}_{\mu\nu\rho\,I} &= - \frac1{3!} \sqrt{|{\cal G}|} \epsilon_{\mu\nu\rho\sigma\kappa\lambda} \eta_{IJ} {\cal M}^{JK} \,{\cal H}^{\sigma\kappa\lambda}{}_K \,,
\label{eq:6DSelfDuality}
\end{split}
\end{equation}
where ${\cal G}_{\mu\nu}$ is the external metric which is used to raise/lower the external indices on the field strengths and $|{\cal G}|$ is its determinant.

For details about the ExFT actions and field equations, we refer to \cite{Abzalov:2015ega,Musaev:2015ces}. In appendices \ref{A:SL5} and \ref{A:SO55}, we collect/derive the details of the dictionaries between the ExFT fields and the original IIA/IIB supergravity fields.

\section{Half-maximal AdS vacua from ExFT} \label{s:AdSReview}
Generic supersymmetric AdS vacua of 10-/11-dimensional SUGRA have non-trivial fluxes. Since ExFT unifies fluxes and geometry into generalised tensor fields, it leads to a natural description of supersymmetric AdS vacua that is largely analogous to special holonomy spaces in Riemannian geometry, as shown in \cite{Malek:2017njj} for the case of 16 supercharges, and in \cite{Ashmore:2016qvs} for 8 supercharges. Thus, having a supersymmetric AdS$_D \times M$ vacuum is equivalent to the existence of a nowhere vanishing set of generalised tensor fields on $M$ subject to certain algebraic compatibility conditions and differential conditions. These conditions ensure that $M$ admits appropriate Killing spinors for the AdS$_D$ vacuum. As shown in \cite{Malek:2017njj}, for supersymmetric AdS$_6$ and AdS$_7$ vacua the relevant generalised tensors are $d-1$ generalised vector fields $J_u \in \Gamma\left({\cal R}_1\right)$, with $u = 1, \ldots, d-1$, and a generalised tensor field $\hK \in \Gamma\left({\cal R}_{D-4}\right)$, where $d = 11 - D$ and $D$ denotes the dimension of the AdS vacuum. Here we denote by ${\cal R}_p$ the generalised vector bundle whose fibres are $R_p$, as listed in \eqref{eq:7DTH} and \eqref{eq:6DTH}. These generalised tensors must satisfy the algebraic conditions
\begin{equation}
 \begin{split}
  \J_u \wedge \J_v - \frac1{d-1} \delta_{uv} \J_w \J^w &= 0 \,, \\
  \J_u \wedge \J^u \wedge \hK > 0 \,, \\
  \hK \otimes \hK \vert_{R_c} = 0 \,,
 \end{split} \label{eq:AlgConditions}
\end{equation}
with the ExFT $\wedge$ product defined in \eqref{eq:7Dwedge}, \eqref{eq:7Dwedge2} and \eqref{eq:6Dwedge1}, \eqref{eq:6Dwedge2}, the $u, v = 1, \ldots, d-1$ indices raised and lowered by $\delta_{uv}$ and $R_c = \emptyset$ for $D = 7$ and $R_c = \mathbf{1}$ for $D = 6$. This set of generalised tensors $\J_u$ and $\hK$ defines a $\GH = \SO{d-1}$ structure, because it is stabilised by $\SO{d-1} \subset E_{d(d)}$. This ensures the existence of well-defined spinors on $M_{int}$ carrying 16 supercharges\footnote{In 6 dimensions, the above description is equivalent to having 16 non-chiral supercharges. It is also possible to have a chiral set of 16 supercharges in 6 dimensions, which requires having a different set of generalised tensors \cite{Malek:2017njj}. However, there are no chirally supersymmetric AdS$_6$ vacua, and so we will not comment further on this possibility.}, and we will therefore also call the set $\J_u$, $\hK$ satisfying \eqref{eq:AlgConditions} a ``half-maximal structure''. The commutant of $\GH$ within the maximal compact subgroup of $E_{d(d)}$ is itself given by $\SO{d-1}_R$ which acts as a R-symmetry group, rotating the well-defined spinors into each other, and similarly the generalised vector fields $\J_u$. As we will show in section \ref{s:GenMetric}, one can express the generalised metric, i.e. the scalar fields on $M$, in terms of a $\SO{d-1}_R$-invariant combination of the half-maximal structure, $\J_u$ and $\hK$.

To ensure that the well-defined spinors are Killing spinors of the supersymmetric AdS$_{6,7}$ vacua, the half-maximal structure $\J_u$, $\hK$ must satisfy the following differential constraints \cite{Malek:2017njj}
\begin{equation}
 \begin{split}
  \gL_{\J_u} \J_v &= - \Lambda_{uvw} J^w \,, \\ 
  \gL_{\J_u} \hK &= 0 \,, \\
  d\hK &= \left\{ \begin{array}{c}
  \frac{1}{3!3\sqrt{2}} \epsilon^{uvw} \Lambda_{uvw} \J_x \wedge \J^x \,, \quad \textrm{when D = 7} \,, \\
  \frac{1}{9} \epsilon_{uvwx} \Lambda^{uvw} \J^x \,, \quad \, \textrm{ when D = 6} \,,
 \end{array} \right.
\end{split} \label{eq:DiffConditions}
\end{equation}
where the generalised Lie derivatives, $\gL_{\J_u} \J_v$, $\gL_{J_u} \hK$, and the $d\hK$ operator are as defined in equations \eqref{eq:gen_diff}, \eqref{eq:gen_diff_gen}, \eqref{eq:7Dd} and \eqref{eq:6Dd}. For $D = 7$, i.e. in the $\SL{5}$ ExFT, the explicit expressions for the generalised Lie derivatives appearing in the first two equations of \eqref{eq:DiffConditions} are
\begin{equation}
 \begin{split}
  \gL_{J_u} J_v{}^{ab} &= \frac12 J_u{}^{cd} \partial_{cd} J_v{}^{ab} - 2 J_v{}^{c[b} \partial_{cd} J_u{}^{a]d} + \frac12 J_v{}^{ab} \partial_{cd} J_u{}^{cd} \,, \\
  \gL_{J_u} \hK^a &= \frac12 J_u{}^{bc} \partial_{bc} \hK^a - \hK^b \partial_{bc} J_u{}^{ac} + \frac12 \hK^a \partial_{bc} J_u{}^{bc} \,,
 \end{split}
\end{equation}
while for $D = 6$, i.e. in the $\SO{5,5}$ ExFT, they are
\begin{equation}
 \begin{split}
  \gL_{J_u} J_v{}^M &= J_u{}^N \partial_N J_v{}^M - J_v{}^N \partial_N J_u{}^M + \frac12 \left(\gamma_I\right)^{MN} \left( \gamma^I \right)_{PQ} J_v{}^P \partial_N J_u{}^Q \,, \\
  \gL_{J_u} \hK^I &= J_u{}^M \partial_M \hK^I + \frac12 \hK^J \left( \gamma_J \right)_{MN} \left( \gamma^I \right)^{NP} \partial_P J_u{}^M \,.
 \end{split}
\end{equation}

The objects $\Lambda_{uvw}$ appearing in \eqref{eq:DiffConditions} are totally antisymmetric constants which imply that the $\J_u$'s generate a $\SU{2}_R$ algebra with respect to the generalised Lie derivative and that the $\hK$ is invariant under this $\SU{2}_R$ symmetry \cite{Malek:2017njj}. The cosmological constant, $\Lambda$, of the AdS$_{6,7}$ vacuum is encoded in $\Lambda_{uvw}$ as
\begin{equation}
 \Lambda_{uvw} \Lambda^{uvw} \sim - \Lambda \,,
\end{equation}
up to numerical factors which we will fix in sections \ref{s:AdS7} and \ref{s:AdS6} by comparing with known supersymmetric AdS$_{6,7}$ vacua. From \eqref{eq:DiffConditions}, we see that $\Lambda_{uvw}$ breaks the $\SO{d-1}_R$ symmetry of the half-maximal structure to $\SU{2}_R$, the R-symmetry of the supersymmetric AdS$_{6,7}$ vacua.

Moreover, \eqref{eq:DiffConditions} implies that the vector fields, $\J_u$, generate, via the generalised Lie derivative, a $\SU{2}_R \subset \SO{d-1}_R$ rotation on the $J_u$'s themselves and leave $\hK$ invariant. As we will make explicit in the next section, the generalised metric $\gM$ is constructed from $\SO{d-1}_R$-invariant combinations of $\J_u$ and $\hK$ and thus
\begin{equation}
 \gL_{J_u} \gM = 0 \,. \label{eq:GenKilling}
\end{equation}
Thefore, the $J_u$ are generalised Killing vector fields of the background. As made explicit in appendices \ref{A:SL5} and \ref{A:SO55} generalised vector fields consist of formal sums of spacetime vector fields and differential forms. Equation \eqref{eq:GenKilling} implies that either the spacetime vector fields in $\J_u$ are Killing vector fields of the spacetime metric and leave the SUGRA field strengths invariant \cite{Malek:2017njj}, or that some of the $\J_u$ contain a vanishing spacetime vector field component and consist of only exact differential forms. We call a generalised Killing vector of the latter type a trivial Killing vector field. As discussed in more detail in \cite{Malek:2017njj}, for AdS$_7$ vacua we see that the $\SU{2}_R$ symmetry must be generated by the three spacetime vector fields of $\J_u$, $u = 1, \ldots, 3$. On the other hand, for AdS$_6$ vacua three of the $\J_u$'s contain spacetime Killing vector fields that generate the $\SU{2}_R$ symmetry, while the fourth generalised vector field
\begin{equation}
 J_T \equiv \frac{1}{3!} \epsilon^{uvwx} \Lambda_{uvw} J_x \,,
\end{equation}
is given by
\begin{equation}
 J_T = \frac32 d\hK \,,
\end{equation}
which implies that it is a trivial generalised Killing vector field. In fact, it satisfies
\begin{equation}
 \gL_{J_T} = 0 \,,
\end{equation}
when acting on any generalised tensor. We will make use of these general properties of $J_u$ and $\hK$ when constructing supersymmetric AdS$_{6,7}$ vacua in section \ref{s:AdS7} and \ref{s:AdS6}.

Finally, one can define the following generalised tensor fields from the half-maximal structure $\J_u$ and $\hK$ which will be useful to us
\begin{equation}
\J_u \wedge \J_v = \delta_{uv} \K \,, \qquad \K \wedge \hK = \kappa^{D-2} \,, \qquad \hJ_u = J_u \wedge \hK \,, \label{eq:Kkappa}
\end{equation}
where $\K \in \Gamma\left({\cal R}_2\right)$ and $\kappa$ is a scalar density of weight $\frac{1}{D-2}$.

\subsection{Minimal consistent truncation} \label{s:RevConsTruncation}
One benefit of constructing or describing half-maximal AdS$_D$ vacua by the structures $\J_u$ and $\hK$ is that we immediately obtain a ``minimal'' consistent truncation around the vacuum to a half-maximal $D$-dimensional gauged SUGRA containing only the gravitational supermultiplet \cite{Malek:2017njj}. This is therefore a proof and an explicit realisation of the (half-maximal subcase of the) conjecture that such a consistent truncation exists for any supersymmetric warped AdS vacuum of 10-/11-dimensional SUGRA \cite{Gauntlett:2007ma}. Moreover, the usually highy non-linear truncation Ansatz is given by a simple linear factorisation Ansatz on the ExFT structures. If we denote by $Y^M$ the internal coordinates and by $x^\mu$ the $D$-dimensional external coordinates, then the truncation Ansatz (of the purely internal fields from the $D$-dimensional perspective) is given by \cite{Malek:2016bpu,Malek:2017njj}
\begin{equation}
 \begin{split}
  {\cal J}_u(x,Y) &= X^{-1}(x)\, J_u(Y) \,, \\
  {\cal \hK}(x,Y) &= X^2(x)\, \hK(Y) \,.
 \end{split} \label{eq:TruncAnsatz}
\end{equation}
Here $X(x)$ is the scalar field of the $D$-dimensional half-maximal SUGRA. For each value of the scalar field $X(x) > 0$, ${\cal J}_u(x,Y)$ and $\hK(x,Y)$ satisfy the algebraic conditions \eqref{eq:AlgConditions} and thus a half-maximal structure. This guarantees that the theory obtained after truncation is half-maximally supersymmetric. However, for $X \neq 1$, the differential conditions \eqref{eq:DiffConditions} defining the AdS vacuum are no longer satisfied. Therefore, at $X \neq 1$, the theory will not have a supersymmetric AdS vacuum. Finally, as shown in \cite{Malek:2017njj}, the differential conditions \eqref{eq:DiffConditions} ensure that the truncation Ansatz \eqref{eq:TruncAnsatz} is consistent.

We will show in section \ref{s:GenMetric} how to construct the generalised metric from the half-maximal structure. By constructing the generalised metric of ${\cal J}_u$ and ${\cal \hK}$ and using the dictionary between ExFT and SUGRA, given in appendices \ref{A:EFT-IIAdictionary} and \ref{A:IIBGM}, we thus obtain the non-linear truncation Ansatz for the internal supergravity fields.

For the fields of the ExFT tensor hierarchy, the truncation Ansatz is as follows \cite{Malek:2017njj}. For the ExFT vector fields, we have
\begin{equation}
 {\cal A}_{\mu}(x,Y) = A_\mu{}^u(x)\, \J_u(Y) \,.
\end{equation}
In $D = 7$, the truncation Ansatz for the remaining fields is
\begin{equation}
 \begin{split}
  {\cal B}_{\mu\nu}(x,Y) &= - B_{\mu\nu}(x)\, \K(Y) \,, \\
  {\cal C}_{\mu\nu\rho}(x,Y) &= C_{\mu\nu\rho}(x)\, \hK(Y) \,, \\
  {\cal D}_{\mu\nu\rho\sigma}(x,Y) &= D_{\mu\nu\rho\sigma}{}^u(x)\, \hJ_u(Y) \,,
 \end{split}
\end{equation}
where $A_\mu{}^u(x)$, $B_{\mu\nu}(x)$, $C_{\mu\nu\rho}(x)$ and $D_{\mu\nu\rho\sigma}{}^u(x)$ are the fields of the 7-dimensional half-maximal gravitational supermultiplet. In particular, $A_\mu{}^u$ are the 3 vector fields, $B_{\mu\nu}$ are the 2-forms, $C_{\mu\nu\rho}$ are the 3-forms dual to $B_{\mu\nu}$, and $D_{\mu\nu\rho\sigma}{}^u$ are the 4-forms dual to the vector fields. The duality relations between these half-maximal gauged SUGRA fields comes from the duality relations \eqref{eq:D7DualityRelations} between the ExFT field strengths \eqref{eq:ExFTFieldStrength}, \eqref{eq:ExFTFieldStrengths}. Finally, the truncation Ansatz for the external 7-D ExFT metric is
\begin{equation}
 {\cal G}_{\mu\nu}(x,Y) = G_{\mu\nu}(x)\, \kappa^2(Y) \,,
\end{equation}
with $G_{\mu\nu}(x)$ the metric of the half-maximal gauged SUGRA.

Similarly, in $D=6$, the truncation Ansatz for the tensor hierarchy field is
\begin{equation}
 \begin{split}
  {\cal B}_{\mu\nu}(x,Y) &= B_{\mu\nu}(x)\, \hK(Y) - \tilde{B}_{\mu\nu}(x)\, \K(Y) \,, \\
  {\cal C}_{\mu\nu\rho}(x,Y) &= C_{\mu\nu\rho}{}^u(x) \, \hJ_{u}(Y) \,.
 \end{split} \label{eq:6DAnsatz}
\end{equation}
Now, $A_\mu{}^u$ are the 4 vector fields of the gravitational supermultiplet, while $B_{\mu\nu}$ is its 2-form. $\tilde{B}_{\mu\nu}$ is the dual 2-form and $C_{\mu\nu\rho}{}^u$ are the 3-forms dual to the vector fields. Once again, the relationship between these objects arises from the duality relation \eqref{eq:6DSelfDuality} between the ExFT field strengths \eqref{eq:ExFTFieldStrength}, \eqref{eq:ExFTFieldStrengths}. Using the truncation Ansatz \eqref{eq:6DAnsatz} and differential conditions \eqref{eq:DiffConditions}, we find that the field strengths factorise as
\begin{equation}
 \begin{split}
  {\cal F}_{\mu\nu}(x,Y) &= \tilde{F}_{\mu\nu}{}^u(x)\, J_u(Y) \,, \\
  {\cal H}_{\mu\nu\rho}(x,Y) &= \tilde{F}_{\mu\nu\rho}(x)\, \hK(Y) - \tilde{G}_{\mu\nu\rho}(x)\, \K(Y) \,,
 \end{split}
\end{equation}
where
\begin{equation}
 \begin{split}
  \tilde{F}_{\mu\nu}{}^u &= 2 \partial_{[\mu} A_{\nu]}{}^u + \Lambda^{uvw} A_{\mu\,v}\, A_{\nu\,w} - \frac19 \epsilon^{uvwx} \Lambda_{vwx}\, B_{\mu\nu} \,, \\
  \tilde{F}_{\mu\nu\rho} &= 3 \partial_{[\mu} B_{\nu\rho]} \,, \\
  \tilde{G}_{\mu\nu\rho} &= 3 \partial_{[\mu} \tilde{B}_{\nu\rho]} + 3 A_{[\mu}{}^u \partial_{\nu} A_{\rho]\,u} + \Lambda_{uvw}\, A_\mu{}^u A_\nu{}^v A_\rho{}^w - \frac19 \epsilon_{uvwx} \Lambda^{uvw}\, C_{\mu\nu\rho}{}^x \,,
 \end{split} \label{eq:6dFieldStrengthMinimal}
\end{equation}
and similarly for the higher field strengths of the ExFT. We will use this to derive the duality relations between $B_{\mu\nu}$ and $\tilde{B}_{\mu\nu}$ explicitly in section \ref{s:GenMetricSO55}. Similar to $D=7$, the truncation Ansatz for the external ExFT metric  is
\begin{equation}
 {\cal G}_{\mu\nu}(x,Y) = \sqrt{2}\, G_{\mu\nu}(x)\, \kappa^2(Y) \,,
\end{equation}
with $G_{\mu\nu}(x)$ the 6-dimensional gauged SUGRA metric.

\section{Consistent truncations with matter multiplets} \label{s:VecTruncation}
As shown in \cite{Malek:2017njj}, half-maximal consistent truncations with $N$ vector multiplets require a further reduction of the structure group to $\SO{d-1-N} \subset \SO{d-1} \subset \EG{d}$, as well as differential conditions on the tensors defining the $\SO{d-1-N}$ structure. In order to have a $\SO{d-1-N}$ structure, we require $d-1+N$ generalised vector fields, $J_a$, where $a = 1, \ldots, d-1+N$, satisfying
\begin{equation}
\J_a \wedge \J_b = \eta_{ab} K \,, \label{eq:VecAlgCon}
\end{equation}
in addition to the $\hK$ as in \eqref{eq:AlgConditions}. Here $\eta_{AB}$ is a constant $\SO{d-1,N}$ invariant metric and $K$ is defined as in \eqref{eq:Kkappa},
\begin{equation}
K = \frac1{d-1} J_u \wedge J^u \,.
\end{equation}
Therefore, given the $d-1$ generalised vector fields defining the half-maximal structure of the AdS vacuum \eqref{eq:AlgConditions}, we must have a further $N$ generalised vector fields, one for each vector multiplet. Labelling these extra generalised vector fields by $\bar{u} = 1, \ldots, N$, the algebraic conditions \eqref{eq:VecAlgCon} become
\begin{equation}
\begin{split}
\J_{\bar{u}} \wedge \J_u &= 0 \,, \\
\J_{\bar{u}} \wedge \J_{\bar{v}} &= - \delta_{\bar{u}\bar{v}} K \,.
\end{split} \label{eq:VecExtraAlgCon}
\end{equation}
Since these algebraic conditions must hold point-wise, it is easy to show that we can only have $N \leq d-1$ vector multiplets in a consistent truncation.

Moreover, for the truncation around the supersymmetric AdS vacuum to be consistent, the $\SO{d-1-N}$ structure must satisfy the differential conditions
\begin{equation}
\begin{split}
\gL_{J_a} J_b &= - f_{ab}{}^c\, J_c \,, \\
\gL_{J_a} \hK &= 0 \,,
\end{split} \label{eq:VecDiffCon}
\end{equation}
where $f_{abc} = f_{ab}{}^d \eta_{dc}$ are totally antisymmetric structure constants with $f_{uvw} = \Lambda_{uvw}$ and $f_{uv\bar{w}}=0$. Here we are considering a special case of the more general conditions given in \cite{Malek:2017njj} because we want to ensure that the truncation contains a supersymmetric AdS vacuum. The differential conditions \eqref{eq:VecDiffCon} can be thought of as the higher-dimensional analogue of the conditions imposed on the embedding tensor of 6-/7-dimensional half-maximal gauged SUGRA in \cite{Louis:2015mka,Karndumri:2016ruc}.

For what follows, it's useful to note that the first condition of \eqref{eq:VecDiffCon} implies that the extra generalised vector fields form a representation under the R-symmetry group 
\begin{equation}
\gL_{J_u} J_{\bar{v}} = -f_{u\bar{v}}{}^{\bar{w}}J_{\bar{w}}\,.
\end{equation}
Together with the fact that there can be only $N \leq d - 1$ vector multiplets, this will allow us to fully classify the possible consistent truncations with vector multiplets in sections \ref{s:VecTruncationAdS7} and \ref{s:VecTruncationAdS6}.

\subsection{Truncation Ansatz} \label{s:VecTruncationAnsatz}
As shown in \cite{Malek:2017njj}, given the $d-1+N$ vector fields satisfying \eqref{eq:VecAlgCon} and \eqref{eq:VecDiffCon}, we obtain a consistent truncation by expanding the fields of the ExFT as follows.

For the scalar sector, we expand the background $\SO{d-1}$ structure in terms of the $\SO{d-1-N}$ structure as
\begin{equation}
\begin{split}
{\cal J}_u(x,Y) &= X^{-1}(x)\, b_u{}^a(x)\, J_a(Y) \,, \\
{\cal \hK}(x,Y) &= X^2(x)\, \hK(Y) \,.
\end{split} \label{eq:VecScalarAnsatz}
\end{equation}
The fields $b_u{}^a$ must satisfy
\begin{equation}
b_u{}^a b_v{}^b \eta_{ab} = \delta_{uv} \,,
\end{equation}
and are identified up to $\SO{d-1}$ rotations acting on the $u, v$ indices. Therefore, they parameterise the coset space
\begin{equation}
b_u{}^a \in \frac{\SO{d-1,N}}{\SO{d-1}\times \SO{N}}\,,
\end{equation}
and together with $X \in \mathbb{R}^+$ they form the scalar manifold of half-maximal gauged SUGRA coupled to $N$ vector multiplets
\begin{equation}
M_{scalar} =  \frac{\SO{d-1,N}}{\SO{d-1}\times \SO{N}} \times \mathbb{R}+ \,.
\end{equation}
Using the formulae of section \ref{s:GenMetric}, in which we show how to construct the generalised metric from the half-maximal structure, we can then translate the above truncation Ans\"{a}tze into the non-linear truncation Ans\"{a}tze of the internal SUGRA fields.

For $D=7$, the remaining fields of the ExFT are expanded as
\begin{equation}
\begin{split}
{\cal A}_{\mu}(x,Y) &= A_\mu{}^a(x)\, J_a(Y) \,, \\
{\cal B}_{\mu\nu}(x,Y) &= -B_{\mu\nu}(x)\, \K(Y) \,, \\
{\cal G}_{\mu\nu}(x,Y) &= G_{\mu\nu}(x)\, \kappa^2(Y) \,,
\end{split} \label{eq:VecD7Ansatz}
\end{equation}
where $A_\mu{}^a$ are the $3+N$ vector fields, $B_{\mu\nu}$ are the two-form fields and $G_{\mu\nu}$ the metric of the seven-dimensional half-maximal gauged SUGRA.

For $D=6$, the other fields of the ExFT are expanded as
\begin{equation}
 \begin{split}
  {\cal A}_{\mu}(x,Y) &= A_\mu{}^a(x)\, J_a(Y) \,, \\
  {\cal B}_{\mu\nu}(x,Y) &= B_{\mu\nu}(x)\, \hK(Y) - \tilde{B}_{\mu\nu}(x)\, \K(Y) \,, \\
  {\cal C}_{\mu\nu\rho}(x,Y)&= C_{\mu\nu\rho}{}^a(x)\, \hJ_a(Y)\,, \\
  {\cal G}_{\mu\nu}(x,Y) &= \sqrt{2}\, G_{\mu\nu}(x)\, \kappa^2(Y) \,,
 \end{split} \label{eq:VecD6Ansatz}
\end{equation}
Here $G_{\mu\nu}$ is the metric, $A_\mu{}^a$ are the $4+N$ vector fields, $B_{\mu\nu}$ are the two-form fields and their duals $\tilde{B}_{\mu\nu}$ of the six-dimensional half-maximal gauged SUGRA. $C_{\mu\nu\rho}{}^a$ are the 3-form fields dual to the $A_\mu{}^a$, which appear via St\"{u}ckelberg coupling in the field strength of $\tilde{B}_{\mu\nu}$. To see this, one can compute the ExFT field strengths \eqref{eq:ExFTFieldStrengths}. Using the truncation Ansatz \eqref{eq:VecD6Ansatz} and the differential conditions \eqref{eq:DiffConditions} we find
\begin{equation}
 \begin{split}
  {\cal F}_{\mu\nu}(x,Y) &= \tilde{F}_{\mu\nu}{}^a(x)\, J_a(Y) \,, \\
  {\cal H}_{\mu\nu\rho}(x,Y) &= \tilde{F}_{\mu\nu\rho}(x)\, \hK(Y) - \tilde{G}_{\mu\nu\rho}(x)\, \K(Y) \,,
 \end{split}
\end{equation}
where
\begin{equation}
 \begin{split}
  \tilde{F}_{\mu\nu}{}^a &= 2 \partial_{[\mu} A_{\nu]}{}^a + f^a{}_{bc} A_\mu{}^b A_\nu{}^c + \frac23 \Lambda^a B_{\mu\nu} \,, \\
  \tilde{F}_{\mu\nu\rho} &= 3 \partial_{[\mu} B_{\nu\rho]} \,, \\
 \tilde{G}_{\mu\nu\rho} &= 3 \partial_{[\mu} \tilde{B}_{\nu\rho]} + 3 A_{[\mu}{}^a \partial_{\nu} A_{\rho]}{}^b \eta_{ab} + \Lambda_{uvw}\, A_\mu{}^u A_\nu{}^v A_\rho{}^w + \frac23 \Lambda_a\, C_{\mu\nu\rho}{}^a \,,
\end{split} \label{eq:6dFieldStrengthGeneral}
\end{equation}
where we defined
\begin{equation}
 \Lambda^a = \left( \Lambda^u,\, \Lambda^{\bar{u}} \right) = \left( -\frac1{3!} \epsilon^{uvwx} \Lambda_{vwx},\, 0 \right) \,.
\end{equation}
Clearly $F_{\mu\nu}{}^a$ are the field strengths of the 6-dimensional half-maximal gauged SUGRA whose gauge group is determined by the structure constants $f_{abc}$ and $F_{\mu\nu\rho}$ is the field strength of the two-form $B_{\mu\nu}$ of the gauged SUGRA. Using the ExFT/SUGRA dictionary of appendix \ref{A:EFT-IIB-dictionary-TH}, we can use the above formulae to read off the consistent truncation Ans\"{a}tze for the SUGRA fields.

\section{Generalised metric from the half-maximal structure} \label{s:GenMetric}
To obtain expressions for the AdS vacua and their consistent truncations in terms of SUGRA fields, we need to know how the SUGRA fields are encoded in the ExFT objects used in the truncation Ans\"{a}tze of sections \ref{s:RevConsTruncation} and \ref{s:VecTruncationAnsatz}. The SUGRA fields with at least one external leg are encoded in the ExFT tensor hierarchy fields ${\cal A}_\mu$, ${\cal B}_{\mu\nu}$, etc. and can be determined in the usual fashion via the SUGRA / ExFT dictionary, which we give for the $\SO{5,5}$ case in appendix \ref{A:EFT-IIB-dictionary-TH}. However, the purely internal SUGRA fields are encoded in the generalised metric of ExFT, via the dictionary we give in appendices \ref{A:EFT-IIAdictionary} and \ref{A:IIBGM}, and therefore we must know how to obtain a generalised metric from the half-maximal structure.

Firstly, it is clear that one can construct a generalised metric from $J_u$ and $\hK$. Just like on a $d$-dimensional manifold, a Riemannian metric defines a $\SO{d} \subset \GL{d}$ structure, a generalised metric defines a (generalised) $H_d \subset E_{d(d)}$ structure, where $H_d$ is the maximal compact subgroup of $E_{d(d)}$. On the other hand, $\J_u$ and $\hK$ define a $\GH = \SO{d-1} \subset H_d$ structure and, thus, $\J_u$ and $\hK$ provide \emph{more} information than the generalised metric. In ExFT, the generalised metric parameterises the coset space
\begin{equation}
\gM_{MN} \in \frac{\EG{d}}{H_d} \,.
\end{equation}
Since $\J_u$ and $\hK$ are by construction invariant under $\GH = \SO{d-1} \subset H_d$, we must construct $\gM_{MN}$ using an $\SO{d-1}_R$-invariant combination of $\J_u$ and $\hK$. Therefore, the generalised metric must be given by
\begin{equation}
 \gM_{MN} = A\, \kappa^{6-2D} \hJ_{u\,M} \hJ^{u}{}_N + B\, \kappa^{4-D} \hK_{MN} + C\, \epsilon^{u_1\ldots u_{d-1}} \left( J_{u_1} \ldots J_{u_{d-1}} \right)_{MN} \,. \label{eq:GenMetricSchem}
\end{equation}
The factors of $\kappa$ are chosen so that $\gM_{MN}$ has no weight under generalised diffeomorphisms and $A$, $B$ and $C$ are coefficients which are fixed by requiring $\gM_{MN}$ to be an element of $\EG{d}$. The final term schematically denotes an appropriate product of $\left(R_1\right)^{d-1} \longrightarrow R_1 \otimes R_1$. In the following subsections we will give the explicit expressions for the case of $\SL{5}$ ExFT and $\SO{5,5}$ ExFT.

\subsection{Generalised metric in $\SL{5}$ ExFT} \label{s:GenMetricSL5}
In $\SL{5}$ ExFT, the generalised metric is often used either in the $R_1 = \mbf{10}$ representation or its dual representation, or in the fundamental representation, $R_2 = \mbf{5}$, of $\SL{5}$. The two are related by \cite{Berman:2011cg}
\begin{equation}
 \gM_{ab,cd} = 2 \gM_{a[c}\gM_{d]b} \,,
\end{equation}
where $a, b = 1, \ldots, 5$ denote fundamental $\SL{5}$ indices. It will be useful to have explicit expressions for both representations.

The generalised metric in the $\mbf{10}$ representation of $\SL{5}$ is given as in \eqref{eq:GenMetricSchem} which now explicitly becomes
\begin{equation}
 \begin{split}
  \gM_{ab,cd} = A \, \kappa^{-8} \hJ_{u\,ab} \hJ^{u}{}_{cd} + B\, \kappa^{-3} \epsilon_{abcde} \hK^e + C\, \kappa^{-3} \epsilon^{uvw} \epsilon_{abefg} \epsilon_{cdhij} \J_{u}{}^{ef} \J_{v}{}^{hi} \J_w{}^{gj} \,,
 \end{split}
\end{equation}
where $\hJ_{u\,ab}$ is defined as in \eqref{eq:Kkappa}, explicitly
\begin{equation}
 \hJ_{u\,ab} = \frac14 \epsilon_{abcde} \J_u{}^{cd} \hK^e \,.
\end{equation}
Requiring this to be an $\SL{5}$ element fixes $A = 8 \sigma^2$, $B = -\sigma^2$ and $C = -\frac{\sigma^3}{6\sqrt{2}}$, up to a coefficient $\sigma$. Note that the minimal consistent truncation \eqref{eq:TruncAnsatz} corresponds precisely to a rescaling $\sigma \longrightarrow \sigma\, X$. $\sigma$ is determined by the differential conditions \eqref{eq:DiffConditions} and can therefore be fixed by comparison of AdS vacua obtained from the half-maximal structures to known AdS$_7$ vacua, for example the maximally supersymmetric AdS$_7 \times S^4$ vacua of 11-d SUGRA. This way we find $\sigma = 1$. Thus, the generalised metric and its inverse in the $\mbf{10}$ and $\obf{10}$ representations are given by
\begin{equation}
 \begin{split}
  \gM_{ab,cd} &= 8 \, \kappa^{-8} \hJ_{u\,ab} \hJ^{u}{}_{cd} - \kappa^{-3} \epsilon_{abcde} \hK^e - \frac{1}{6\sqrt{2}} \kappa^{-3} \epsilon^{uvw} \epsilon_{abefg} \epsilon_{cdhij} \J_{u}{}^{ef} \J_{v}{}^{hi} \J_w{}^{gj} \,, \\
  \gM^{ab,cd} &= 2 \, \kappa^{-2} \J_{u}{}^{ab} \J^{u,cd} - \kappa^{-2} \epsilon^{abcde} \K_e - \frac{2\sqrt{2}}{3} \kappa^{-12} \epsilon^{uvw} \epsilon^{abefg} \epsilon^{cdhij} \hJ_{u\,ef} \hJ_{v\,hi} \hJ_{w\,gj} \,.
 \end{split} \label{eq:SL510GM}
\end{equation}

Similarly, one can show that the generalised metric and its inverse in the $\mbf{5}$ and $\obf{5}$ representations of $\SL{5}$ are given by
\begin{equation}
 \begin{split}
  \gM_{ab}& = \kappa^{-4} \left( \K_a \K_b + \frac{4\sqrt{2}}{3} \kappa^{-5} \, \epsilon^{uvw} \hJ_{u,ac} \hJ_{v,bd} \J_w{}^{cd} \right) \,, \\
  \gM^{ab}& = \kappa^{-6} \left( \hat{\K}^a \hat{\K}^b+  \frac{2\sqrt{2}}{3} \, \epsilon^{uvw}  \J_u{}^{ac} \J_v{}^{bd} \hJ_{w,cd} \right) \,.
 \end{split} \label{eq:SL55GM}
\end{equation}

\subsection{Generalised metric in $\SO{5,5}$ ExFT} \label{s:GenMetricSO55}
In $\SO{5,5}$ ExFT, the generalised metric is often used either in the spinor or vector representation of $\SO{5,5}$. In the fundamental representation, the generalised metric ${\cal M}_{IJ}$ must satisfy
\begin{equation}
 \gM_{IK} \gM_{JL} \eta^{KL} = \eta_{IJ} \,,
\end{equation}
where $I = 1, \ldots, 10$ labels the $\mbf{10}$ representation of $\SO{5,5}$ and $\eta_{IJ}$ is the constant $\SO{5,5}$-invariant metric with which the $I, J$ indices are raised/lowered. The generalised metric in the $\mbf{16}$ is related to that in the $\mbf{10}$ by
\begin{equation}
 \gM_{MP} \gM_{NQ} \left(\gamma_I\right)^{MN} \gM^{IJ} = \left(\gamma^J\right)_{PQ} \,,
\end{equation}
where $M = 1, \ldots, 16$ label the $\mbf{16}$ representation and $\left(\gamma_I\right)^{MN}$ and $\left(\gamma_I\right)_{MN}$ are the $\SO{5,5}$ $\gamma$-matrices satisfying
\begin{equation}
 \left( \gamma_I\right)^{MP} \left( \gamma_J \right)_{NP} + \left( \gamma_J \right)^{MP} \left( \gamma_I \right)_{NP} = 2\, \eta_{IJ} \delta^{M}_P \,.
\end{equation}

We thus find the generalised metric and its inverse in the $\mbf{16}$ are given by
\begin{equation}
 \begin{split}
  \gM_{MN} &= \frac{1}{\sqrt{2}} \left(4\, \kappa^{-6}\, \hJ^u{}_M \hJ_{u\,N} - \kappa^{-2} \left(\gamma^I\right)_{MN} \hK_I \right. \\
  & \quad \left. -\frac1{4!}\, \kappa^{-6} \epsilon^{uvwx} \left(\gamma_I\right)_{MP} \left(\gamma_J\right)_{NQ} \left(\gamma^{IJ}\right)^{S}{}_R \J_u{}^P \J_v{}^Q \J_w{}^R \hJ_{x,S} \right) \,, \\
  \gM^{MN} &= \frac{1}{\sqrt{2}} \left( 2\, \kappa^{-2} \J_u{}^M \J^{u\,N} - \kappa^{-2} \left(\gamma_I\right)^{MN} \K^I \right. \\
  & \left. \quad - \frac{2}{4!} \kappa^{-10} \epsilon_{uvwx} \left(\gamma_I\right)^{MP} \left(\gamma_J\right)^{NQ} \left(\gamma^{IJ}\right)^{S}{}_R \hJ^u{}_P \hJ^v{}_Q \J_w{}^R \hJ_{x,S} \right) \,,
 \end{split} \label{eq:SO55GenMetric1}
\end{equation}
where $\hJ_{u\,M}$ is defined in \eqref{eq:Kkappa}, and is given explicitly by
\begin{equation}
 \hJ^u{}_M =\frac12 \left(\gamma^I\right)_{MN} \hK_I \J^{u\,N} \,.
\end{equation}
Similarly, the generalised metric in the $\mbf{10}$ is given by
\begin{equation}
 \gM_{IJ} = \left( \frac{1}{4!} \epsilon^{uvwx} \left(\gamma_{IK} \right)_{M}{}^{N} \left( \gamma_J{}^K \right)_{P}{}^{Q} J_u{}^M \hJ_{v,N} \J_w{}^P \hJ_{x,Q} + \kappa^{-4}\, \K_I \K_J + \kappa^{-4} \hK_I \hK_J \right) \,. \label{eq:SO55GenMetric2}
\end{equation}
Just as in $\SL{5}$, there is a scaling degree of freedom which is generated by the minimal consistent truncation \eqref{eq:TruncAnsatz}. Thus, the coefficients above can only be defined once those in the differential conditions \eqref{eq:DiffConditions} are fixed, or vice versa. Once one set of coefficients is fixed, the other can be obtained either by comparison with known AdS vacua, by a careful analysis of the ExFT BPS equations or by reducing the ExFT action to that of gauged SUGRA upon applying the consistent truncation.

We can now explicitly compute the relationship between $F_{\mu\nu\rho}$ and $\tilde{F}_{\mu\nu\rho}$ in \eqref{eq:6dFieldStrengthGeneral}. Using the expression for the generalised metric \eqref{eq:SO55GenMetric2} and the scalar truncation Ansatz \eqref{eq:VecScalarAnsatz}, we find
\begin{equation}
\begin{split}
\langle \gM^{IJ} \rangle \hK_J &= X^{-4}(x)\, K^I(Y) \,, \\
\langle \gM^{IJ} \rangle \K_J &= X^4(x)\, \hK^I(Y) \,,
\end{split}
\end{equation}
where $\langle \gM_{IJ} \rangle$ denotes the generalised metric with the truncation Ansatz plugged in, i.e. that computed from ${\cal J}_u$, ${\cal K}$ and ${\cal \hK}$ of \eqref{eq:VecScalarAnsatz}. Therefore, the twisted self-duality equation \eqref{eq:6DSelfDuality} becomes
\begin{equation}
\begin{split}
\tilde{G}_{(3)} &= X^{-4} \star_6 \tilde{F}_{(3)} \,.
\end{split} \label{eq:6dDuality}
\end{equation}

\section{AdS$_7$ vacua from massive IIA supergravity} \label{s:AdS7}
As shown in \cite{Malek:2017njj} and reviewed in section \ref{s:AdSReview}, supersymmetric AdS$_7$ vacua are characterised by three nowhere-vanishing generalised vector fields $\J_u \in \Gamma\left({\cal R}_1\right)$, transforming as a triplet of $\SO{3}_R$, and a nowhere-vanishing generalised tensor $\hK \in \Gamma\left({\cal R}_3\right)$, transforming as a singlet of $\SO{3}_R$. The differential conditions involve a constant totally antisymmetric 3-index tensor $\Lambda_{uvw}$ which therefore takes the form
\begin{equation}
 \Lambda_{uvw} = \sqrt{-c}\, \epsilon_{uvw} \,,
\end{equation}
where the constant $c$ is related to minus the seven-dimensional cosmological constant. The precise relation between $c$ and the cosmological constant, or, equivalently, the AdS$_7$ radius, can be found from the ExFT BPS equations and using the formula for the generalised metric \eqref{eq:SL510GM}, or by comparison to known AdS$_7$ vacua of 10-/11-dimensional SUGRA. By comparing to the AdS$_7 \times S^4$ vacuum of 11-dimensional SUGRA, we find $\Lambda_{uvw} = 2\, \sqrt{2}\,R^{-1} \epsilon_{uvw}$, where $R$ is the AdS$_7$ radius. Plugging this into the differential conditions \eqref{eq:DiffConditions}, they become
\begin{equation}
 \begin{split}
  \gL_{\J_u} \J_v &= -\frac{2\sqrt{2}}{R}\, \epsilon_{uvw} \J^w\,, \\
  \gL_{\J_u} \hK &= 0 \,, \\
  d\hK &= \frac{2}{R}\, \K \,,
 \end{split} \label{eq:SL5DC}
\end{equation}
where $\K$ is defined via
\begin{equation}
 \J_u \wedge \J_v = \delta_{uv} \, \K \,.
\end{equation}

\subsection{Half-maximal structure}
Here we are interested in studying AdS$_7$ vacua of massive IIA supergravity. As we discussed in section \ref{s:AdSReview}, the $\SO{d-1}_R = \SO{3}_R$ symmetry must be generated by spacetime Killing vectors. This suggests that the vacua are of the form AdS$_7 \times S^2 \times I$, with the Killing vectors on $S^2$ generating the $\SU{2}_R$ symmetry. As we explain in appendix \ref{A:IIASL5}, the generalised vector fields $\J_u$ and tensor $\hK$ are formal sums of internal spacetime vector fields and differential forms as follows
\begin{equation}
\begin{split}
\J_u &= V_u + \lambda_u + \sigma_u + \phi_u \,, \\
\hK &= \homega_{(0)} + \homega_{(2)} + \homega_{(3)} \,,
\end{split} \label{eq:JhKdecomp}
\end{equation}
where $V_u$, $\lambda_u$, $\sigma_u$ and $\phi_u$ are the vector, 1-form, 2-form and scalar parts of $\J_u$, while $\homega_{(p)}$ are the $p$-forms appearing in $\hK$. Similarly, $\K = \frac13 \J_u \wedge \J^u \in \Gamma\left({\cal R}_2\right)$ is a formal sum of differential forms
\begin{equation}
\K = \bomega_{(0)} + \bomega_{(1)} + \bomega_{(3)} \,, \label{eq:Kdecomp}
\end{equation}
where $\bomega_{(p)}$ are $p$-forms.

In terms of the above, the wedge products \eqref{eq:7Dwedge}, \eqref{eq:7Dwedge2} appearing in the algebraic conditions \eqref{eq:AlgConditions} become
\begin{equation}
\begin{split}
\J_u \wedge \J_v &= 2 \imath_{V_{(u}} \lambda_{v)} - 2 \left( \lambda_{(u} \phi_{v)} + \imath_{V_{(u}} \sigma_{v)} \right) - 2 \lambda_{(u} \wedge \sigma_{v)} \,, \\
\hK \wedge \K &= \homega_{(0)} \bomega_{(3)} + \homega_{(1)} \wedge \bomega_{(2)} + \homega_{(3)} \bomega_{(0)} \,,
\end{split} \label{eq:mIIAAlgConditions}
\end{equation}
while the quadratic algebraic constraint on $\hK$ is automatically fulfilled for $\SL{5}$.

The differential operators appearing in the differential conditions \eqref{eq:SL5DC} are modified as described in \cite{Ciceri:2016dmd,Cassani:2016ncu} to capture the Roman's mass, $m$, of massive IIA SUGRA. We explain in detail how to do this in appendix \ref{A:SL5Massive}. Including the Roman's mass, and thus using equations \eqref{eq:DefGenLie}, \eqref{eq:DefGenLie5}, \eqref{eq:Defd}, the differential operators appearing in the conditions \eqref{eq:SL5DC} become
\begin{equation}
\begin{split}
\gL_{J_u} J_v &= L_{V_u} V_v + L_{V_u} \lambda_v + L_{V_u} \sigma_v + L_{V_u} \phi_v \\
& \quad + \imath_{V_v} \left( m \lambda_u - d\phi_u \right) - \imath_{V_v} \left( d\lambda_u \right) - \imath_{V_v} \left( d\sigma_u \right) + \phi_v \left( d\lambda_u \right) + \lambda_v \wedge \left( m \lambda_u - d\phi_u \right) \,, \\
\gL_{J_u} \hK &= L_{V_u} \homega_{(0)} + L_{V_u} \homega_{(2)} + L_{V_u} \omega_{(3)} \\
& \quad - \homega_{(0)} \left( d\lambda_u \right) - \homega_{(0)} \left( d\sigma_u \right) - \homega_{(2)} \wedge \left( m \lambda_u - d\phi_u \right) \,, \\
d\hat{K} &= - d\homega_{(0)} + d \homega_{(2)} \,. \label{eq:mIIADiffConditions2}
\end{split}
\end{equation}

To describe supersymmetric AdS$_7$ vacua, we must therefore find the vector fields and differential forms satisfying the above algebraic and differential conditions. In doing so, we will use the differential equations
\begin{equation}
 \gL_{J_u} J_v = - \frac{2\sqrt{2}}{R} \epsilon_{uvw} J^w \,, \qquad \gL_{J_u} \hK = 0 \,,
\end{equation}
as a guiding principle. These imply that the $J_u$'s must transform as a triplet under $\SU{2}_R$ and $\hK$ as a singlet under $\SU{2}_R$, which as we discussed before is generated by the Killing vector fields on $S^2$. Therefore, we will construct $J_u$ out of spacetime tensors on $S^2 \times I$ that are triplets of $\SU{2}_R$ as generated by the Killing vectors on $S^2$, and similarly $\hK$ out of differential forms that are singlets of $\SU{2}_R$.

In fact, the above decomposition \eqref{eq:JhKdecomp}, \eqref{eq:Kdecomp} of the generalised tensors in terms of vector fields and differential forms on the internal space is only true locally, because the generalised tangent bundles are twisted by the internal gauge potentials of the IIA supergravity, in this case the three-form potential $C$, two-form potential, $B$, and one-form potential, $A$. The gauge potentials mix the different components of the generalised tensors, for example, if $A = B = 0$ but $C \neq 0$, then
\begin{equation}
 \sigma_u = \hsigma_u + \imath_{V_u} C\,, \qquad \omega_{(3)} = \homega_{(3)} + \omega_{(0)}\,C \,, \qquad  \bomega_{(3)} = \hat{\bomega}_{(3)} - \bomega_{(0)}\, C \,, \label{eq:3FormTwist}
\end{equation}
where $\hsigma_u$, $\homega_{(3)}$, $\hat{\bomega}_{(3)}$ are the globally well-defined 2-forms and 3-forms, respectively, while $\sigma_u$, $\omega_{(3)}$ and $\bomega_{(3)}$ are only local 2-forms and 3-forms. Therefore, to construct the $J_u$ and $\hK$ we must understand what the possible form of the gauge potentials is. Since their field strengths must be invariant under the $\SU{2}_R$ symmetry, the gauge potentials must take the form
\begin{equation}
 dB = R^3\, f(z)\, vol_{S^2} \wedge dz \,, \qquad dA = R^2\, l(z)\, vol_{S^2} \,,
\end{equation}
for some functions $f(z)$ and $l(z)$, where $z$ labels the coordinate on the interval $I$ and $vol_{S^2}$ is the volume form on $S^2$, see also appendix \ref{A:S2} for our $S^2$ conventions. $C$ is always pure gauge since the internal space is three-dimensional. Moreover, we can choose a gauge such that
\begin{equation}
 B = R^3\, F(z)\, vol_{S^2} \,,
\end{equation}
with $\frac{d F(z)}{dz} = f(z)$, so that $B$ is constructed from well-defined differential forms on $S^2$ and $I$. On the other hand, $A$ cannot be written in terms of well-defined differential forms on $S^2$ since it necessarily breaks the R-symmetry. This implies that we can automatically cater for the twisting by the two-form potential by writing the most general $J_u$ and $\hK$ built of out spacetime tensors on $S^2$ and $I$. On the other hand, the twist by $A$ will break the $\SU{2}_R$ symmetry and, therefore, we will keep track of it explicitly.

In particular, we will write $\phi_u = \hphi_u + \imath_{V_u} A$ and $\sigma_u = \hsigma_u + \lambda_u \wedge A$ and $\homega_{(3)} = \comega_{(3)} + \homega_{(2)} \wedge A$, where $\hphi_u$, $\hsigma_u$ and $\homega_{(3)}$ are spacetime tensors on $S^2 \times I$ that respect the $\SU{2}_R$ symmetry. In terms of the hatted objects, the differential operators appearing in the differential conditions become
\begin{equation}
\begin{split}
\gL_{J_u} J_v &= L_{v_u} V_v + L_{V_u} \lambda_v + L_{V_u} \sigma_v + L_{V_u} \hphi_v \\
& \quad + \imath_{V_v} \left( m \lambda_u + \imath_{V_u} dA - d\hphi_u \right) - \imath_{V_v} \left( d\lambda_u \right) - \imath_{V_v} \left( d\hsigma_u - \lambda_u \wedge dA \right) + \phi_v \left( d\lambda_u \right) \\
& \quad + \lambda_v \wedge \left( m \lambda_u + \imath_{V_u} dA - d\phi_u \right) + \left( L_{V_u} \lambda_v - \imath_{V_v} d\lambda_u \right) \wedge A + \imath_{[V_u,V_v]} A \,, \\
\gL_{J_u} \hK &= L_{V_u} \homega_{(0)} + L_{V_u} \homega_{(2)} + L_{V_u} \comega_{(3)} + L_{V_u} \homega_{(2)} \wedge A - \homega_{(0)} d\lambda_u \wedge A \\
& \quad - \homega_{(0)} \left( d\lambda_u \right) - \homega_{(0)} \left( d\hsigma_u - \lambda_u \wedge dA \right) - \homega_{(2)} \wedge \left( m \lambda_u + \imath_{V_u} dA - d\hphi_u \right) \,, \\
d\hat{K} &= - d\homega_{(0)} + d \homega_{(2)} \,. \label{eq:mIIADiffConditionsTwisted}
\end{split}
\end{equation}

The most general $\J_u$ we can construct that is compatible with the $\SU{2}_R$ symmetry is
\begin{equation}\label{eq:J_u-ansatz-AdS7}
 \begin{split}
  \J_u &= \frac{2\sqrt{2}}{R} v_u + \frac{R}{4} \left( k(z)\, y_u\, dz + g(z)\, dy_u + r(z)\, \theta_u \right) - \frac{R}{2} p(z)\,y_u \\
  & \quad + \frac{R^3}{16\sqrt{2}} \left( n(z)\,y_u\,vol_{S^2} + h(z)\, \theta_u \wedge dz + v(z)\, dy_u \wedge dz \right) \\
  & \quad + \frac{2\sqrt{2}}{R} \imath_{v_u} A + \frac{R}{4} \left( k(z)\, y_u\, dz + g(z)\, dy_u + r(z)\, \theta_u \right) \wedge A \,,
 \end{split}
\end{equation}
where $k(z)$, $g(z)$, $p(z)$, $n(z)$, $h(z)$, $v(z)$ and $r(z)$ are at this stage arbitrary functions of $z$, the coordinate on $I$, $y_u$ are a triplet of functions on $S^2$, $v_u$ are the Killing vector fields on $S^2$ and $\theta_u$ are 1-forms on $S^2$. The objects on $S^2$ are defined in appendix \ref{A:S2}. The algebraic conditions impose
\begin{equation}
 \begin{split}
  \J_u &= \frac{2\sqrt{2}}{R} v_u + \frac{R}{4} \left( -\frac{h(z)}{p(z)} y_u\, dz + g(z)\, dy_u \right) - \frac{R}{2} p(z)\,y_u \\
  & \quad + \frac{R^3}{16\sqrt{2}} \left( p(z)\,g(z)\,y_u\,vol_{S^2} + h(z)\, \theta_u \wedge dz + v(z)\, dy_u \wedge dz \right) \\
  & \quad + \frac{2\,\sqrt{2}}{R} \imath_{v_u} A + \frac{R}{4} \left( - \frac{h(z)}{p(z)} y_u\, dz + g(z)\, dy_u \right) \wedge A \,,
 \end{split} \label{eq:AdS7JWAlg}
\end{equation}
such that $K$ defined via
\begin{equation}
 \J_u \wedge \J_v = \delta_{uv}\, K \,,
\end{equation}
is given by
\begin{equation}
 K = - \frac{R^2}4 h(z)\,dz + \frac{R^4\,g(z)\,h(z)}{32\sqrt{2}} vol_{S^2} \wedge dz \,.
\end{equation}
Furthermore, the most general $\hK$ constructed from R-symmetry singlets is given by
\begin{equation}
 \begin{split}
  \hK &= \frac{R}{2}\,s(z) + \frac{R^3}{16\sqrt{2}} \left( g(z)\,s(z) - t(z) \right) vol_{S^2} + R^5\, u(z) vol_{S^2} \wedge dz \\
  & \quad + \frac{R^3}{16\sqrt{2}} \left(g(z)\,s(z) - t(z) \right) vol_{S^2} \wedge A \,.
 \end{split} \label{eq:AdS7hK}
\end{equation}
The algebraic condition $\J_u \wedge \J^u \wedge \hK > 0$ then becomes
\begin{equation}
 \frac{R^5}{64\sqrt{2}}\, h(z)\,t(z)\, vol_{S^2} \wedge dz > 0 \,.
\end{equation}
While this suggests that we must have $h(z)\, t(z) > 0$, this is not true at the endpoints of the interval parameterised by $z$. There we can in fact have $h(z)\,t(z) = 0$. Thus, we must impose
\begin{equation}
 h(z)\, t(z) \geq 0 \,, \label{eq:AdS7Reg}
\end{equation}
with possible equality at the boundary. This ensures that the metric is non-singular everywhere. For holographic applications, we also want to impose that the internal space is compact by requiring that the $S^2$ shrinks at the endpoints of $I$, which will further refine \eqref{eq:AdS7Reg}. However, to determine the precise conditions, we must first construct the SUGRA fields of the AdS$_7$ solution.

With \eqref{eq:AdS7JWAlg} and \eqref{eq:AdS7hK}, the differential conditions \eqref{eq:mIIADiffConditionsTwisted} reduce to
\begin{equation}
 \begin{split}
  m \lambda_u + \imath_{V_u} dA - d\hphi_u &= 0 \,, \\
  d\lambda_u &= 0 \,, \\
  d\hsigma_u - \lambda_u \wedge dA &= 0 \,, \\
  d\omega_{(0)} + \frac2R \bomega_{(1)} &= 0 \,, \\
  d\omega_{(2)} - \frac2R \bomega_{(3)} &= 0 \,,
 \end{split}
\end{equation}
where, as we discussed above, $R$-symmetry implies that
\begin{equation}
dA = R^2 l(z)\, vol_{S^2} \,. \label{eq:7DdA}
\end{equation}
Explicitly, the differential conditions imply the following set of ODEs
\begin{equation}
 \dot{g} = - \frac{h}{p} \,, \qquad 2\,p\,\dot{p}= m\,h \,, \qquad \dot{s} = h \,, \qquad \dot{t} = - \frac{h\,s}{p} \,, \qquad l = - \frac{p}{4 \sqrt{2}} - \frac{m\,g}{8\sqrt{2}} \,. \label{eq:7DGenODE}
\end{equation}

Note that the functions $u(z)$ and $v(z)$ do not appear in the differential conditions. This is due to the fact that they can be removed by gauge transformations of the gauge potentials $A$ and $C$, as can be seen from \eqref{eq:3FormTwist} and \eqref{eq:J_u-ansatz-AdS7}. Thus, we can, and will, set $u = v = 0$ without loss of generality.

Using the ODEs \eqref{eq:7DGenODE} and having set $u = v = 0$ by gauge transformations, the half-maximal structures simplify to
\begin{equation}
 \begin{split}
  \J_u &= \frac{2\,\sqrt{2}}{R} v_u + \frac{R}{4} d \left(g\,y_u\right) - \frac{R}{2}\, p \,dy_u + \frac{R^3}{16\sqrt{2}} p \left( d\left(g\,\theta_u\right) - g\,y_u\, vol_{S^2}\right) \\
  & \quad + \frac{2\,\sqrt{2}}{R} \imath_{v_u} A + \frac{R}{4} d\left(g\,y_u\right) \wedge A \,, \\
  \hK &= \frac{R}{2}\,s + \frac{R^3}{16\sqrt{2}} \left( g\, s - t \right) vol_{S^2} + \frac{R^3}{16\sqrt{2}} \left( g\,s - t \right) vol_{S^2} \wedge A \,,
 \end{split} \label{eq:AdS7StrucG}
\end{equation}
with
\begin{equation}
dA = - \frac{R^2}{4\sqrt{2}} \left( p + \frac{m}{2}\,g\right) vol_{S^2} \,.
\end{equation}

Moreover, we can redefine the $z$ coordinate to set $h(z)$ to anything we like. There are two convenient choices that help us solve the ODEs \eqref{eq:7DGenODE}.

\paragraph{Choice 1} The first is to take $h(z) = p(z)$ so that we can integrate the equation $\dot{g} = - \frac{h}{p}$ to set $g = - z$, where we absorb the integration constant by shifting $z$. Then, the remaining ODEs are solved by
\begin{equation}
 s = - \dot{t} \,, \qquad p = - \ddot{t} \,, \qquad l = \frac{\ddot{t}}{4\sqrt{2}} - \frac{m\,g}{8\sqrt{2}} \,.
\end{equation}
where $t(z)$ must satisfy
\begin{equation}\label{eq:condition-tddd-AdS7}
 \dddot{t} = - \frac{m}{2} \,.
\end{equation}
Finally, with this gauge, the regularity condition \eqref{eq:AdS7Reg} becomes
\begin{equation}
 t(z)\, \ddot{t}(z) \leq 0 \,. \label{eq:AdS7RegG}
\end{equation}

\paragraph{Choice 2} The second choice is to simply take $h(z) = 1$ and integrate the equation $\dot{s} = 1$ to set $s = z$ without loss of generality. The remaining ODEs now become
\begin{equation}
 \begin{split}
  \dot{g} &= - \frac{1}{p} \,, \qquad 2\,p\,\dot{p} = m \,, \qquad \dot{t} = -\frac{z}{p} \,, \qquad l = - \frac{p}{4\sqrt{2}} - \frac{m\,g}{8\sqrt{2}} \,.
 \end{split}
\end{equation}
The functions $g$ and $t$ are therefore determined in terms of $p$, its integral and its derivatives, and where $p$ satisfies
\begin{equation}
 \frac{\partial p^2}{\partial z} = m \,.
\end{equation}
The regularity condition \eqref{eq:AdS7Reg} becomes
\begin{equation}
 t(z) \geq 0 \,,
\end{equation}
with equality only possible at $\partial I$.

To compare to the literature, especially the form of AdS$_7$ solutions given in \cite{Apruzzi:2015zna}, it is worthwhile to introduce
\begin{equation}
q = \frac{p}{\sqrt{2}} \,, \qquad \by = \frac{9}{4} z \,, \qquad \sqrt{\beta} = \frac{81}{4\sqrt{2}} t \,,
\end{equation}
which now satisfy
\begin{equation}
\frac{\partial q^2}{\partial \by} = \frac92 m \,, \qquad q = - 4\,\by\, \frac{\sqrt{\beta}}{\beta'} \,, \label{eq:betacond}
\end{equation}
where $'$ is our shorthand notation for $\frac{\partial}{\partial \by}$. Moreover, we note the following identities
\begin{equation}
\frac{t}{p} = - \frac{1}{81} \frac{\beta'}{\by} \,, \qquad p\,t = -\frac{32}{81} \frac{\beta\,\by}{\beta'} \,, \qquad z^2 + 2\,p\,t = \frac{16}{81} \by \left( \by - \frac{4\beta}{\beta'} \right) \,, \label{eq:AdS7Identities}
\end{equation}
which will allow us to recover precisely the description of AdS$_7$ vacua given in \cite{Apruzzi:2015zna}. 

\subsection{The supersymmetric AdS$_7$ vacua}
It is now straightforward to compute the SUGRA fields of the supersymmetric AdS$_7$ vacua. We first plug $\J_u$ and $\hK$, given in \eqref{eq:AdS7StrucG}, into the generalised metric. We then use the ExFT / IIA dictionary worked out in \cite{Malek:2015hma} and summarised in appendix \ref{A:EFT-IIAdictionary} to read off the supergravity fields. In string frame, the warp factor of the AdS$_7$ part of the metric is given by \cite{Malek:2016bpu,Malek:2017njj}
\begin{equation}
 f_7 = |g_{int}|^{-1/5} \kappa^{2}\, e^{4\psi/5} \,,
\end{equation}
where $|g_{int}|$ is the determinant of the internal metric, $\psi$ is the dilaton and $\kappa^5 = \frac13 J_u \wedge J^u \wedge \hK$.

Without fixing $h(z)$ we therefore find the following SUGRA fields in string frame
\begin{equation}
 \begin{split}
  ds_{10}^2 &= \sqrt{\frac{t}{p}} ds_{AdS_7}^2 + \frac{R^2}{8} \sqrt{\frac{t}{p}} \left( \frac{p\,t}{s^2 + 2\,p\,t} ds_{S^2}^2 + \frac{h^2}{p\,t} dz^2 \right) \,, \\
  e^{\psi} &= \frac{2}{R} \left( \frac{t}{p} \right)^{3/4} \frac{1}{\sqrt{s^2 + 2\,p\,t}} \,, \\
  B &= \frac{R^2}{8\sqrt{2}} \left( - g + \frac{s\,t}{s^2 + 2\,p\,t} \right) vol_{S^2} \,,
 \end{split} \label{eq:AdS7GenGauge}
\end{equation}
with 2-form field strength $F_2 = dA - m\,B_2$ and 3-form field strength $H = dB$ given by
\begin{equation}
 \begin{split}
  F_2 &= - \frac{R^2}{8\sqrt{2}} \left( 2\,p + \frac{m\,s\,t}{s^2 + 2\,p\,t} \right) vol_{S^2} \,, \\
  H_3 &= \frac{R^2}{8\sqrt{2}} \frac{h\,t}{p} \left( \frac{3\,p}{s^2 + 2\,p\,t} - \frac{m\,s\, t}{\left(s^2 + 2\,p\,t\right)^2} \right) vol_{S^2} \wedge dz \\
  & = \frac{2}{R} \left( 3 \left(\frac{t}{p}\right)^{-1/4}  - \frac{m\,s}{s^2 + 2\,p\,t} \left(\frac{t}{p}\right)^{3/4} \right) vol_{M_3} \,,
 \end{split}
\end{equation}
where $vol_{M_3}$ denotes the volume form on the internal manifold with the metric \eqref{eq:AdS7GenGauge}. Note that we have the opposite sign convention for our $B$ field to \cite{Apruzzi:2013yva} so that the Bianchi identity of $F_2$ is
\begin{equation}
dF_2 = - m H_3 \,.
\end{equation}

\paragraph{Choice 1} With the choice $h(z) = p(z)$, the expressions for the AdS$_7$ vacua \eqref{eq:AdS7GenGauge} reduce to
\begin{equation}
 \begin{split}
  ds_{10}^2 &= \sqrt{- \frac{t}{\ddot{t}}}\, ds_{AdS_7}^2 + \frac{R^2}{8} \sqrt{- \frac{\ddot{t}}{t}} \left( \frac{t^2}{\dot{t}^2 - 2\, \ddot{t}\, t} ds_{S^2}^2 + d\z^2 \right) \,, \\
  e^{\psi} &= \frac{2}{R} \left(- \frac{t}{\ddot{t}}\right)^{3/4} \frac{1}{\sqrt{\dot{t}^2 - 2\, \ddot{t} \, t}} \,, \\
  B &= \frac{R^2}{8\sqrt{2}} \left( \z - \frac{\dot{t}\, t}{ \dot{t}^2 - 2 \, \ddot{t} \, t} \right) vol_{S^2} \,,
 \end{split} \label{eq:AdS7Gauge1}
\end{equation}
with field strengths
\begin{equation}
 \begin{split}
  F_2 &= \frac{R^2}{8\sqrt{2}} \left( 2\, \ddot{t} + \frac{m\, \dot{t}\, t}{ \dot{t}^2 - 2 \, \ddot{t} \, t} \right) vol_{S^2} \,, \\
  H_3 &= \frac{R^2}{8\sqrt{2}} \left( \frac{m\,t^2\,\dot{t}}{\left(\dot{t}^2 - 2\,t\,\ddot{t}\right)^2} - \frac{ t\, \ddot{t}}{\dot{t}^2 - 2\,t\,\ddot{t}} \right) vol_{S^2} \wedge dz \\
  &= \frac{2}{R} \left[ 3 \left(-\frac{t}{\ddot{t}}\right)^{-1/4} + \frac{m\,\dot{t}}{\dot{t}^2 - 2\,t\,\ddot{t}} \left( - \frac{t}{\ddot{t}}\right)^{3/4} \right] vol_{M_3} \,,
 \end{split}
\end{equation}
where the function $t$ satisfies
\begin{equation}
 \begin{split}
  \dddot{t} = - \frac{m}{2} \,, \qquad t \geq 0 \,, \label{eq:tcond}
 \end{split}
\end{equation}
with $t = 0$ at $\partial I$ so that the internal space has no boundaries. For every such function $t$ there is a supersymmetric AdS$_7$ vacuum of massive IIA supergravity. This matches the infinite family of supersymmetric AdS$_7$ vacua of \cite{Apruzzi:2013yva} when we set the AdS radius to $R = 2$, and where our variables are related to those of \cite{Cremonesi:2015bld} by a rescaling
\begin{equation}
 t = \frac{4\,\sqrt{2}}{81} \alpha \,, \qquad z = 2 \sqrt{2}\,\pi\, \bar{z} \,,
\end{equation}
where we denote the ``$z$'' coordinate of \cite{Cremonesi:2015bld} by $\bar{z}$ to distinguish it from our $z$ coordinate.

\paragraph{Choice 2} With the choice $h(z) = 1$ and using \eqref{eq:AdS7Identities}, the AdS$_7$ vacua \eqref{eq:AdS7GenGauge} are given by
\begin{equation}
 \begin{split}
  ds_{10}^2 &= \frac19 \sqrt{-\frac{\beta'}{\by}} ds_{AdS_7}^2 + \frac19 \sqrt{-\frac{\beta'}{\by}} R^2 \left( \frac{\beta/4}{4\,\beta - \beta'\,\by} ds_{S^2}^2 - \frac1{16} \frac{\beta'd\by^2}{\beta\,\by} \right) \,, \\
  e^{\psi} &= R^{-1} \frac{\left(-\beta'/\by\right)^{5/4}}{6\sqrt{4\,\beta - \beta'\by}} \,, \\
  H_3 &= \frac{18}{R} \left(-\frac{\by}{\beta'}\right)^{1/4} \left( 1 - \frac{m}{108\,\by} \frac{\left(\beta'\right)^2}{4\,\beta - \beta'\by} \right) vol_{M_3} \,, \\
  F_2 &= \frac{R^2\,\by}{4} \frac{\sqrt{\beta}}{\beta'} \left( 4 + \frac{m}{18\,\by} \frac{\left(\beta'\right)^2}{4\,\beta - \beta'\by} \right) vol_{S^2} \,,
 \end{split} \label{eq:AdS7Gauge2}
\end{equation}
Here $vol_{M_3}$ is the internal volume form with respect to the full internal metric in \eqref{eq:AdS7Gauge2}, and $\beta$ satisfies the ODE
\begin{equation}
 \frac{\partial q^2}{\partial \by} = \frac92 m \,, \quad \textrm{with }\, q = - 4\,\by\, \frac{\sqrt{\beta}}{\beta'} \,.
\end{equation}
This matches the AdS vacua in the coordinates of \cite{Apruzzi:2015zna} when the AdS radius is set to $R = 2$.

\section{AdS$_6$ vacua from IIB supergravity} \label{s:AdS6}
As we reviewed in section \ref{s:AdSReview} and was shown in \cite{Malek:2017njj}, supersymmetric AdS$_6$ vacua are described in ExFT by four nowhere-vanishing generalied vector fields $\J_u \in \Gamma\left({\cal R}_1\right)$, transforming as a $\mbf{4}$ of $\SO{4}$, and a nowhere-vanishing generalised tensor $\hK \in \Gamma\left({\cal R}_3\right)$ which is invariant under $\SO{4}$.

Upon defining $\Lambda^u = -\frac{1}{3!} \epsilon^{uvwx} \Lambda_{vxw}$, the differential conditions \eqref{eq:DiffConditions} become
\begin{equation}
 \begin{split}
  \gL_{\J_u} \J_v &= - \epsilon_{uvwx} \J^w \Lambda^x \,, \\
  \gL_{\J_u} \hK &= 0 \,, \\
  d\hK &= \frac23 \Lambda^u \J_u \,,
 \end{split} \label{eq:SO55DiffConditions}
\end{equation}
We can use a $\SO{4}_R$ rotation to write, without loss of generality,
\begin{equation}
\Lambda_u = \left( 0, 0, 0, \frac{3}{\sqrt{2}\,R} \right) \,, \label{eq:SO55Lambda}
\end{equation}
with $R$ the AdS$_6$ radius. The numerical coefficient in front of $R$ have been fixed by comparing the solutions with known supersymmetric AdS$_6$ vacua of IIB \cite{DHoker:2016ujz}.

$\Lambda_{uvw}$ breaks the $\SO{4}$ symmetry to the $\SO{3}_R$ R-symmetry of AdS$_6$ vacua. Let us therefore write $u = \left(A, 4\right)$ with $A = 1, 2, 3$ labelling the vector representation of $\SO{3}_R$. With respect to $\left(A, 4\right)$ the differential conditions become
\begin{equation}
\begin{split}
\gL_{\J_A} \J_B &= - \frac{3}{\sqrt{2}\,R}\, \epsilon_{ABC} \J^C \,, \\
\gL_{\J_A} \J_4 &= 0 \,, \\
\gL_{\J_A} \hK &= 0 \,, \\
d\hK &= \frac{\sqrt{2}}{R}\, \J_4 \,. \label{eq:SO55RsymmDC}
\end{split}
\end{equation}
Note that the conditions $\gL_{\J_4} \J_u = 0$ and $\gL_{\J_4} \hK = 0$ are automatically satisfied by $\J_4 \propto d\hK$ \cite{Malek:2017njj}.

\subsection{Half-maximal structure}\label{sec:Half-max-str-AdS6}
We will now construct the half-maximal structures on the internal space that yields an AdS$_6$ vacuum. To do this, we will guide ourselves by the differential equations \eqref{eq:SO55RsymmDC} determining the AdS vacuum. Recall from section \ref{s:AdSReview} that these imply that the $\J_u$'s are generalised Killing vector fields and therefore either consist of a Killing vector field plus a compensating gauge transformation, or consist of a trivial gauge transformation. The latter can be written as $dB$ for some $B \in \Gamma\left({\cal R}_2\right)$ and will always generate a vanishing generalised Lie derivative on any vector field. We see from \eqref{eq:SO55RsymmDC} that $\J_4$ generates such a trivial gauge transformation, while $\J_A$ must generate the $\SU{2}_R$ symmetry of the AdS vacuum and therefore have non-vanishing vector components which generate this symmetry. The generalised tenors $\hK$ and $\J_4$ must be invariant under this R-symmetry.

To generate the $\SU{2}_R$ symmetry we take the internal space to contain an $S^2$ and on the remaining two-dimensional space, the Riemann surface $\Sigma$, we introduce coordinates $x^\alpha$, $\alpha = 1, \ldots, 2$. We will raise/lower $\alpha$ in a Northwest/Southeast convention by the $\SL{2}$-invariants $\epsilon_{\alpha\beta} = \pm 1$ and $\epsilon^{\alpha\beta} = \pm 1$ with
\begin{equation}
 \epsilon^{\alpha\gamma} \epsilon_{\beta\gamma} = \delta^\alpha_\beta \,.
\end{equation}
Thus we write
\begin{equation}
 x^\alpha = \epsilon^{\alpha\beta} x_\beta \,, \qquad x_\alpha = x^\beta \epsilon_{\beta\alpha} \,.
\end{equation}

In IIB SUGRA with the conventions in appendix \ref{A:IIBSO55}, the $J_u$'s and $\hK$ become formal sums of spacetime vector fields and differential forms as follows
\begin{equation}
\begin{split}
\J_u &= V_u + \lambda_u^{\alpha} + \sigma_u \,, \\
\hK &= \homega_{(0)}^\alpha + \homega_{(2)} + \homega_{(4)}^\alpha \,,
\end{split}
\end{equation}
where $V_u$, $\lambda_u{}^{\alpha}$ and $\sigma_u$ denote the vector, 1-form and 3-form parts of $\J_u$, while $\omega_{(p)}$ are $p$-forms appearing in $\hK$. With our conventions \ref{A:IIBSO55}, the wedge products and tensor products appearing in the algebraic conditions \eqref{eq:AlgConditions} become
\begin{equation}
\begin{split}
\J_u \wedge \J_v &= \sqrt{2} \left( \imath_{V_{(u}} \lambda_{v)}^{\alpha} + \lambda_{(u}^{\alpha} \wedge \sigma_{v)} + \left( - \imath_{V_{(u}} \sigma_{v)} - \frac12 \lambda_{u\,\beta} \wedge \lambda_v^\beta \right) \right) \,, \\
\hK \otimes \hK \vert_{R_c} &= \homega_{(2)} \wedge \homega_{(2)} + 2\, \homega_{(0)\alpha}\, \homega_{(4)}^\alpha \,, \\
\hK \wedge \K &= \homega_{(2)} \wedge \bomega_{(2)} + \homega_{(0)\alpha}\, \bomega_{(4)}^\alpha + \bomega_{(0)\alpha}\, \homega_{(4)}^\alpha \,,
\end{split} \label{eq:IIBAlgConditions}
\end{equation}
where we defined $K = \frac14 \J_u \wedge J^u = \bomega_{(0)}^\alpha + \bomega_{(2)} + \bomega_{(4)}^\alpha$. Moreover, the differential operators appearing in the differential conditions \eqref{eq:SO55DiffConditions} become
\begin{equation}
\begin{split}
\gL_{\J_u} \J_v &= L_{V_u} V_v + L_{V_u} \sigma_v + L_{V_u} \lambda_v^\alpha \\
& \quad - \imath_{V_v} d\lambda_u^\alpha - \imath_{V_v} d\sigma_u - \lambda_{v\,\beta} \wedge d\lambda_u^\beta \,, \\
\gL_{J_u} \hK &= L_{V_u} \homega_{(0)}^\alpha + L_{V_u} \homega_{(2)} + L_{V_u} \homega_{(4)}^\alpha \\
& \quad + \homega_{(0)\beta}\, d\lambda_{u}^\beta - \homega_{(0)}^\alpha\, d\sigma_{u} - \homega_{(2)} \wedge d\lambda_u^\alpha \,, \\
d\hat{K} &= - \sqrt{2}\, d\homega_{(2)} + \sqrt{2}\, d \homega_{(0)}^\alpha \,.
\end{split} \label{eq:IIBDiffConditions}
\end{equation}

As discussed above, the $\J_A$'s will need to be formed out of vector fields and differential forms forming $\SU{2}_R$ triplets, while $\J_4$ and $\hK$ will need to be constructed from $\SU{2}_R$-invariant vector fields and differential forms. We will now construct the most general $\J_A$ and $\hK$, up to gauge transformations, which transform as an $\SU{2}$-triplet and singlet, and satisfy their algebraic conditions. We then calculate $\J_4$ from $d\hK$ and impose its algebraic condition $\J_4 \wedge \J_4 = \frac13 \J_A \wedge \J_A$ and $\J_4 \wedge \J_A = 0$ and finally solve the remaining differential conditions
\begin{equation}
 \begin{split}
  \gL_{\J_A} \J_B &= - \frac{3}{\sqrt{2}\,R} \epsilon_{ABC} \J^C \,, \\
  \gL_{\J_A} \J_4 &= 0 \,, \\
  \gL_{\J_A} \hK &= 0 \,.
  \end{split} \label{eq:SO55FinalStep}
\end{equation}

Just like for AdS$_7$ vacua, we must first ascertain whether the gauge fields of the supergravity can be chosen in a way that respects the $\SU{2}_R$ symmetry and will thus naturally appear in the most general structures we write down, or whether the gauge fields necessarily break the $\SU{2}_R$ symmetry and need to be included by hand as a ``twist'' term. Since we are considering IIB SUGRA with an internal four-manifold, we will only have 3-form field strengths $dC^{\alpha}$ which must necessarily be $\SU{2}_R$ singlets. Therefore, they must be given by
\begin{equation}
 dC^\alpha = b^\alpha \wedge vol_{S^2} \,,
\end{equation}
for some 1-forms $b^\alpha$ on $\Sigma$. Therefore, we can choose a gauge such that locally
\begin{equation}
 C^\alpha = c^\alpha vol_{S^2} \,,
\end{equation}
for functions $c^\alpha$ on $\Sigma$ which are $\SU{2}_R$ singlets. Hence, the gauge potentials can be chosen to be $\SU{2}_R$ invariant and will naturally appear in the most general structures we write down. This is in contrast with the mIIA AdS$_7$ vacua we studied in section \ref{s:AdS7}, where the R-R 1-form potential had to be included via a ``twist'' term.

The most general $\J_A$ we can construct as an $\SU{2}_R$-triplet is
\begin{equation}
 \begin{split}
  \J_A &= \frac{1}{\sqrt{2}} \left( \frac{3}{R}\, v_A + 4\,c_6\,R\, k^\alpha dy_A + 4\,c_6\,R\, y_A m^{\alpha} + n^\alpha \theta_A + \frac{16\,c_6^2\,R^3}{3} y_A h \wedge vol_{S^2} \right. \\
  & \quad \left. + \frac{16\,c_6^2\,R^3}{3} l\, \theta_A \wedge vol_\Sigma + f dy_A \wedge vol_\Sigma \right) \,, 
 \end{split} \label{eq:SO55JAnsatz}
\end{equation}
where $v_A$ are the Killing vectors, $\theta_A$ are 1-forms and $vol_{S^2}$ is the volume form on $S^2$ as defined in appendix \ref{A:S2} and
\begin{equation}
 vol_\Sigma = \frac12 \epsilon_{\alpha\beta} dx^\alpha \wedge dx^\beta \,.
\end{equation}
$l$, $k^\alpha$ and $n^\alpha$ are at this stage arbitrary functions on $\Sigma$, while $h = h_\alpha dx^\alpha$ and $m^{\alpha} = m^\alpha{}_{\beta}\, dx^\beta$ are 1-forms on $\Sigma$. $c_6$ is a constant. It and the other numerical coefficients in front of $R$ have been introduced for later convenience.
We can further simplify \eqref{eq:SO55JAnsatz} by using generalised diffeomorphisms, i.e. a combination of diffeomorphisms and gauge transformations: we can use the generalised vector field
\begin{equation}
 V = \chi \wedge vol_{S^2} \,,
 \end{equation}
where $\chi$ is a one-form on $\Sigma$ satisfying
\begin{equation}
 d\chi = - \frac{R}{3} f vol_\Sigma \,,
\end{equation}
to remove the term in $J_A$ that depends on the function $f$ by acting with $\gL_{V} \J_A$. In fact by working out the explicit twisting of the generalised tangent bundle by gauge potentials, e.g. using appendix E of \cite{Ashmore:2015joa}, one sees that this generalised diffeomorphism corresponds to a gauge transformation of the R-R 4-form.

We now impose the algebraic conditions \eqref{eq:AlgConditions} such that the functions appearing in $J_A$ are now no longer all independent. As a result, we find
\begin{equation}
 \begin{split}
  \J_A &= \frac{1}{\sqrt{2}} \left(\frac{3}{R} \, v_A + 4\,c_6\,R \left( y_A m^{\alpha} + k^\alpha dy_A \right) + \frac{16\,c_6^2\,R^3}{3} \left( |m|\, \theta_A \wedge vol_\Sigma - y_A k^{\beta} m_{\beta} \wedge vol_{S^2} \right) \right) \,,
 \end{split}
\end{equation}
where $|m| = \frac12 m_{\alpha\beta} m^{\alpha\beta}$.

Next, we construct $\hK$ such that is an $\SU{2}_R$-invariant and satisfies $\hK \wedge \hK = 0$ and $J_A \wedge J^A \wedge \hK > 0$. We find the unique combination
\begin{equation}
 \begin{split}
 \hat{K} &= \frac{1}{\sqrt{2}} \left( 4\, c_6\, p_\alpha + \frac{2\,c_6 R^2}{3} q_\alpha vol_{S^2}\wedge vol_\Sigma - \frac{16\,c_6^2 R^2}{3} \left(r + p_\beta k^\beta\right) vol_2 + \frac{p_\beta q^\beta}{r + p_\gamma k^\gamma} vol_\Sigma \right)  \,, 
 \end{split} \label{eq:SO55KAnsatz}
\end{equation}
in terms of $r$, $p_\alpha$ and $q_\alpha$ which are so far arbitrary functions of $x^\alpha$. However, just as for $\J_A$ we can use gauge transformations to further simplify this expression. A particular class of gauge transformations corresponds to shifts of $\hK$ by $d$-exact terms,
\begin{equation}
 \hK \sim \hK + dQ \,,
\end{equation}
where $Q \in \Gamma\left({\cal R}_3\right)$. Taking
\begin{equation}
 Q = Q^{\alpha\beta} vol_{S^2} \wedge dx_\beta \,,
\end{equation}
with $\partial_\beta Q^{\alpha\beta} \sim q^\alpha$ (with appropriate coefficients) we see that we can remove the functions $q^\alpha$ in \eqref{eq:SO55KAnsatz}. Thus, we are left with the general $\hK$ up to gauge transformations given by
\begin{equation}
 \begin{split}
  \hat{K} &= \frac{1}{\sqrt{2}} \left( 4\,c_6\, p_\alpha - \frac{16\,c_6^2 R^2}{3} \left(r + p_\beta k^\beta\right) vol_2 \right) \,, 
 \end{split} \label{eq:SO55KSimple}
\end{equation}
The algebraic condition $J_u \wedge J^u \wedge \hK >0 $ is equivalent to $J_A \wedge J^A \wedge \hK > 0$ once we impose the remaining algebraic conditions. Therefore, we require
\begin{equation}
 J_A \wedge J^A \wedge \hK = 128\,c_6^4 R^4 r\, |m|\, vol_{S^2} \wedge vol_\Sigma > 0 \,, \label{eq:Ad6Reg}
\end{equation}
which implies that $r\,|m| \geq 0$ with equality at the points on $\Sigma$ where the $S^2$ degenerates.
From $\hK$ we find
\begin{equation}
 \J_4 = \frac{R}{\sqrt{2}} d \hK = \frac{1}{\sqrt{2}} \left( 4\,c_6 R\, dp^\alpha - \frac{16\, c_6^2 R^3}{3} d \left( r + p_\beta k^\beta \right) \wedge vol_{S^2} \right) \,.
\end{equation}
The algebraic conditions
\begin{equation}
 \J_4 \wedge \J_4 = \frac13 \J_A \wedge \J^A \,, \qquad \J_4 \wedge \J_A = 0 \,,
\end{equation}
now impose
\begin{equation}
 \begin{split}
  m_{\alpha} \wedge dp^\alpha &= 0 \,, \\
  m^\alpha \wedge m^\beta &= dp^\alpha \wedge dp^\beta \,, \\
  dr + p_\alpha dk^\alpha &= 0 \,.
 \end{split} \label{eq:SO55DiffCond1}
\end{equation}
Note that the final condition can be used to simplify the expression of $\J_4$
\begin{equation} 
 \J_4 = \frac{1}{\sqrt{2}} \left( 4\,c_6 R\, dp^\alpha - \frac{16\, c_6^2 R^3}{3} k_\beta\, dp^\beta \wedge vol_{S^2} \right) \,.
\end{equation}

Finally, we are left to solve the differential conditions \eqref{eq:SO55FinalStep}. Using the explicit expression of the generalised Lie derivative \eqref{eq:IIBDiffConditions} and the fact that $\J_A$, $\J_4$ and $\hK$ are $\SU{2}$ triplets, singlet and singlets, respectively, these equations reduce to 
\begin{equation}
 \begin{split}
  \imath_{V_A} d\lambda_B^\alpha &= 0 \,, \\
  \imath_{V_A} d\sigma_B + \lambda_{A\,\alpha} \wedge d\lambda_B^\alpha &= 0 \,, \\
  \lambda_{4\,\alpha} \wedge d\lambda_A^\alpha &= 0 \,, \\
  \homega_{(0)\alpha} d\lambda_A^\alpha &= 0 \,, \\
  \homega_{(0)}^\alpha d\sigma_A + \homega_{(2)} \wedge d\lambda_A^\alpha &= 0 \,.
 \end{split}
\end{equation}
For our $\J_A$'s and $\hK$ these further simplify to
\begin{equation}
 \begin{split}
  d\lambda_A^\alpha &= d\sigma_A = 0 \,,
 \end{split}
\end{equation}
which implies $m^\alpha = - dk^\alpha$.

Thus, we find that
\begin{equation}
 \begin{split}
  \J_A &=  \frac{1}{\sqrt{2}} \left(\frac{3}{R} v_A + 4\, c_6 R\, d\left( k^\alpha\, y_A \right) + \frac{8\,c_6^2 R^3}{3} d \left( k^\alpha \theta_A \wedge dk_\alpha \right) \right) \,, \\
  \J_4 &= \frac{1}{\sqrt{2}} \left( 4\,c_6 R\, dp^\alpha - \frac{16\, c_6^2 R^3}{3}\, k_\beta\, dp^\beta \wedge vol_{S^2} \right) \,, \\
  \hK &= \frac{1}{\sqrt{2}} \left( 4\, c_6\, p_\alpha - \frac{16\, c_6^2 R^2}{3}\, \left( r + p_\beta k^\beta \right) vol_{S^2} \right) \,, \label{eq:AdS6StructuresSummary}
 \end{split}
\end{equation}
where $k^\alpha$ and $p^\alpha$ are any $\SL{2}$-doublets of functions on $\Sigma$ subject to the differential conditions
\begin{equation}
 dk^\alpha \wedge dk^\beta = dp^\alpha \wedge dp^\beta \,, \qquad dk^\alpha \wedge dp_\alpha = 0 \,, \label{eq:AdS6DC}
\end{equation}
and $r$ is defined up to an integration constant by
\begin{equation}
 dr = - p_\alpha\, dk^\alpha \,.
\end{equation}
The condition \eqref{eq:Ad6Reg} implies
\begin{equation}
 r |dk| vol_\Sigma \wedge vol_{S^2} > 0 \,,
\end{equation}
where $|dk| = \frac12 \partial_\alpha k_\beta \partial^\alpha k^\beta$. This seems to suggest that $r |dk| > 0$ but care needs to be taken at the boundaries of $\Sigma$. Instead, we must have
\begin{equation}
 r |dk| \geq 0 \,, \label{eq:AdS6RegS}
\end{equation}
with equality only possible at the boundaries of $\Sigma$. In fact, as discussed in \cite{DHoker:2016ujz,DHoker:2017mds}, and as will become appart from the explicit SUGRA solution given in section \ref{s:AdS6Sol} in order for the internal four-manifold not to have a boundary, we must have
\begin{equation}
 r = 0 \,, \qquad |dk| = 0 \,, \label{eq:AdS6Compactness}
\end{equation}
on $\partial \Sigma$.

At this stage, one might wonder how the quadratic differential conditions \eqref{eq:AdS6DC} can underlie supersymmetric AdS vacua, which ought to be described by a first-order BPS equation. The answer is that we still have residual diffeomorphism symmetry on the Riemann surface $\Sigma$ that can be used to turn \eqref{eq:AdS6DC} into first-order differential equations. We will show how to do this after calculating the supergravity fields from the structures.

We conclude this section by giving the explicit expressions for the objects $\K = \frac14 \J_u \wedge \J^u$ and $\kappa^4 = \K \wedge \hK$, which appear in the truncation Ansatz \eqref{eq:VecD6Ansatz}. They are given by
\begin{equation}
 \begin{split}
  \K &= -8\,\sqrt{2}\,c_6^2 R^2\, |dk| \left( vol_\Sigma + \frac{4\,c_6 R^2}{3} vol_{S^2} \wedge vol_\Sigma \right) \,, \\
  \kappa^4 &= \frac{128}{3} c_6^4 R^4\,r\,|dk|\, vol_{S^2} \wedge vol_\Sigma \,.
 \end{split}
\end{equation}

\subsection{The AdS$_6$ vacua} \label{s:AdS6Sol}
We will now compute the supergravity background corresponding to the half-maximal structures \eqref{eq:AdS6StructuresSummary}. The supergravity fields are encoded in the generalised metric \eqref{eq:SO55GenMetric1}, \eqref{eq:SO55GenMetric2} as detailed in appendix \ref{A:IIBGM}. Moreover, the AdS$_6$ part of the metric is warped by the factor \cite{Malek:2017njj}
\begin{equation}
 f_6 = |g_{int}|^{-1/4} \kappa^2 \,.
\end{equation}

Thus, we find the following background
\begin{equation}
 \begin{split}
  ds^2 &= \frac{4\,r^{5/4}\,\Delta^{1/4}\,c_6 R^2}{3^{3/4} |dk|^{1/2}} \left[ \frac{3}{r} ds_{AdS_6}^2 + \frac{|dk|^2}{\Delta} ds_{S^2}^2 + \frac{1}{4\,r^2} dk^\alpha \otimes dp_\alpha \right] \,, \\
  C_{(2)}{}^{\alpha} &= - \frac{4\,c_6 R^2}{3}\, vol_{S^2} \left( k^\alpha + \frac{r\, p_\gamma\, \partial_\beta k^\gamma\, \partial^\beta p^\alpha}{\Delta} |dk| \right) \,, \\
  H_{\alpha\beta} &= \frac{1}{\sqrt{3\,\Delta}} \left( \frac{|dk|}{\sqrt{r}}\, p_\alpha p_\beta + 3 \sqrt{r}\, \partial_\gamma k_\alpha \partial^\gamma p_\beta \right) \,,
 \end{split}\label{eq:AdS6Vacuum} 
\end{equation}
where
\begin{equation}
 \Delta = 3\, r |dk|^2 + |dk| p_\gamma p_\delta \partial_\sigma k^\gamma \partial^\sigma p^\delta \,, \qquad |dk| = \frac12 \partial_\alpha k_\beta \partial^\alpha k^\beta \,.
\end{equation}
The solutions are completely determined by the two pairs of functions $p^\alpha$ and $k^\alpha$ on $\Sigma$ satisfying
\begin{equation}
 dk^\alpha \wedge dk^\beta = dp^\alpha \wedge dp^\beta \,, \qquad dk^\alpha \wedge dp_\alpha = 0 \,. \label{eq:AdS6DCRep}
\end{equation}
$r$ is defined in terms of these functions as
\begin{equation}\label{eq:r-def-AdS6}
 dr = - p_\alpha dk^\alpha \,.
\end{equation}
In order to have a compact internal space, we must require that the $S^2$ shrinks on the boundary of $\Sigma$ while the warp factor and the metric on $\Sigma$ remain non-singular. From the explicit metric \eqref{eq:AdS6Vacuum}, one can easily see that this requires
\begin{equation}
 r = |dk| = 0 \,,
\end{equation}
on $\partial \Sigma$.

We will now show that the differential equations for $k^\alpha$ and $p^\alpha$ can be turned into first-order PDEs by coordinate choices. In particular, we can always use diffeomorphisms to make the metric on $\Sigma$ conformally flat. From \eqref{eq:AdS6Vacuum} we see that this requires
\begin{equation}
 \partial_1 k^\alpha \partial_1 p_\alpha = \partial_2 k^\alpha \partial_2 p_\alpha \,, \qquad \partial_1 k^\alpha \partial_2 p_\alpha = 0 \,.
\end{equation}
Together with \eqref{eq:AdS6DC}, and imposing the condition \eqref{eq:AdS6RegS}, the differential conditions become the Cauchy-Riemann equations
\begin{equation}
 dk^\alpha = I \cdot dp^\alpha \,, \label{eq:CR}
\end{equation}
where $I_\alpha^{\beta} = \delta_{\alpha\gamma} \epsilon^{\gamma\beta}$ is a complex structure on $\Sigma$. Therefore, $p^\alpha$ and $k^\alpha$ are the real and imaginary parts of two holomorphic functions on $\Sigma$
\begin{equation}
 f^\alpha = - p^\alpha + i\, k^\alpha \,.
\end{equation}

We now recover the description of supersymmetric AdS$_6$ vacua of \cite{DHoker:2016ujz} by identifying our holomorphic functions with the ${\cal A}_{\pm}$ of \cite{DHoker:2016ujz} via
\begin{equation}
 {\cal A}_{\pm} = i\, f^1 \pm f^2 \,. \label{eq:HolMatch}
\end{equation}
We present a dictionary between our objects and those of \cite{DHoker:2016ujz}, as well as \cite{Hong:2018amk}, in appendix \ref{A:AdS6Match}. As discussed in \cite{DHoker:2016ysh,DHoker:2017mds,DHoker:2017zwj} these local solutions can be extended to globally regular solutions by including a boundary of the Riemann surface on which the holomorphic functions $f^\alpha$ have poles, and by introducing $\SL{2}$ monodromies.

\section{Minimal consistent truncations} \label{s:ConsTruncation}
As shown in \cite{Malek:2017njj} and reviewed in section \ref{s:RevConsTruncation}, given the half-supersymmetric structures describing an AdS vacua, one can automatically construct a consistent truncation around it containing a gravity multiplet and a scalar. This method was applied in \cite{Malek:2018zcz} to construct the minimal consistent truncations around the supersymmetric AdS$_7$ and AdS$_6$ vacua, for the case where only the scalar fields of the lower-dimensional gauged SUGRA are turned on and are constant, agreeing with the consistent truncations found in \cite{Passias:2015gya} and \cite{Hong:2018amk} for the AdS$_7$ and AdS$_6$ vacua, respectively. Furthermore, as described in section \ref{s:RevConsTruncation}, using the exceptional field theory tensor hierarchy and the dictionaries in appendix \ref{A:EFT-IIB-dictionary-TH}, one can construct the uplift of all the fields of the minimal half-maximal gauged supergravity, including the $p$-forms. In the following, we summarise the results for the the minimal consistent truncations around AdS$_7$ and AdS$_6$. For the latter, we show explicitly how to construct the full ten-dimensional uplift, including all the fields of the 6-dimensional gauged SUGRA. This result will be generalised in section \ref{s:VecTruncationAdS6} to construct uplifts of half-maximal gauged supergravities around AdS$_6$ including matter multiplets.

\subsection{AdS$_7$} \label{s:AdS7ConsTruncation}
We can now use \eqref{eq:TruncAnsatz}, \eqref{eq:VecD7Ansatz} to construct the consistent truncation Ansatz of IIA SUGRA around the supersymmetric AdS$_7$ vacua of section \ref{s:AdS7} to the pure 7-dimensional half-maximal $\SU{2}$ gauged SUGRA \cite{Townsend:1983kk}. Here we will consider the truncation Ansatz where only the scalar fields of the 7-dimensional gauged SUGRA have been turned on and are constant. Thus, we compute the generalised metric of ${\cal J}_u(x,Y)$ and ${\cal \hK}(x,Y)$ given in \eqref{eq:TruncAnsatz} and use the ExFT/IIA dictionary of appendix \ref{A:EFT-IIAdictionary} to find the supergravity expressions. This way, we obtain the truncation Ansatz in string frame
\begin{equation}
 \begin{split}
  ds_{10}^2 &= X^{1/2} \sqrt{\frac{t}{p}} ds_7^2 + \frac{R^2}{8} \sqrt{\frac{t}{p}} \left[ X^{5/2} \frac{p\,t}{X^5\,s^2 + 2\,p\,t} ds_{S^2}^2 +  X^{-5/2} \frac{h^2}{p\,t} d\z^2 \right] \,, \\
  e^{\psi} &= \frac{2}{R} X^{5/4} \left( \frac{t}{p} \right)^{3/4} \frac{1}{\sqrt{X^5\,s^2 + 2\,p\,t}} \,, \\
  B &= \frac{R^2}{8\sqrt{2}} \left( - g + \frac{X^5\,s\,t}{X^5\,s^2 + 2\,p\,t} \right) vol_{S^2} \,,
\end{split} \label{eq:AdS7CTG}
\end{equation}
and field strengths
\begin{equation}
 \begin{split}
  F_2 &= - \frac{R^2}{8\sqrt{2}} \left( 2\,p + \frac{X^5\,m\,s\,t}{X^5\,s^2+ 2\,p\,t} \right) vol_{S^2} \,, \\
  H_3 &= \frac{2}{R} \left(\frac{t}{p}\right)^{-1/4} X^{-5/4} \left( 3 - \frac{t}{p} \frac{m\,s}{X^5\,s^2 + 2\,p\,t} \right) vol_{\tilde{M}_3} \\
  & \quad + \frac{2}{R} \left( \frac{t}{p} \right)^{-1/4} X^{-5/4} \left(1-X^5\right) \left( 1 - \frac{4\,p\,t}{X^5\,s^2 + 2\,p\,t} + \frac{t}{p} \frac{m\,s}{X^5\,s^2 + 2\,p\,t} \right) vol_{\tilde{M}_3} \,,
 \end{split}
\end{equation}
where $vol_{\tilde{M}_3}$ denotes the volume form of the internal 3-manifold with the metric \eqref{eq:AdS7CTG}.

Let us now evaluate the truncation Ansatz for our two gauge choices.
\paragraph{Choice 1} With $h(z) = p(z)$, the truncation Ansatz becomes
\begin{equation}
\begin{split}
ds_{10}^2 &= X^{1/2} \sqrt{-\frac{t}{\ddot{t}}} ds_7^2 + \frac{R^2}{8} \sqrt{-\frac{\ddot{t}}{t}} \left[ X^{5/2} \frac{t^2}{X^5\,\dot{t}^2 - 2\,t\ddot{t}} ds_{S^2}^2 +  X^{-5/2} d\z^2 \right] \,, \\
e^{\psi} &= \frac{2}{R} X^{5/4} \left( -\frac{t}{\ddot{t}} \right)^{3/4} \frac{1}{\sqrt{ X^5\, \dot{t}^2 - 2\,t\,\ddot{t}}} \,, \\
B &=  \frac{R^2}{8\sqrt{2}} \left( z - \frac{X^5\,t\,\dot{t}}{X^5\,\dot{t}^2 - 2\,t\,\dot{t}} \right) vol_{S^2} \,,
\end{split} \label{eq:AdS7CT1}
\end{equation}
and field strengths
\begin{equation}
 \begin{split}
  F_2 &= \frac{R^2}{8\sqrt{2}} \left( 2\,\ddot{t} + \frac{X^5\,m\,t\,\dot{t}}{X^5\,\dot{t}^2 - 2\,t\,\ddot{t}} \right) vol_{S^2} \,, \\
  H_3 &= \frac{2}{R} \left(-\frac{\ddot{t}}{t}\right)^{1/4} X^{-5/4} \left( 3 - \frac{t}{\ddot{t}} \frac{m\,\dot{t}}{X^5\,\dot{t}^2 - 2\,t\,\ddot{t}} \right) vol_{\tilde{M}_3} \\
  & \quad + \frac{2}{R} \left(-\frac{\ddot{t}}{t}\right)^{1/4} X^{-5/4} \left(1-X^5\right) \left( 1 + \frac{4\,t\,\ddot{t}}{X^5\,\dot{t}^2 - 2\,t\,\ddot{t}} + \frac{t}{\ddot{t}} \frac{m\,\dot{t}}{X^5\,\dot{t}^2 - 2\,t\,\ddot{t}} \right) vol_{\tilde{M}_3} \,.
 \end{split}
\end{equation}

\paragraph{Choice 2} We now take $h(z) = 1$ and find
\begin{equation}
\begin{split}
ds_{10}^2 &= \frac{1}{9} \sqrt{-\frac{\beta'}{\by}} X^{1/2} ds_7^2 + \frac{R^2}9 \sqrt{-\frac{\beta'}{\by}} \left[ \frac{X^{5/2}\, \beta/4}{4\,\beta - X^5\,\by\,\beta'} ds_{S^2}^2 -  \frac1{16} X^{-5/2} \frac{\beta'\,d\by^2}{\beta\,\by} \right] \,, \\
e^{\psi} &= R^{-1} X^{5/4} \frac{\left(-\beta'/\by\right)^{5/4}}{6\sqrt{4\,\beta- X^5\,\beta'\by}} \,, \\
F_2 &= \frac{R^2\,\by}{4} \frac{\sqrt{\beta}}{\beta'} \left( 4 + \frac{X^5\,m}{18\,\by} \frac{\left(\beta'\right)^2}{4\,\beta - X^5\,\beta'\by} \right) vol_{S^2} \,, \\
H_3 &= \frac{2}{R} \left(- \frac{\beta'}{\by}\right)^{-1/4} X^{-5/4} \left( 9 - \frac{m}{12\by} \frac{\left(\beta'\right)^2}{4\,\beta - X^5\,\beta'\by} \right)  vol_{\tilde{M}_3} \\
& \quad + \frac{6}{R} \left( -\frac{\beta'}{\by} \right)^{-1/4} X^{-5/4} \left(1-X^5\right) \left( 1 - \frac{8\,\beta}{4\,\beta - X^5\,\beta'\,\by} + \frac{m}{36\by} \frac{\left(\beta'\right)^2}{4\,\beta - X^5\,\beta'\by} \right) vol_{\tilde{M}_3} \,.
\end{split}
\end{equation}

The truncation Ansatz is completely determined by the function $t(z)$ satisfying \eqref{eq:tcond} for gauge choice 1 and $\beta(\by)$ satisfying \eqref{eq:betacond} for choice 2, and corresponds to the truncation Ansatz found in \cite{Passias:2015gya} in the coordinates of \cite{Cremonesi:2015bld} and \cite{Apruzzi:2015zna}, respectively. Upon truncation, $X$ becomes the scalar field of the minimal 7-dimensional gauged SUGRA \cite{Townsend:1983kk} and all of the supersymmetric AdS vacua correspond to the same vacuum of the 7-dimensional theory.

\subsection{AdS$_6$} \label{s:AdS6ConsTruncation}
We can similarly use \eqref{eq:TruncAnsatz} to find the minimal consistent truncation corresponding to the supersymmetric AdS$_6$ vacua of IIB SUGRA described in section \ref{s:AdS6}. For example, the internal fields can be read off from the generalised metric \eqref{eq:SO55GenMetric1}, while the remaining fields can be determined using the truncation Ansatz \eqref{eq:6DAnsatz}. Recall that the AdS vacua are characterised in terms of two holomorphic functions $f^\alpha$, with real/imaginary parts $k^\alpha$, $p^\alpha$, and a real function $r$ defined through \eqref{eq:r-def-AdS6}.

As before, we will denote by $X$ the scalar field and $A_A$, $A_4$ the $\SU{2}_R$ and $\UO$ gauge fields of the 6-dimensional gauged SUGRA, the so-called pure $\mathrm{F}(4)$ gauged SUGRA \cite{Romans:1985tw}, obtained from the consistent truncation. In terms of these objects, we find that the metric in Einstein frame, the axio-dilaton and the 2-forms are given by
\begin{equation}
\begin{split}
 ds^2 &= \frac{4\,r^{5/4}\,\bar\Delta^{1/4}\,c_6 R^2}{3^{3/4} |dk|^{1/2}} \left[ \frac{3}{R^2\,r} ds_6^2 + \frac{X^2\,|dk|^2}{\bar\Delta} ds_{\tilde{S}^2}^2 + \frac{1}{X^2\,r^2} dk^\alpha \otimes dp_\alpha \right] \,, \\
 H_{\alpha\beta} &= \frac{1}{\sqrt{3\,\bar\Delta}} \left( \frac{X^4\,|dk|}{\sqrt{r}}\, p_\alpha p_\beta + 3 \sqrt{r}\, \partial_\gamma k_\alpha \partial^\gamma p_\beta \right) \,, \\
 C_{(2)}{}^\alpha&=-\frac{4\,c_6 R^2}{3} \left( k^\alpha + \frac{X^4\, r\, p_\gamma\, \partial_\beta k^\gamma\, \partial^\beta p^\alpha}{\bar{\Delta}} |dk| \right) \tvol + 2 \sqrt{2}\,c_6 R\, A{}^A\wedge\left(y_A dk^\alpha+k^\alpha \tD y_A \right)\\
 &\qquad+2 \sqrt{2}\,c_6 R\, A^4\wedge dp^\alpha + 4\,c_6 B\, p^\alpha- 3\, c_6 k^\alpha \epsilon_{ABC} y^A A^B\wedge A^C\,,
\end{split} \label{eq:AdS6Trunc}
\end{equation}
where
\begin{equation}
 \begin{split}
  \bar\Delta &= 3\, r\, |dk|^2 + X^4\, |dk|\, p_\gamma p_\delta \partial_\sigma k^\gamma \partial^\sigma p^\delta \,.
 \end{split}
\end{equation}
Moreover,
\begin{equation}
 \tD y^A = dy^A + \frac{3}{\sqrt{2}\,R} \epsilon^{ABC} A_B y_C \,,
\end{equation}
is the $\SU{2}_R$ covariant derivative of $y^A$, in terms of which the $\SU{2}_R$ covariant $S^2$ metric and $S^2$ volume form are defined as
\begin{equation}
 \begin{split}
  ds_{\tilde{S}^2} &= \delta_{AB} \tD y^A \otimes \tD y^B \,, \\
  \tvol &= \frac{1}{2}\epsilon_{ABC}y^A\tilde Dy^B\wedge\tilde Dy^C \,.
 \end{split}
\end{equation}

After applying a gauge transformation, the two-forms can equivalently be written as
\begin{equation}
 C_{(2)}{}^\alpha=- \frac{4\,c_6 R^2}{3} \left( k^\alpha + \frac{X^4\, r\, p_\gamma\, \partial_\beta k^\gamma\, \partial^\beta p^\alpha}{\bar{\Delta}} |dk| \right) \tvol
+2\sqrt{2}\,c_6 R\, \left( k^\alpha \tilde F_{(2)}{}^A y_A 
+ \tilde{F}_{(2)}{}^4 p^\alpha \right) \,, \label{eq:MinimalC2F}
\end{equation}
where
\begin{equation}
 \begin{split}
  \tilde{F}_{(2)}{}^A &= dA^A + \frac{3}{2\sqrt{2}\,R} \epsilon^{ABC} A_B \wedge A_C \,, \\
  \tilde{F}_{(2)}{}^4 &= dA^4 + \frac{\sqrt{2}}{R} B \,,
 \end{split} \label{eq:tilde-F-2forms}
\end{equation}
are the 2-forms of the 6-dimensional gauged SUGRA as defined in \eqref{eq:6dFieldStrengthMinimal} and using equations \eqref{eq:SO55DiffConditions}, \eqref{eq:SO55Lambda}.

The five-form field strength can easily be computed from \eqref{eq:6DAnsatz} and using its self-duality. We find
\begin{equation}
 \begin{split}
  F_{(5)} &= F_{(2,3)} + F_{(3,2)} + F_{(4,1)} \,,
 \end{split}
\end{equation}
where $F_{(p,q)}$ are the parts of the 5-form with $p$ external and $q$ internal legs, appropriately $\SU{2}_R$-covariantised. Explicitly, they are given by
\begin{equation}
 \begin{split}
  F_{(2,3)} &= \frac{8 \sqrt{2}\,c_6^2 R^3\, |dk|}{3} \left( \tilde{F}_{(2)}{}^A \wedge \tTh_A \wedge vol_\Sigma + \frac{X^4\,r\,|dk|}{\bar{\Delta}} p_\alpha \left( y_A\, \tilde{F}_{(2)}{}^A \wedge dp^\alpha - \tilde{F}_{(2)}{}^4 \wedge dk^\alpha \right) \wedge \tvol \right) \,, \\
  F_{(3,2)} &= 16\,c_6^2 R^2 |dk| \left( \frac{r^2\,|dk|}{\bar{\Delta}} \tilde{F}_{(3)} \wedge \tvol + X^{-4} \star_6 \tilde{F}_{(3)} \wedge vol_\Sigma \right) \,, \\
  F_{(4,1)} &= 8\sqrt{2}\,c_6^2 R\,X^2 \left( - r \star_6 \tilde{F}_{(2)}{}^A \wedge \tD y_A + p_\alpha \left( \star_6 \tilde{F}_{(2)}{}^4 \wedge dp^\alpha + y_A \star_6 \tilde{F}_{(2)}{}^A \wedge dk^\alpha \right) \right) \,,
 \end{split}
\end{equation}
where $F_{(2,3)}$ and $F_{(3,2)}$ can be read off directly from \eqref{eq:6dFieldStrengthMinimal} and $F_{(4,1)}$ can be obtained from $F_{(2,3)}$ by self-duality of the 5-form field strength. Above $\star_6$ refers to the Hodge dual of the metric of the six-dimensional gauged SUGRA, and $\tilde{F}_{(3)}$ is, as defined in \eqref{eq:6dFieldStrengthMinimal}, the field strength of the 2-form potential
\begin{equation}
 \tilde{F}_{(3)} = dB_{(2)} \,.
\end{equation}
Moreover, we have used \eqref{eq:6dDuality} to replace $\tilde{G}_{(3)}$ by $X^{-4} \star_6 \tilde{F}_{(3)}$. A non-trivial check of the truncation Ansatz is that the component $F_{(3,2)}$ is self-dual.

In deriving these relations, we used the fact that the 10-dimensional Hodge dual is related to the 6-dimensional Hodge dual and the Hodge dual on $S^2$ and $\Sigma$ as
\begin{equation}
 \begin{split}
  \star_{10} F_{(2)} \wedge \Theta_A \wedge vol_\Sigma &= \frac{f_6}{f_\Sigma} \star_{S^2} \Theta_A \wedge \star_{6} F_{(2)} \,, \\
  \star_{10} F_{(2)} \wedge \omega \wedge vol_{S^2} &= \frac{f_6}{f_{S^2}} \star_{\Sigma} \omega \wedge \star_{6} F_{(2)} \,,
 \end{split}
\end{equation}
for any 1-form $\omega \in \Omega^{(1)}\left(\Sigma\right)$ and where
\begin{equation}
 f_6 = \frac{3}{R^2\,r} \,, \qquad f_\Sigma = \frac{|dk|}{X^2\,r^2} \,, \qquad f_{S^2} = \frac{X^2|dk|^2}{\bar{\Delta}} \,,
\end{equation}
denote the relative factors of the 6-dimensional, $S^2$ and Riemann surface metric in \eqref{eq:AdS6Trunc}. Also
\begin{equation}
 \star_\Sigma dk^\alpha = - |dk|\, dp^\alpha \,, \qquad \star_\Sigma dp^\alpha = |dk|\, dk^\alpha \,.
\end{equation}

After the consistent truncation, all the 10-dimensional AdS vacua correspond to the same vacuum of the 6-dimensional gauged SUGRA. Our truncation Ansatz includes the previously-found consistent truncation of a particular AdS$_6$ vacuum in this family \cite{Jeong:2013jfc} as a particular example. This arises by using the form of the holomorphic function given in \cite{Hong:2018amk}.

\section{Consistent truncations with vector multiplets for AdS$_7$}\label{s:VecTruncationAdS7}
Here we will now search for consistent truncation with vector multiplets around the supersymmetric AdS$_7$ vacua of massive IIA SUGRA. There are in fact many 7-dimensional half-maximal gauged SUGRAs that contain supersymmetric AdS$_7$ vacua \cite{Louis:2015mka} and could, in principle, arise as a consistent truncation of 10-dimensional SUGRA. We will see that in fact only the pure $\SU{2}$ gauged SUGRA \cite{Townsend:1983kk} and coupled to one vector multiplet can be uplifted, where in the latter case the Roman's mass must vanish.

As we discussed above, we can only have $N \leq 3$ vector multiplets in a consistent truncation and the corresponding generalised vector fields must form representations of the $\SU{2}_R$ symmetry group generated by the $\J_u$ of the AdS$_7$ vacua. Therefore, we must consider generalised vector fields that are singlets or triplets under $\SU{2}_R$, and satisfy the algebraic conditions \eqref{eq:VecAlgCon} as well as the differential conditions \eqref{eq:VecDiffCon}. Doublets under $\SU{2}_R$ do not lead to $f_{abc}$ of the form required in \eqref{eq:VecDiffCon}. Moreover, plugging in the form of the $\J_u$ for the AdS$_7$ vacua, we have
\begin{equation}
 \gL_{J_{u}} J_{\bar{v}} = 2\sqrt{2} R^{-1} L_{v_u} J_{\bar{v}} \,,
\end{equation}
where on the right-hand side we have the usual three-dimensional Lie derivative generated by $v_u$ acting on the vector, scalar, 1-form and 2-form parts of $J_{\bar{u}}$ separately. This implies that the $J_{\bar{u}}$ must form a representation of $\SU{2}_R$ under the Lie derivative generated by the $\SU{2}_R$ Killing vector fields on $S^2$.

In the following, we choose the gauge $h(z) = p(z)$ so that the AdS vacua are described by a cubic function $t(z)$.

\subsection{Singlets under $\SU{2}_R$}
For the $\J_{\bar{u}}$ to form singlets under $\SU{2}_R$, they must take the general form
\begin{equation}
 J_{\bar{u}} = f_{\bar{u}}(z)\, \partial_z + g_{\bar{u}}(z)+l_{\bar{u}}(z)\iota_{\partial_z}A + h_{\bar{u}}(z)\, dz + k_{\bar{u}}(z)\, vol_{S^2}+r_{\bar{u}}(z)\,dz\wedge A \,.
\end{equation}
Plugging the above parametrisation into the algebraic conditions \eqref{eq:VecExtraAlgCon}, we find they can be solved by only one generalised vector field which is unique (up to an overall sign which just amounts to a redefinition of the scalar field in the truncation Ansatz)
\begin{equation}
 \J_{\bar{1}} = \frac{R}{2}\,\ddot{t} +\frac{R}{4}\,dz+ \frac{R^3}{16\sqrt{2}}\,\ddot{t}\,z\, vol_2+\frac{R}{4}\,dz\wedge A \,.
\end{equation}
Therefore, the algebraic conditions already restrict us to having at most 1 vector multiplet that transforms as a singlet under $\SU{2}_R$. However, we must now also check the differential conditions \eqref{eq:VecDiffCon} but find that
\begin{equation}
 \gL_{J_{\bar{1}}} \hK = - m \frac{R^4\,t}{32\sqrt{2}} vol_2 \wedge dz \neq 0 \textrm{ unless } m = 0 \,.
\end{equation}
Therefore, if the Roman's mass is non-vanishing, it is impossible to have a consistent truncation with singlet vector multiplets.

For vanishing Roman's mass the existence of this consistent truncation is not surprising. In this case the vacuum lifts to a AdS$_7 \times S^4$ solution of 11-dimensional SUGRA, where the $S^4$ is written as a $S^3$ fibred over an interval. It is known that there is a maximally supersymmetric consistent truncation of 11-dimensional SUGRA around this vacuum with gauge group $\SO{5}$. This truncation of 11-dimensional SUGRA can be further consistently truncated to a consistent truncation with gauge group $\SU{2} \times \UO \subset \SO{5}$ by keeping only the singlets under a Cartan $\UO \subset \SU{2}_L \subset \SU{2}_L \times \SU{2}_R \subset \SO{5}$. Moreover, the generators of $\SU{2} \times \UO$ are independent of one of the four internal coordinates, which can be identified with the Hopf fibre of $S^3$ when writing $S^4$ as a $S^3$ fibred over an interval, see for example \cite{Lee:2014mla} for an explicit realisation of the $\SO{4}$ generators on $S^3$, albeit in $\mathrm{O}(3,3)$ generalised geometry. Thus we find a consistent truncation of IIA SUGRA giving rise to $\SU{2} \times \UO$ gauge group, which corresponds precisely to the above setup.

\subsection{Triplets under $\SU{2}_R$}
We repeat the above analysis but consider $\J_{\bar{u}}$ with $\bar{u} = 1, \ldots, 3$ forming a triplet under $\SU{2}_R$, which implies they must take the general form \eqref{eq:J_u-ansatz-AdS7}. The algebraic conditions \eqref{eq:VecExtraAlgCon} then lead to (up to an overall sign)
\begin{equation}
\begin{split}
 \J_{\bar{u}} &=\frac{2\sqrt{2}}{R}v_{\tilde{u}}-\epsilon\,\frac{R}{2}\,\ddot{t}\,y_{\bar{u}}+\frac{2\sqrt{2}}{R}\iota_{v_{\bar{u}}}A-\frac{R}{4}(z\,dy_{\bar{u}}+\epsilon\,y_{\bar{u}}dz)\\
 &\qquad +\frac{R^3}{16\sqrt{2}}\,\ddot{t}\,(\theta_{\bar{u}}\wedge dz-\epsilon\,z\,y_{\bar{u}}Vol_{S^2})-\frac{R}{4}\,(z\,dy_{\bar{u}}+\epsilon\,y_{\bar{u}}dz)\wedge A \,,
 \end{split}
\end{equation}
where $\epsilon=\pm 1$. Finally, one needs to check the differential conditions  \eqref{eq:VecDiffCon}. However, we find
\begin{equation}\label{eq:LJubarKhat-AdS7}
\begin{split}
 \gL_{J_{\bar{u}}} \hK &=\frac{R^2}{8}\,(1-\epsilon)\,\dot{t}\,dy_{\bar{u}}\wedge dz-\frac{R^4 \left(m\,\epsilon\,t + 2\,\dot{t}\,\ddot{t}\,\right) }{32\sqrt{2}}\,y_{\bar{u}}\, vol_{S^2} \wedge dz\\
&\qquad +\frac{R^2}{8}\,(1-\epsilon)\,\dot{t}\,dy_{\bar{u}} \wedge dz \wedge A \,.
\end{split}
\end{equation}
Looking at the two form part of \eqref{eq:LJubarKhat-AdS7} we observe that it can only vanish when $\epsilon=1$, since $\dot{t}$ cannot vanish for non-zero Roman's mass. In this case, \eqref{eq:LJubarKhat-AdS7} vanishes if the condition 
\begin{equation}
(m\,t+2\,\dot{t}\,\ddot{t}\,)=0\,,
\end{equation}
is satisfied. However, by taking a $z$-derivative of this condition we find that it implies
\begin{equation}
\ddot{t}=0\,,
\end{equation}
which can never be satisfied for $m\neq 0$ due to the condition  \eqref{eq:condition-tddd-AdS7}. We therefore conclude that, if the Roman's mass is non-vanishing, consistent truncations with a $SU(2)_R$ triplet of vector multiplets do not exist.

Moreover, even when $m = 0$, the truncation is only consistent if $\epsilon = 1$ and $\dot{t}\, \ddot{t} = 0$ and hence requires $\ddot{t} = 0$, or $\epsilon = - 1$ and $\dot{t} = \ddot{t} = 0$. However, from \eqref{eq:AdS7RegG} we see that for the AdS$_7$ solution to be non-singular requires $t\,\ddot{t} \leq 0$ with equality only allowed at $\partial I$. Therefore, if $\ddot{t} = 0$, the AdS$_7$ solutions would be badly singular, as is also apparent by direct inspection of \eqref{eq:AdS7Gauge1}. Therefore, there are no consistent truncations around AdS$_7$ vacua of IIA with a triplet of vector multiplets.
 
\section{Consistent truncations with vector multiplets for AdS$_6$}\label{s:VecTruncationAdS6}
We now turn to consistent truncations with vector multiplets around AdS$_6$ vacua of IIB.  In principle there are a large number of 6-dimensional half-maximal gauged SUGRAs (containing vector multiplets) that contain supersymmetric AdS$_6$ vacua \cite{Karndumri:2016ruc}, and which could thus arise from a consistent truncation of AdS$_6$ vacua of IIB. Here we will now address the question of which of these 6-dimensional gauged SUGRAs can be uplifted to IIB.

Since we can only keep $N \leq 4$ vector multiplets in a consistent truncation and the generalised vector fields corresponding to the vector multiplets must transform as representations under the $\SU{2}_R$ we have the following possibilities:
\begin{itemize}
	\item up to 4 singlets,
	\item a triplet,
	\item a triplet plus singlet.
\end{itemize}
Once again, doublets under $\SU{2}_R$ are forbidden by \eqref{eq:VecDiffCon}.

Just as for AdS$_7$ vacua, the form of the generalised Lie derivative simplifies when plugging in the form of the $\J_u$ for the AdS$_6$ vacua. We find
\begin{equation}
 \begin{split}
  \gL_{J_{A}} J_{\bar{u}} &= \frac{3}{\sqrt{2}\,R} L_{v_A} J_{\bar{u}} \,, \\
  \gL_{J_4} J_{\bar{u}} &= 0 \,,
 \end{split}
\end{equation}
where in the first equation on the right-hand side we have the usual four-dimensional Lie derivative generated by $v_u$ acting on the vector, 1-form and 3-form parts of $J_{\bar{u}}$ separately. This implies that the $J_{\bar{u}}$ must form a representation of $\SU{2}_R$ under the Lie derivative generated by the $\SU{2}_R$ Killing vector fields on $S^2$.

\subsection{One singlet under $\SU{2}_R$}
We first consider a single vector multiplet whose corresponding generalised vector field satisfies the differential conditions
\begin{equation}
 \begin{split}
  \gL_{J_A} J_{\bar{1}} &= 0 \,, \\
  \gL_{J_{\bar{1}}} \hK &= 0 \,.
 \end{split} \label{eq:SingletDiff}
\end{equation}
Note the algebraic conditions \eqref{eq:VecAlgCon} together with the above immediately imply that
\begin{equation}
 \gL_{J_{\bar{1}}} J_{a} = 0 \,,
\end{equation}
while $J_4 \propto d\hK$ implies
\begin{equation}
 \gL_{J_4} J_{\bar{1}} = 0 \,.
\end{equation}
The corresponding consistent truncation will lead to a half-maximal gauged SUGRA with one vector multiplet and gauge group $\SU{2} \times \UO$.

The most general Ansatz we can write for a generalised vector field that transforms as a singlet under $\SU{2}_R$ is
\begin{equation}
 \J_{\bar{1}} =\frac{1}{\sqrt{2}}\left( w(z) +4\,R\,c_6\, n^\alpha(z) + \frac{16\,R^3\,c_6^2}{3}\,l(z) \wedge vol_{S^2} \right)\,,
\end{equation}
where $w(z)$ is a vector field on $\Sigma$ and $n^\alpha(z)$ is an $\SL{2}$-doublet of 1-forms on $\Sigma$ and $l(z)$ is a 1-forms on $\Sigma$. The algebraic conditions \eqref{eq:VecExtraAlgCon} now impose that
\begin{equation}
 \begin{split}
  w(z) &= 0 \,, \qquad l(z) =  k_\alpha(z) n^\alpha(z) \,,
 \end{split}
\end{equation}
and further imposes on $n_\alpha$ that
\begin{equation}
 \begin{split}
  n_\alpha \wedge dk^\alpha &= n_\alpha \wedge dp^\alpha = 0 \,, \\
  n_\alpha \wedge n^\alpha &= - dk_\alpha \wedge dk^\alpha \,.
 \end{split} \label{eq:nAlgCondition}
\end{equation}
Thus, the generalised vector field simplifies to 
\begin{equation}
J_{\bar{1}} = \frac{1}{\sqrt{2}}\left(4\,R\,c_6\, n^\alpha + \frac{16\,R^3\,c_6^2}{3}\, k_\alpha n^\alpha \wedge vol_{S^2}\right) \,.
\end{equation}

The conditions \eqref{eq:nAlgCondition} fix $n^\alpha$ up to one degree of freedom. The explicit form of $n^\alpha$ depends on the precise relation between $dk^\alpha$ and $dp^\alpha$. For example, if we impose the Cauchy-Riemann equations \eqref{eq:CR}, then $n^\alpha$ can be nicely expressed in terms of the holomorphic function $f^\alpha = - p^\alpha + i\, k^\alpha$ and complex coordinate $z = x_1 + i\, x_2$ on $\Sigma$
\begin{equation}
 n^\alpha = \frac12 g\, \partial f^\alpha\, d\bar{z} + \frac12 \bar{g}\, \bar{\partial} \bar{f}^\alpha dz \,. \label{eq:nSingleParam}
\end{equation}
Here $g \in \UO$ is the single degree of freedom left in $n^\alpha$.

The differential condition
\begin{equation}
 \gL_{J_{\bar{1}}} \hK = 0 \,,
\end{equation}
imposes that we must have
\begin{equation}\label{eq:nDifCondition}
 dn^\alpha = 0 \,.
\end{equation}
If we impose the Cauchy-Riemann equations then together with \eqref{eq:nSingleParam} this becomes
\begin{equation}
 \partial \left( g \partial f^\alpha \right) - c.c = 0 \,, \label{eq:AdS61VecDiffCond}
\end{equation}
where $c.c.$ stands for complex conjugate. \eqref{eq:AdS61VecDiffCond} is an equivalent conditions to having a consistent truncation.

\subsubsection{Uplift formulae} \label{s:SingletUplift}
By computing the generalised metric using \eqref{eq:SO55GenMetric1}, \eqref{eq:SO55GenMetric2} and using the ExFT / IIB SUGRA dictionary \eqref{eq:SO55GenMetricParam1}, \eqref{eq:SO55GenMetricParam2}, we can read off the consistent truncation Ansatz for the purely internal components of the metric, 2-form, 4-form and axio-dilaton. The components with some external legs can be read off from the ExFT fields of the tensor hierarchy, ${\cal A}_\mu$ and ${\cal B}_{\mu\nu}$, and using their IIB parameterisation given in section \ref{A:EFT-IIB-dictionary-TH}. Moreover, we can also compute the field strengths of IIB supergravity from the ExFT field strengths \eqref{eq:ExFTFieldStrengths}, which become \eqref{eq:6dFieldStrengthGeneral} upon plugging in the truncation Ansatz.

It is now straightforward to read off the uplift formulae for the consistent truncation including a vector multiplet by using the ExFT/IIB dictionary \ref{A:IIBtensor}. The result is best expressed in terms of the scalar fields
\begin{equation}
m_a = \left( m_A ,\, m_4,\, m_5 \right) \,,
\end{equation}
which satisfy
\begin{equation}
m_a \eta^{ab} m_b = -1 \,.
\end{equation}
Therefore, they parameterise the coset space
\begin{equation}
m_a \in \frac{\SO{4,1}}{\SO{4}} \,,
\end{equation}
and are the scalar fields of the half-maximal gauged SUGRA. They are related to the $b_u{}^a$ of the truncation Ansatz \eqref{eq:VecScalarAnsatz} up to $\SO{3}$ transformations. In particular, they satisfy
\begin{equation}
m^a m^b = \delta^{uv} b_u{}^a b_v{}^b - \eta^{ab} \,,
\end{equation}
so that $m_a$ and $b_u{}^a$ parameterise the same coset space $\frac{\SO{4,1}}{\SO{4}}$. Moreover, we define
\begin{equation}
m \cdot y \equiv m_A\, y^A \,,
\end{equation}
and the $\SU{2}$-covariant derivative in the $\mbf{3}$ representation of $\SU{2}$
\begin{equation}
 \begin{split}
  \tD y^A &= dy^A + \frac{3}{\sqrt{2}\,R} \epsilon^{ABC} A_B\, y_C \,, \\
  \tD m^A &= dm^A + \frac{3}{\sqrt{2}\,R} \epsilon^{ABC} A_B\, m_C \,, \\
 \end{split}
\end{equation}
Similarly, we define the $\SU{2}$-covariantised 1-forms
\begin{equation}
 \tTh_A = \epsilon_{ABC}\, y^B\, \tD y^C \,,
\end{equation}
and the $\SU{2}$-covariantised volume on $S^2$
\begin{equation}
 \tvol = \frac12 \epsilon_{ABC} y^A\, \tD y^B \wedge \tD y^C \,.
\end{equation}

In all our uplift formulae, we will throughout impose the Cauchy-Riemann equations \eqref{eq:CR} on $k^\alpha$ and $p^\alpha$, so that $n^\alpha$ is given by \eqref{eq:nSingleParam}, although one can use the above method to derive the uplift formulae in a different gauge as well. Then, with the above conventions, the metric is given by
\begin{equation}
 \begin{split}
  ds^2 &= \frac{4\,c_6\,R^2\,r^{5/4} |dk|^{3/2}}{3^{3/4}\bar{\Delta}^{3/4}} \left[ \frac{3\, \bar{\Delta}}{R^2\,r\,|dk|^2} ds_6^2 + X^2 \left( \delta_{AB} \tD y^A \otimes \tD y^B + w \otimes w - \frac{1}{r^2} p_\alpha p_\beta n^\alpha \otimes n^\beta \right) \right. \\
  & \left. + \frac{\tilde{\Delta}}{X^2\,r^2\,|dk|^2} dk^\alpha \otimes dp_\alpha - \frac{3}{X^2\,r} n_\alpha \otimes \left( m_4\, dk^\alpha - m \cdot y\, dp^\alpha \right) \right] \,,
 \end{split} \label{eq:SingletMetric}
\end{equation}
where
\begin{equation}
 \begin{split}
  \bar{\Delta} &= X^4 |dk|\, p_\alpha\,p_\beta \left( m_5 \partial_\gamma k^\alpha \partial^\gamma p^\beta + n^{\alpha\gamma} \left( \left(m \cdot y\right) \partial_\gamma p^\beta - m_4 \partial_\gamma k^\beta \right) \right) \\
  & \quad + 3 \,r\,|dk|^2 \left(m_5^2 - m_4^2 - \left(m \cdot y\right)^2\right) \,, \\
  \tilde{\Delta} &= 3\, r\, m_5 |dk|^2 + X^4 |dk| p_\alpha p_\beta \partial_\gamma k^\alpha \partial^\gamma p^\beta \,, \\
  w &= m_A\, \tD y^A + \frac{1}{3\,r^2} p_\alpha\, \sigma^\alpha \,, \\
  \sigma^\alpha &= 3\,r \left( m_5\, n^\alpha - m_4\, dp^\alpha - m \cdot y\, dk^\alpha \right) - X^4 p^\alpha\, p_\beta\, {\star_2 n^\beta} \,.
 \end{split}
\end{equation}
Here $\star_2 n^\alpha$ denotes the Hodge dual of $n^\alpha$ with respect to the flat metric on the Riemann surface.

The axio-dilaton is given by
\begin{equation}
 \begin{split}
  H^{\alpha\beta} &= \frac{X^4\, p^\alpha\,p^\beta\, |dk|}{\sqrt{3\,r\,\bar{\Delta}}} + \sqrt{\frac{3\,r}{\bar{\Delta}}} \left( m_5\, \partial_\gamma k^\alpha \partial^\gamma p^\beta + n^{\alpha\gamma} \left( m \cdot y\, \partial_\gamma p^\beta - m_4 \partial_\gamma k^\beta \right) \right) \,,
 \end{split}
\end{equation}
and the 2-form by
\begin{equation}
 \begin{split}
  C_{(2)}{}^\alpha &= - \frac{4\,c_6 R^2}{3} \tvol \left( k^\alpha + L^\alpha \right) - \frac{4\,c_6 R^2}{3} \frac{|dk|^2}{\bar{\Delta}} \sigma^\alpha \wedge \tTh_A\, m^A \\
  & \quad + 2 \sqrt{2}\, c_6 R \left( k^\alpha\, \tilde{F}_{(2)}{}^A y_A + p^\alpha\, \tilde{F}_{(2)}{}^4 + A^{\bar{1}} \wedge n^\alpha \right) \,,
 \end{split} \label{eq:2formSinglet}
\end{equation}
where we have defined the $\SL{2}$-doublet function
\begin{equation}
 \begin{split}
  L^\alpha &= \frac{X^4\, r\,|dk|}{\bar{\Delta}} p_\beta \left[ m_5 \partial_\gamma k^\beta \partial^\gamma p^\alpha + n^{\beta\gamma} \left(m \cdot y\, \partial_\gamma p^\alpha - m_4 \partial_\gamma k^\alpha \right) \right] \,.
 \end{split}
\end{equation}
Moreover,
\begin{equation}
 \begin{split}
  \tilde{F}_{(2)}{}^A &= dA^A + \frac{3}{2\sqrt{2}\,R} \epsilon^{ABC}\, A_B \wedge A_C \,, \\
  \tilde{F}_{(2)}{}^4 &= dA^4 + \frac{\sqrt{2}}{R} B\,, \\
  \tilde{F}_{(2)}{}^{\bar{1}} &= dA^{\bar{1}} \,,
 \end{split}
\end{equation}
are the 6-dimensional two-form field strengths as defined in \eqref{eq:6dFieldStrengthGeneral}, using \eqref{eq:SO55DiffConditions} and \eqref{eq:SO55Lambda}. In constructing $C_{(2)}{}^\alpha$ from the truncation Ansatz \eqref{eq:VecScalarAnsatz} and \eqref{eq:VecD6Ansatz}, we have performed a gauge transformation to write the 2-form in terms of the field strengths of the 6-dimensional gauged SUGRA, just like we did in the minimal case in going from \eqref{eq:AdS6Trunc} to \eqref{eq:MinimalC2F}. When $n^\alpha$ is exact (it must always be closed), i.e. $n^\alpha = d\chi^\alpha$, e.g. if $H^{(1)}\left(\Sigma\right) = 0$, we can perform a further gauge transformation to write the 2-form as
\begin{equation}
 \begin{split}
  C_{(2)}{}^\alpha &= - \frac{4\,c_6 R^2}{3} \tvol \left( k^\alpha + L^\alpha \right) - \frac{4\,c_6 R^2}{3} \frac{|dk|^2}{\bar{\Delta}} \sigma^\alpha \wedge \tTh_A\, m^A \\
  & \quad + 2 \sqrt{2}\, c_6 R \left( k^\alpha\, \tilde{F}_{(2)}{}^A y_A + p^\alpha\, \tilde{F}_{(2)}{}^4 + \chi^\alpha\, \tilde{F}_{(2)}{}^{\bar{1}} \right) \,.
 \end{split} \label{eq:2formSingletExact}
\end{equation}
The self-dual 5-form of IIB supergravity is given by
\begin{equation}
 F_{(5)} = F_{(1,4)} + F_{(2,3)} + F_{(3,2)} + F_{(4,1)} + F_{(5,0)} \,,
\end{equation}
with
\begin{equation}
 \begin{split}
  F_{(1,4)} &= \frac{16\,c_6^2 R^4}{3} \frac{|dk|^3\,r}{\bar{\Delta}} \epsilon^{ABC} y_A\, m_B\, \tD m_C \wedge \tvol \wedge vol_\Sigma \,, \\
  F_{(2,3)} &= \frac{8\sqrt{2}\,c_6^2 R^3\,|dk|}{3} \left[ \tilde{F}_{(2)}{}^A \wedge \tTh_A \wedge vol_\Sigma \right. \\
  & \quad \left. + \frac{|dk|}{\bar{\Delta}} \left( 6\,r\,|dk| \left( \frac12 \left( \left( m \cdot y \right) y_A + m_A \right) \tilde{F}_{(2)}{}^A  + m_4\, \tilde{F}_{(2)}{}^4 + m_5\, \tilde{F}_{(2)}{}^{\bar{1}} \right) \right. \right. \\
  & \quad \left. \left. + X^4\,p_\alpha\,p_\beta \left( n^{\alpha\gamma} \left( \partial_\gamma k^\beta \tilde{F}_{(2)}{}^4 - \partial_\gamma p^\beta\, y_A\, \tilde{F}_{(2)}{}^A \right) + \partial_\gamma k^\alpha \partial^\gamma p^\beta\, \tilde{F}_{(2)}{}^{\bar{1}} \right) \right) \wedge vol_\Sigma \wedge \tTh_B\, m^B \right. \\
  & \quad + \frac{X^4\,r\,|dk|}{\bar{\Delta}} p_\alpha \left( \tilde{F}_{(2)}{}^A \wedge \left( y_A \lambda^\alpha + m_A\, {\star_2 n^\alpha} \right) - \tilde{F}_{(2)}{}^4 \wedge \rho^\alpha + \tilde{F}_{(2)}{}^{\bar{1}} \wedge \tsigma^\alpha \right) \wedge \tvol \,, \\
  F_{(3,2)} &= 16\, c_6^2 R^2 \frac{r^2 |dk|^2}{\bar{\Delta}} \tilde{F}_{(3)} \wedge \left( \left( m_5^2 - m_4^2 - \left( m\cdot y\right)^2 \right) \tvol - \omega \wedge \tTh_B m^B \right) \\
  & \quad + 16\, c_6^2 R^2 |dk| X^{-4} \left(\star_6 \tilde{F}_{(3)}\right) \wedge vol_\Sigma \,, \\
  F_{(4,1)} &= \star_{10} F_{(2,3)} \,, \\
  F_{(5,0)} &= 48\, c_6^2\, r\, \epsilon^{ABC} y_A\, m_B \, \star_6 \tD m_A \,,
 \end{split}
\end{equation}
where
\begin{equation}
 \begin{split}
  \lambda^\alpha &= m_5\, dp^\alpha - m_4\, n^\alpha \,, \\
  \rho^\alpha &= m_5\, dk^\alpha - \left( m \cdot y \right) n^\alpha - m_4\, {\star_2 n^\alpha} \,,
 \end{split}
\end{equation}
and
\begin{equation}
 {\star_2 \sigma}^\alpha = m_5\, {\star_2 n^\alpha} + \left( m \cdot y \right) dp^\alpha - m_4\, dk^\alpha \,,
\end{equation}
is the Hodge dual of $\sigma^\alpha$ with respect to the flat metric on $\Sigma$. $F_{(p,q)}$ are the $\SU{2}_R$ covariantised components of the 5-form field strength with $p$ external and $q$ internal legs. $\star_{10}$ refers to the Hodge dual operator with respect to the full 10-dimensional metric \eqref{eq:SingletMetric}, while $\star_6$ refers to the Hodge dual operator of the metric of the six-dimensional gauged SUGRA whose line element is $ds_{6}^2$. $\tilde{F}_{(3)}$ is as defined in \eqref{eq:6dFieldStrengthMinimal} the field strength of the two-form
\begin{equation}
 \tilde{F}_{(3)} = dB_{(2)} \,,
\end{equation}
In the above, we have used \eqref{eq:6dDuality} to replace $\tilde{G}_{(3)}$ by $X^{-4} \star_6 \tilde{F}_{(3)}$.  The self-duality of the five-form relates the components $F_{(p,q)}$ to $F_{(6-p,4-q)}$.In particular, it implies that $F_{(3,2)}$ should be self-dual, which can easily be checked using \eqref{eq:SingletMetric}. This provides a non-trivial check of the truncation Ansatz. Moreover, we have used the self-duality of the 5-form to compute $F_{(5,0)}$ and $F_{(4,1)}$ from $F_{(1,4)}$ and $F_{(2,3)}$ rather than using the truncation Ansatz of section \ref{s:VecTruncationAnsatz}.

\subsection{Multiple singlets under $\SU{2}_R$} \label{s:2Singlets}
We next consider the situation where we have $N\le 4$ vector multiplets transforming as singlets under $\SU{2}_R$. The corresponding consistent truncation will lead to a half-maximal gauged SUGRA with gauge group $\SU{2} \times G$, where as we will see we can only have $G = \UO$ or $G = \UO^2$. Following the same logic as in the case for one single vector multiplet, the most general solution to the algebraic conditions \eqref{eq:VecAlgCon} and differential conditions
\begin{equation}
 \gL_{J_A} J_{\bar{u}} = 0 \,, \qquad \gL_{J_{\bar{u}}} \hK = 0 \,, \qquad \gL_{J_{\bar{u}}} J_{\bar{v}} = - f_{\bar{u}\bar{v}}{}^{\bar{w}} J_{\bar{w}} \,,
\end{equation}
is
\begin{equation}
J_{\bar{u}} = \frac{1}{\sqrt{2}}\left(4\,R\,c_6\, n_{\bar{u}}{}^\alpha + \frac{16\,R^3\,c_6^2}{3}\, k_\alpha n_{\bar{u}}{}^\alpha \wedge vol_{S^2}\right)  \,, \qquad\text{with }\bar{u}=1,\ldots,N\,,
\end{equation}
where the $n_{\bar{u}}{}^\alpha$ have to satisfy
\begin{equation}
 \begin{split}
  n_{\bar{u}\,\alpha} \wedge dk^\alpha &= n_{\bar{u}\,\alpha} \wedge dp^\alpha = 0 \,, \\
  n_{\bar{u}\,\alpha} \wedge n_{\bar{v}}{}^\alpha &= - \delta_{\bar{u}\bar{v}}\, dk_\alpha \wedge dk^\alpha \,,
 \end{split}\label{eq:nMultAlgCondition}
\end{equation}
as well as
\begin{equation}
 d\,n_{\bar{u}\,\alpha} = 0 \,. \label{eq:nMultDiffCondition}
\end{equation}
As in the one vector multiplet case, we can solve the algebraic conditions \eqref{eq:nMultAlgCondition} by
\begin{equation}
 n_{\bar{u}}{}^\alpha = \frac12 g_{\bar{u}}\, \partial f^\alpha\, d\bar{z} + \frac12 \bar{g}_{\bar{u}}\, \bar{\partial} \bar{f}^\alpha dz \,, \label{eq:nSingleParam-2vec}
\end{equation}
with $g_{\bar{u}} \in \UO$. But the second of the conditions \eqref{eq:nMultAlgCondition} now imposes that
\begin{equation}
g_{\bar{u}}\bar{g}_{\bar{v}}+\bar{g}_{\bar{u}}g_{\bar{v}}= 2\, \delta_{\bar{u}\bar{v}} \,.
\end{equation}
It is easy to check that these conditions can only be solved when $N\leq 2$, which implies that consistent truncations with $N=3,\,4$ vector multiplets which are singlets under $\SU{2}_R$ cannot exist. For the case $N=2$, the condition is solved by
\begin{equation}
 g_{\bar{2}}= \pm i \,g_{\bar{1}}\,, \qquad g_{\bar{1}} \in \UO \,.
\end{equation} 
Without loss of generality, we can take $g_{\bar{2}} = i\, g_{\bar{1}}$, by suitably redefining the scalar fields of the truncation Ansatz. In this case, the differential conditions \eqref{eq:nMultDiffCondition} implies
\begin{equation}
 \partial \left( g_{\bar{1}} \partial f^\alpha \right) = 0 \,,
\end{equation}
which admits non-trivial solutions only in the cases where 
\begin{equation}
 \partial f^2=\lambda\,\partial f^1\,,
\end{equation}
with $\lambda$ a constant. In this case,
\begin{equation}
g_{\bar{1}}=e^{i\,c}\,\frac{\bar{\partial}\bar{f}^1}{\partial f^1}=e^{i\,c}\,\frac{\bar{\partial}\bar{f}^2}{\partial f^2}\,,
\end{equation}
where $c$ is a real integration constant.

Recall that $J_4 \propto d\hK$ immediately implies
\begin{equation}
 \gL_{J_4} J_{\bar{u}} = 0 \,,
\end{equation}
while one can also easily check that
\begin{equation}
 \gL_{J_{\bar{u}}} J_{\bar{v}} = 0 \,.
\end{equation}
Therefore, the consistent truncation leads to a $\SU{2} \times \UO^2$ gauged SUGRA.

One can then wonder whether AdS$_6$ vacua described by two holomorphic functions satisfying the relation $\partial f^2=\lambda\,\partial f^1$ exist. Firstly, we see that this rules out having $\SL{2}$ monodromies. Moreover, we observe that, in this situation, 
\begin{equation}
|dk|=\frac12 i\,\partial f^\alpha\,\bar{\partial}\bar{f}_\alpha=\frac12 i(\lambda-\bar\lambda)\,|\partial f^1|^2\,. \label{eq:2Singletdk}
\end{equation}
However, as explained in section \ref{s:AdS6Sol} and \cite{DHoker:2016ysh,DHoker:2017mds,DHoker:2017zwj}, any globally regular vacuum must be described by functions satisfying the condition $r \geq 0$ and $|dk|\geq 0$, with equality on the boundary of the Riemann surface $\Sigma$. The latter ensures that the total space has no boundary. For \eqref{eq:2Singletdk} this condition implies that $\lambda\neq\bar\lambda$ and that $\partial f^1 = \partial f^2 = 0$ on the boundary of $\Sigma$. However, since $\Sigma$ is compact, we must have $\partial f^1 = \partial f^2 = 0$ everywhere. Therefore, although the differential and algebraic conditions for consistent truncations with two vector multiplets can be locally solved, there are no half-supersymmetric compactifications to AdS$_6$ vacua with an internal space without boundaries that allow such a consistent truncation.

\subsubsection{Uplift formulae for two singlets under $\SU{2}_R$} \label{s:2SingletsUplift}
As we discussed above, a consistent truncation with two vector multiplets and gauge group $\SU{2}$ around an AdS$_6$ vacua of IIB SUGRA necessarily requires the internal space to have a boundary. Although this is not particularly interesting from a holographic perspective, we can nonetheless use the formalism described in \cite{Malek:2017njj} to derive the consistent truncation. For simplicity, we will only give the truncation Ansatz which preserves the full $\SO{5,2}$ symmetry of the AdS vacuum since this is sufficient for a wide variety of applications. Therefore, we will consider the case where only the scalar fields of the six-dimensional gauged SUGRA are non-zero and depend only on the internal four coordinates. Moreover, as in the case of only one $\SU{2}_R$ singlet, we will impose the Cauchy-Riemann equations \eqref{eq:CR} on $k^\alpha$ and $p^\alpha$ throughout. However, it is straightforward to obtain the uplift formulae in a different gauge.

The scalar manifold of the six-dimensional SUGRA obtained from the consistent truncation is
\begin{equation}
 M_{scalar} = \frac{\SO{4,2}}{\SO{4} \times \SO{2}} \,,
\end{equation}
and can be parameterised by $m_i{}^a$ with $a = 1, \ldots, 6$ labelling the vector representation of $\SO{4,2}$ and $i = 1, 2$ the doublet of $\SO{2}$. The $m_i{}^A$ must satisfy
\begin{equation}
 m_i{}^a m_{j}{}^b \eta_{ab}= - \delta_{ij} \,, \label{eq:SO42ScalarCoset}
\end{equation}
and are related to the $b_u{}^a$ of \eqref{eq:VecScalarAnsatz} by
\begin{equation}
 \delta^{ij} m_i{}^a m_j{}^b = \delta^{uv} b_u{}^a b_v{}^b - \eta^{ab} \,.
\end{equation}
Moreover, we can decompose $\SO{4,2} \longrightarrow \SO{3} \times \SO{2}$ such that
\begin{equation} 
 \mbf{6} \longrightarrow \mbf{\left(3,1\right)} \oplus \mbf{\left(1,1\right)} \oplus \mbf{\left(1,2\right)} \,.
\end{equation}
We accordingly write
\begin{equation}
 m_i{}^a = \left( m_i{}^A,\, m_i,\, \lambda_i{}^{\bar{u}} \right) \,,
\end{equation}
where $\lambda_i{}^{\bar{u}}$ are constrained by \eqref{eq:SO42ScalarCoset}, i.e.
\begin{equation}
 \lambda_i{}^{\bar{u}} \lambda_j{}^{\bar{v}} \delta_{\bar{u}\bar{v}} = m_i{}^A m_j{}^B \delta_{AB} + m_i m_j + \delta_{ij} \,.
\end{equation}

The uplift formulae can be conveniently formulated in terms of
\begin{equation}
 \begin{split}
  n_i{}^{\alpha} &= \lambda_i{}^{\bar{u}}\, n_{\bar{u}}{}^{\alpha} \,, \\
  \omega_i{}^{\alpha} &= \left(m_i \cdot y\right) dk^\alpha + m_i\, dp^\alpha \,, \\
  w_i &= m_i{}^A\, dy_A - \frac1r p_\alpha\, \omega_i{}^{\alpha} - \frac1r p_\alpha\, n_i{}^{\alpha} \,, \\
  \sigma &= |\lambda| - \epsilon^{ij} m_i\, \left(m_j \cdot y\right) \,, \\
  \bar{\Delta} &= X^4\, |dk|\, p_\alpha p_\beta \left( \sigma\, \partial_\gamma k^\alpha \partial^\gamma p^\beta - \epsilon^{ij} n_{i}{}^{\alpha\gamma}\, \omega_{j}{}^{\beta}{}_\gamma \right) + 3\, r\, |dk|^2 \left[ |\lambda|^2 \right. \\
  & \quad + \left.\epsilon^{ik} \epsilon^{jl} \left( - \lambda_k{}^{\bar{u}} \lambda_l{}^{\bar{v}} \delta_{\bar{u}\bar{v}} \left( m_i m_j + \left( m_i \cdot y \right) \left( m_j \cdot y\right) \right) + m_i\, m_j \left( m_k \cdot y \right) \left( m_l \cdot y \right) \right) \right]
 \end{split}
\end{equation}
where $|\lambda|$ denotes the determinant of the $2\times 2$ matrix $\lambda_i{}^{\bar{u}}$. Just as in the singlet vector multiplet case, we use the shorthand
\begin{equation}
 m_i \cdot y = m_i{}^{A} \, y_A \,.
\end{equation}

The metric is given by
\begin{equation}
 \begin{split}
  ds^2 &= \frac{4\,c_6\,R^2\,r^{5/4} |dk|^{3/2}}{3^{3/4}\bar{\Delta}^{3/4}} \left[ \frac{3\, \bar{\Delta}}{R^2 r\,|dk|^2} ds_6^2 + X^2 \left( \delta_{AB} dy^A \otimes dy^B + \delta^{ij} w_i \otimes w_j \right) \right. \\
  & \quad \left. + \frac{3}{X^2\,r} \left(  \left( \sigma + 2\,\epsilon^{ij} m_i m_j{}^u y_u \right) dk^\alpha \otimes dp_\alpha + \epsilon^{ij} n_{i\,\alpha} \otimes \omega_j{}^\alpha \right) \right] \,,
 \end{split}
\end{equation}
the axio-dilaton by
\begin{equation}
 H^{\alpha\beta} = \frac{X^4\,p^\alpha\,p^\beta\,|dk|}{\sqrt{3\,r\,\bar{\Delta}}} + \sqrt{\frac{3\,r}{\bar{\Delta}}} \left(\sigma\, \partial_\gamma k^\alpha \partial^\gamma p^\beta - \epsilon^{ij} n_i{}^{\alpha\gamma} \omega_j{}^{\beta}{}_\gamma \right) \,.
\end{equation}
and the 2-form by
\begin{equation}
 \begin{split}
  C_{(2)}{}^\alpha &= - \frac{4\,c_6 R^2}{3} vol_{S^2} \left( k^\alpha + \frac{X^4\,r\,|dk|}{\bar{\Delta}} p_\beta \left[ \sigma\, \partial_\gamma k^\beta \partial^\gamma p^\alpha - \epsilon^{ij} n_i{}^{\beta\gamma}\, \omega_j{}^\alpha{}_\gamma \right] \right) \\
  & \quad + \frac{4 c_6 R^2 |dk|^2}{3\bar{\Delta}}\, \left( X^4\, p^\alpha\, p_\beta \epsilon^{ij} + 3\, r\, \delta^\alpha_\beta\, \delta^{ij} \right) \left( \omega_j{}^\beta + n_j{}^\beta \right) \wedge \Theta_A m_i{}^A  \\
  & \quad - \frac{2 c_6 R^2 |dk|^2}{3\bar{\Delta}} r \left( \epsilon_{ABC} \epsilon^{kl} m_k{}^A m_l{}^B y^C \right) \epsilon^{ij} \left(\omega_i{}^\alpha + n_i{}^\alpha \right) \wedge w_j \,.
 \end{split}
\end{equation}
Since we are considering the subsector of the truncation where only the scalar fields are turned on and are constant, the IIB five-form field strength vanishes
\begin{equation}
 F_{(5)} = 0 \,.
\end{equation}

\subsection{Triplet under $\SU{2}_R$}
We next move to the case where we have $N=3$ vector multiplets transforming as a triplet of $SU(2)_R$, i.e.
\begin{equation}
 \gL_{J_{A}} J_{\bar{B}} = - \frac{3}{\sqrt{2}\,R} \epsilon_{A\bar{B}}{}^{\bar{C}} J_{\bar{C}} \,,
 \label{eq:TripletVec}
\end{equation}
where $\bar{A} = 1, 2, 3$. As we will see shortly, this leads to a $\mathrm{ISO}(3)$ gauged SUGRA. Equation \eqref{eq:TripletVec} implies that the most general ansatz for the fields $J_{\bar{A}}$ must be of the form given in \eqref{eq:SO55JAnsatz}. The algebraic conditions \eqref{eq:VecExtraAlgCon} then fix $\J_{\bar{A}}$ to be
\begin{equation}
J_{\bar{A}}=\frac{1}{\sqrt{2}}\left(\frac{3}{R}v_{\bar{A}}+4\,c_6\,R\,y_{\bar{A}}\,\pi^\alpha+4\,c_ 6\,R\,k^\alpha dy_{\bar{A}}+\frac{16\,c_6^2\, R^3}{3}\,y_{\bar{A}}\,k_\alpha\pi^\alpha\wedge vol_{S^2}-\frac{16\,c_6^2 R^3}{3}|\pi|\theta_{\bar{A}}\wedge vol_\Sigma  \right), \label{eq:6DTripletJ}
\end{equation}
where $|\pi|=\frac{1}{2}\pi^\alpha{}_\beta\pi_\alpha{}^\beta$ and $\pi^\alpha$ is a $\SL{2}$-doublet of one-forms on $\Sigma$ satisfying the algebraic conditions
\begin{equation}
 \begin{split}
  \pi_\alpha \wedge dk^\alpha &= \pi_\alpha \wedge dp^\alpha = 0 \,, \\
  \pi_\alpha \wedge n^\alpha &= - dk_\alpha \wedge dk^\alpha \,.
 \end{split} \label{eq:piAlgCondition}
\end{equation}
Furthermore, in order to satisfy the differential condition
\begin{equation}
\gL_{J_{\bar{A}}}\hat{K}=0\,,
\end{equation}
one needs to impose the conditions
\begin{equation}
\begin{split}
p_\alpha\pi^\alpha &= p_\alpha dk^\alpha\\
d\pi^\alpha &=\frac{1}{r}\,p^\alpha\,\pi_\beta\wedge\pi^\beta\,.
\end{split}\label{eq:piDiffCondition}
\end{equation}
As in the case of the singlet vector multiplet, conditions \eqref{eq:piAlgCondition} can are solved by 
\begin{equation}\label{eq:piSingleParam}
\pi^\alpha=\frac12 g_\pi\, \partial f^\alpha\, d\bar{z} + \frac12 \bar{g}_\pi\, \bar{\partial} \bar{f}^\alpha dz \,,
\end{equation}
where again $g_\pi\in\UO$. In this case, however, the first condition of \eqref{eq:piDiffCondition} fixes the phase $g_\pi$ to
\begin{equation}\label{eq:piSingleParam-g}
g_\pi=i\frac{p_\alpha\bar\partial \bar  f^\alpha}{p_\beta\partial f^\beta}\,,
\end{equation}
thereby fixing the one-forms $\pi^\alpha$ completely. The second equation in \eqref{eq:piDiffCondition} then give an extra differential condition on $f^\alpha$ that has to be satisfied for the vacua to allow consistent truncations with a $SU(2)_R$ triplet of vector multiplets.

Using \eqref{eq:6DTripletJ} and the above differential conditions, we find
\begin{equation}
 \begin{split}
  \gL_{J_{\bar{A}}} J_{\bar{B}} &= - \frac{3\sqrt{2}}{R} \epsilon_{\bar{A}\bar{B}}{}^{\bar{C}} J_{\bar{C}} + \frac{3}{\sqrt{2}\,R} \epsilon_{\bar{A}\bar{B}}{}^C J_{C} \,,  \qquad \gL_{J_4} J_{\bar{A}} = 0 \,,
 \end{split}
\end{equation}
where $\bar{A}$ and $A$ are raised/lowered with $\delta_{\bar{A}\bar{B}}$ and $\delta_{AB}$, respectively. Together with the relations
\begin{equation}
 \begin{split}
  \gL_{J_{A}} J_{B} &= - \frac{3}{\sqrt{2}\,R} \epsilon_{AB}{}^C J_C \,, \\
  \gL_{J_{A}} J_{\bar{B}} &= - \frac{3}{\sqrt{2}\,R} \epsilon_{A\bar{B}}{}^{\bar{C}} J_{\bar{C}} \,, \\
  \gL_{J_{4}} J_a &= \gL_{J_a} J_4 = 0 \,,
 \end{split}
\end{equation}
this implies that the gauge group of the six-dimensional half-maximal gauged SUGRA is $\mathrm{ISO}(3)$.

\subsubsection{Uplift formulae} \label{s:TripletUplift}
Just as for the case of two vector multiplets forming $\SU{2}$ singlets, we will here only give the consistent truncation Ansatz preserving the $\SO{5,2}$ symmetry of the AdS$_6$ vacuum, i.e. where the scalar fields are the only non-zero fields of the six-dimensional half-maximal gauged SUGRA and are constant. The full consistent truncation Ansatz including general values for all gauge fields of the six-dimensional gauged SUGRA can be obtained as discussed above \ref{s:VecTruncationAnsatz} and demonstrated explicitly for the case of a single vector multiplet in section \ref{s:SingletUplift}.

The scalar manifold of the six-dimensional half-maximal gauged SUGRA is
\begin{equation}
 M_{scalar} = \frac{\SO{4,3}}{\SO{4} \times \SO{3}} \times \mathbb{R}^+ \,.
\end{equation}
We will parameterise the coset space $\frac{\SO{4,3}}{\SO{4}\times\SO{3}}$ by
\begin{equation}
 m_I{}^a = \left( m_I{}^A - \lambda_I{}^A ,\, m_I,\, \lambda_I{}^A \right) \,,
\end{equation}
where $I = 1, 2, 3$ and which satisfies
\begin{equation}
 m_I{}^a m_J{}^b \eta_{ab} = -\delta_{IJ} \,.
\end{equation}
The $m_I{}^a$ are related to the $b_u{}^a$ of \eqref{eq:VecScalarAnsatz} by
\begin{equation}
 m_I{}^a m_J{}^b \delta^{IJ} = b_u{}^a b_v{}^b \delta^{uv} - \eta^{ab} \,.
\end{equation}

The uplift formulae can be conveniently expressed in terms of
\begin{equation}
 \begin{split}
  \omega_I &= \left( m_I \cdot y \right) p_\alpha\, dk^\alpha + m_I\, p_\alpha\, dp^\alpha \,, \\
  \sigma^{\pm}_I &= \left( \lambda_I \cdot y \right) p_\alpha\, dp^\alpha \pm m_I\, p_\alpha\, dk^\alpha \,, \\
  \Lambda &= p_\alpha\, p_\beta\, \partial_\gamma k^\alpha \partial^\gamma p^\beta \,, \\
  \bar{\Delta} &= X^4 \Lambda\, |m_I{}^A| |dk| - 3\, r\, |dk|^2 \left( |m_I{}^A|\, |m_I{}^A - 2 \left( \lambda_I \cdot y \right) y^A|\, + \frac14 \left( \epsilon_{ABC} \epsilon^{IJK} m_I y^A m_J{}^B m_K{}^C \right)^2 \right) \,,
 \end{split}
\end{equation}
with
\begin{equation}
 \left( m_I \cdot y \right) = m_I{}^A y_A \,, \qquad \left( \lambda_I\cdot y \right) = \lambda_I{}^A y_A \,.
\end{equation}
The metric, axio-dilaton and 2-form are given by
\begin{equation}
 \begin{split}
  ds^2 &= \frac{4\,c_6 R^2 r^{5/4} |dk|^{3/2}}{3^{3/4}\bar{\Delta}^{3/4}} \left[ \frac{3\, \bar{\Delta}}{R^2 r\,|dk|^2} ds_6^2 \right. \\
  & \quad \left. + X^2 \left( m_I{}^A dy_A - \frac{1}{r} \omega_I \right) \otimes \left( m^{I\,B} dy_B - \frac1{r} \omega^I \right) + \frac{X^2}{r^2} p_\alpha p_\beta dp^\alpha \otimes dp^\beta \right. \\
  & \quad \left. + \frac{3\,|dk|}{X^2 \Lambda\,r} \left( |m_I{}^A| p_\alpha p_\beta \left( dk^\alpha \otimes dk^\beta - dp^\alpha \otimes dp^\beta \right) \right. \right. \\
  & \quad \left. \left. + \frac12 \epsilon_{ABC} \epsilon^{IJK} y^A m_J{}^B m_K{}^C p_\alpha \left( \sigma^+_I \otimes dp^\alpha + dp^\alpha \otimes \sigma^+_I \right) \right) \right] \,, \\
  H^{\alpha\beta} &= \frac{1}{\sqrt{\bar{\Delta}}} \left[ \left( \frac{ X^4\, |dk|}{\sqrt{3\,r}} - \frac{2\,\sqrt{3\,r}}{\Lambda} |m_I{}^A | |dk|^2 \right) p^\alpha p^\beta + \sqrt{3\,r}\, |m_I{}^A| \partial_\gamma k^\alpha \partial^\gamma p^\beta  \right. \\
  & \quad \left. + \frac{\sqrt{3\,r}}{2\,\Lambda} |dk|\, \epsilon^{IJK} \epsilon_{ABC} y^A m_I{}^B m_J{}^C\, \sigma^-_K{}^{\gamma} \left( p^\alpha \partial_\gamma p^\beta + p^\beta \partial_\gamma p^\alpha \right) \right] \,, \\
  \sqrt{\bar{\Delta}} H^{\alpha\beta} C_{(2)\,\beta} &= \frac{\sqrt{3}\,|dk|}{2\sqrt{r}} \left[ -\epsilon^{IJK} m_J{}^A \lambda_{K\,A} \omega_I{}^\gamma \partial_\gamma p^\alpha + 2\,p^\alpha\,|dk| \epsilon^{IJK} m_I m_J{}^A \lambda_K{}^B \left( \delta_{AB} - y_A y_B\right) \right] vol_{\Sigma} \\
  & \quad + \sqrt{3\,r}\,|dk| \epsilon^{IJK}\, m_I{}^A\, dy_A \wedge \Bigg[ \frac{2\,p^\alpha}{\Lambda} p_\beta\, |dk| \left( m_J{}^B \lambda_K{}^C \left( \delta_{BC} - y_B\, y_C \right) dk^\beta \right. \Bigg. \\
  & \quad \Bigg. \left. + 2\,|dk|\, m_J \left( m_K{}^B - \lambda_K{}^B \right) y_B\, dp^\beta \right) + m_K{}^B \left( m_J\, y_B\, dk^\alpha + \lambda_{J B}\, dp^\alpha \right) \Bigg] \\
  & \quad + \frac{\sqrt{3\,r}}{2} \left( - \frac{2\,X^4 |dk|\, p^\alpha}{3\,r} \left(r + k^\beta p_\beta \right) - |m_i{}^u| k_\beta \partial_\gamma p^\beta \partial^\gamma k^\alpha + \frac{4\,p^\alpha\,|dk|^2}{\Lambda} p_\beta k^\beta |m_i{}^u| \right) vol_{S^2} \\
  & \quad -\frac{\sqrt{3\,r}\,|dk|}{2} \left( k^\alpha\,m_I + \frac{2}{\Lambda} p^\alpha\,\sigma^-_{I\,\beta} k_\gamma \partial^\beta p^\gamma \right) \epsilon^{IJK}\, m_J{}^A\, m_K{}^B\, dy_A \wedge dy_B \,,
 \end{split}
\end{equation}
and the five-form vanishes by our assumption that the scalar fields are the only non-vanishing fields and they are constant.

\subsection{Triplet plus singlet under $\SU{2}_R$}
We finally consider the possibility of having consistent truncations with four vector multiplets forming  a triplet and a singlet of $SU(2)_R$, i.e.
\begin{equation}
 \gL_{J_{A}} J_{\bar{B}} = - \frac{3}{\sqrt{2}\,R} \epsilon_{A \bar{B}}{}^{\bar{C}} J_{\bar{C}} \,, \qquad \gL_{J_A} J_{\bar{4}} = 0 \,.
\end{equation}
Since $J_4 \propto d\hK$ we automatically have
\begin{equation}
 \gL_{J_4} J_{a} = 0 \,.
\end{equation}
For a vacuum to allow such consistent truncations around it, it must allow both a truncation with a single vector multiplet, characterised by \eqref{eq:nSingleParam}, and a truncation with a triplet of vector multiplets, characterised by \eqref{eq:piSingleParam}. The resulting gauge group will clearly be $\mathrm{ISO}(3) \times \UO$. Futhermore, in order to have both simultaneously, we need to satisfy the condition
\begin{equation}
J_{\bar{4}}\wedge J_{\bar{A}}=0\,,
\end{equation}
where $\bar{A} = 1, 2, 3$ labels the triplet and $\bar{4}$ the extra singlet. Similar to the case of two singlets, the above condition fixes the phase $g$ that characterises the singlet to be (as before, up to a sign which can be absorbed by a field redefinition of the scalar fields in the truncation)
\begin{equation}\label{eq:g-3+1-AdS6}
g=-i\,g_\pi=\frac{p_\alpha\bar\partial \bar  f^\alpha}{p_\beta\partial f^\beta}\,.
\end{equation}
Therefore, a vacuum allows a consistent truncation with four vector multiplets only in the case where it allows a consistent truncation with a $SU(2)_R$ triplet of vector multiplets and a consistent truncation with a single vector multiplet characterised precisely by the phase \eqref{eq:g-3+1-AdS6}. 

\section{Conclusions} \label{s:Conclusions}
In this paper, we showed how to use ExFT to easily recover the infinite families of supersymmetric AdS$_7$ and AdS$_6$ solutions of massive IIA and IIB SUGRA, respectively, known in the literature \cite{Apruzzi:2013yva,DHoker:2016ujz,DHoker:2017mds,DHoker:2017zwj}. The ExFT description of these vacua allowed us to immediately construct the ``minimal'' consistent truncation of 10-dimensional SUGRA around these solutions \cite{Passias:2015gya,Hong:2018amk,Malek:2018zcz,Jeong:2013jfc}, in which we keep only the gravitational supermultiplet of the lower-dimensional gauged SUGRA. We then analysed whether it is possible to construct consistent truncations around the supersymmetric AdS vacua keeping more modes, which would result in lower-dimensional gauged SUGRAs coupled to vector multiplets. Assuming the method developed in \cite{Malek:2016bpu,Malek:2017njj} is the most general one for constructing consistent truncations with vector multiplets, we found that
\begin{itemize}
	\item there are no consistent truncations with vector multiplets around AdS$_7$ vacua of massive IIA, unless the Roman's mass vanishes. For vanishing Roman's mass, there is a consistent truncation that is itself a truncation (and dimensional reduction) of the maximally supersymmetric consistent truncation of 11-dimensional SUGRA on $S^4$,
	\item there are consistent truncations with vector multiplets of IIB SUGRA around its supersymmetric AdS$_{6}$ solutions. In this case, the holomorphic functions describing the AdS$_6$ solutions must satisfy further differential constraints.
\end{itemize}
In particular, we found that the only consistent truncations with vector multiplets of IIB SUGRA around the supersymmetric AdS$_6$ vacua yield $N \leq 4$ vector multiplets with gauge group $\SU{2} \times \UO$, $\SU{2} \times \UO^2$, $\mathrm{ISO}(3)$ and $\mathrm{ISO}(3) \times \UO$, when the holomorphic functions $f^\alpha$ satisfy the following differential conditions.

\paragraph{Consistent truncation with one vector multiplet}
The differential condition \eqref{eq:AdS61VecDiffCond} is
\begin{equation}
 \partial \left( g\, \partial f^\alpha \right) - c.c. = 0 \,, \label{eq:1VecDiff2}
\end{equation}
for some function $g \in \UO$, where $c.c.$ denotes the complex conjugate. While we will not attempt to find general solutions of \eqref{eq:1VecDiff2} that are holomorphic and satisfy \eqref{eq:AdS6Compactness} and \eqref{eq:AdS6RegS}, it is easy to show that if one of the holomorphic functions is linear in the complex coordinate $z$, i.e. $f^1 = A_0 + A_1\, z$, then the other function must be quadratic, i.e. $f^2 = B_0 + B_1\, z + B_2\, z^2$, where $A_0$, $A_1$, $B_0$, $B_1$ and $B_2$ are constant complex numbers. This implies that the Abelian T-dual to the Brandhuber-Oz solution \cite{Brandhuber:1999np}, which is described by a linear and quadratic holomorphic function \cite{DHoker:2016ujz,Lozano:2018pcp}, admits a consistent truncation with a single vector multiplet, while the non-Abelian T-dual to the Brandhuber-Oz solution \cite{Lozano:2012au}, which is described by a linear and cubic holomorphic function \cite{Hong:2018amk,Lozano:2018pcp}, does not. The consistent truncation Ansatz is given in section \ref{s:SingletUplift} and leads to $\mathrm{F}(4)$ gauged SUGRA coupled to one vector multiplet.

\paragraph{Consistent truncation with two vector multiplets}
The differential condition that the holomorphic functions must satisfy is now
\begin{equation}
 \partial f^2 = \lambda\, \partial f^1 \,,
\end{equation}
for some constant $\lambda$. As we discussed in section \ref{s:2Singlets}, this necessarily implies that the internal space of the AdS$_6$ solutions has a boundary. While such solutions are not interesting from a holographic perspective, we can nonetheless compute the consistent truncation Ansatz, which we have given in \ref{s:2SingletsUplift}, and which leads to $\mathrm{F}(4)$ gauged SUGRA coupled to two Abelian vector multiplets.

\paragraph{Consistent truncation with three vector multiplets}
To allow for a consistent truncation with three vector multiplets, the following differential condition must be satisfied:
\begin{equation}
 d\pi^\alpha =\frac{1}{r}\,p^\alpha\,\pi_\beta\wedge\pi^\beta\,,
\label{eq:ConPiDiffCondition}
\end{equation}
where
\begin{equation}\label{eq:ConPiSingleParam}
 \pi^\alpha=\frac12 i\frac{p_\alpha\bar\partial \bar  f^\alpha}{p_\beta\partial f^\beta}\, \partial f^\alpha\, d\bar{z} - \frac12 i\frac{p_\alpha \partial f^\alpha}{p_\beta\bar{\partial} \bar{f}^\beta}\, \bar{\partial} \bar{f}^\alpha dz \,,
\end{equation}
and $dr = - k_\alpha\, dp^\alpha$ with $p^\alpha$, $k^\alpha$ the real/imaginary parts of the holomorphic functions $f^\alpha = - p^\alpha + i\, k^\alpha$. For any pair of holomorphic functions $f^\alpha$ satisfying the above condition, there is a consistent truncation of IIB SUGRA around that AdS$_6$ solution to 6-dimensional half-maximal $\mathrm{ISO}(3)$ gauged SUGRA. The uplift formulae for the scalar fields is given in section \ref{s:TripletUplift}. It is unclear whether there are globally regular supersymmetric AdS$_6$ solutions satisfying the differential conditions \eqref{eq:ConPiDiffCondition}.

\paragraph{Consistent truncation four vector multiplets}
To admit a consistent truncation with four vector multiplets, the AdS$_6$ vacua must satisfy the differential condition for the triplet, i.e. \eqref{eq:ConPiDiffCondition} with $\pi^\alpha$ as in \eqref{eq:ConPiSingleParam}, as well as
\begin{equation}
 \partial \left( \frac{p_\beta\bar\partial \bar  f^\beta}{p_\gamma\partial f^\gamma} \partial f^\alpha \right) - c.c. = 0 \,.
\end{equation}
For any pair of holomorphic functions $f^\alpha$ satisfying the above, the corresponding AdS$_6$ solution admits a consistent truncation to 6-dimensional half-maximal $\mathrm{ISO}(3) \times \UO$ gauged SUGRA. Once again, it is unclear whether there are such globally regular supersymmetric AdS$_6$ solutions of IIB SUGRA.

\vspace{15pt}
It would be interesting to classify for which Riemann surfaces $\Sigma$ these consistent truncations exist, i.e. for which Riemann surfaces one can have holomorphic functions $f^\alpha$ which satisfy the above differential conditions and lead to closed internal manifolds, thus also satisfying \eqref{eq:AdS6RegS} and \eqref{eq:AdS6Compactness}, or even to find a complete list of such holomorphic functions. For now, we are able to say that the Abelian T-dual of the Brandhuber-Oz solution admits a consistent truncation with one vector multiplet, the non-Abelian T-dual does not, and there are no globally regular solutions that admit a consistent truncation with two vector multiplets. Moreover, the only possible gauge groups in six dimensions are $\SU{2} \times \UO$, $\SU{2} \times \UO^2$, $\mathrm{ISO}(3)$ and $\mathrm{ISO}(3) \times \UO$. This is only a subset of all possible 6-dimensional half-maximal gauged SUGRAs that admit supersymmetric AdS vacua \cite{Karndumri:2016ruc}. The other six-dimensional gauged SUGRAs do not have uplifts to IIB SUGRA.

Our results can be used to uplift the 6-dimensional solutions found in \cite{Gutperle:2017nwo,Gutperle:2018axv} and to complete their holographic study, while they also suggest that there are no IIB uplifts of the 6-dimensional solutions \cite{Karndumri:2014lba} which requires the six-dimensional gauge group $\SU{2} \times \SU{2} \left( \times \UO \right)$. Similarly, we found that of all the 7-dimensional half-maximal gauged SUGRAs that admit a supersymmetric AdS$_7$ vacuum \cite{Louis:2015mka}, only the pure $\SU{2}$ gauged SUGRA \cite{Townsend:1983kk} and it coupled to an Abelian vector multiplet can be uplifted to IIA SUGRA, where in the latter case the Roman's mass is necessarily zero. This suggests that the other 7-dimensional gauged SUGRAs with supersymmetric AdS$_7$ vacua are lower-dimensional artifacts without a clear relation to 10-dimensional SUGRA.

\section*{Acknowledgements}
The authors thank Alex Arvanitakis, Davide Cassani, Yolanda Lozano, Carlos Nu\~{n}ez, Hagen Triendl, Alessandro Tomasiello, Christoph Uhlemann, Oscar Varela and Daniel Waldram for helpful discussions. EM would also like to thank the organisers of the ``2018 USU Workshop on Strings and Black Holes'' for hospitality while part of this work was completed. This project has received funding from the European Research Council (ERC) under the European Union’s Horizon 2020 research and innovation programme (``Exceptional Quantum Gravity'', grant agreement No 740209).

\appendix

\section{$\SL{5}$ ExFT conventions and ExFT/IIA dictionary} \label{A:SL5}

\subsection{Embedding IIA into $\SL{5}$ ExFT} \label{A:IIASL5}
To embed IIA SUGRA in $\SL{5}$ ExFT we decompose $\SL{5} \longrightarrow \GL{4}^+ \longrightarrow \GL{3}^+ \times \mathbb{R}^+$, where $\GL{n}^+ = \SL{n} \times \mathbb{R}^+$. The $\GL{4}$ is the geometric group realised by the internal manifold of a 11-dimensional compactification, which is broken to $\GL{3} \times \mathbb{R}^+$ by reducing to IIA SUGRA. Accordingly, we decompose an object in the fundamental $\SL{5}$ representation as
\begin{equation}
 F^a = \left( F^i ,\, F^4,\, F^5 \right) \,,
\end{equation}
where $a = 1, \ldots, 5$ is the $\SL{5}$ fundamental index and $i = 1, 2, 3$ labels the fundamental of $\GL{3}$.

We will need to decompose the generalised tensors of the half-maximal structure, i.e. generalised vector fields and generalised tensors in the $\obf{5}$ and $\mbf{5}$ representation. A generalised vector field, ${\cal A}^{ab}$, decomposes as
\begin{equation}
 {\cal A}^{i5} = V^{i} \,, \qquad {\cal A}^{ij} = - \epsilon^{ijk} \, \omega_{(1)k} \,, \qquad {\cal A}^{i4} = \frac12 \epsilon^{ijk} \, \omega_{(2)jk} \,, \qquad {\cal A}^{45} = \omega_{(0)} \,,
\end{equation}
a generalised tensor field ${\cal B}_a$ in the $\obf{5}$ as
\begin{equation}
 {\cal B}^i = \frac12 \epsilon^{ijk} \, \omega_{(2)jk} \,, \qquad {\cal B}^4 = \frac1{3!} \epsilon^{ijk} \, \omega_{(3)ijk} \,, \qquad {\cal B}^5 = \omega_{(0)} \,,
\end{equation}
and a generalised tensor field ${\cal C}^a$ in the $\mbf{5}$ as
\begin{equation}
 {\cal C}_i = \omega_{(1)i} \,, \qquad {\cal C}_4 = \omega_{(0)} \,, \qquad {\cal C}_5 = \frac1{3!} \epsilon^{ijk}\, \omega_{(3)ijk} \,,
\end{equation}
where $V$ are spacetime vector fields, $\omega_{(p)}$ are spacetime $p$-forms and $\epsilon^{ijk} = \pm 1$ denotes the three-dimensional alternating symbol, i.e. the tensor \emph{density}.

Just as in the above, we also decompose the $\SL{5}$ ``extended derivatives'' as
\begin{equation}
 \partial_{ab} = \left( \partial_{i5} ,\, \partial_{ij} ,\, \partial_{i4} ,\, \partial_{45} \right) \,,
\end{equation}
These derivatives $\partial_{i5} \neq 0$ are the usual IIA internal spacetime derivatives, and solve the $\SL{5}$ ExFT section conditions
\begin{equation}
 \partial_{[ab} \partial_{cd]} = 0 \,.
\end{equation}

\subsection{IIA parameterisation of the generalised metric} \label{A:EFT-IIAdictionary}
The IIA parameterisation of the $\SL{5}$ generalised metric is given in \cite{Malek:2015hma}. Here we translate the parameterisation given there to the string-frame metric which we use in section \ref{s:AdS7} when describing the supersymmetric AdS$_7$ vacua. The components of the generalised metric $\gM^{ab}$ are parameterised as
\begin{equation}
 \begin{split}
  \gM^{ij} &= |g|^{2/5} e^{2\psi/5} \left( g^{ij} + |g|^{-1}\, B^i\,B^j \right) \,, \\
  \gM^{i4} &=  |g|^{2/5} e^{2\psi/5} \left( - A^i + |g|^{-1} B^i\,C \right) \,, \\
  \gM^{i5} &= -|g|^{-3/5} e^{2\psi/5} B^i \,, \\
  \gM^{44} &= |g|^{2/5} e^{2\psi/5} \left( e^{-2\psi} + g_{ij} A^i A^j + |g|^{-1} C^2 \right) \,, \\
  \gM^{45} &= - |g|^{-3/5} e^{2\psi/5} C \,, \\
  \gM^{55} &= |g|^{-3/5} e^{2\psi/5} \,,
 \end{split}
\end{equation}
where $g_{ij}$ is the internal 3-dimensional IIA string frame metric, $A_i$ is the 1-form, $B_{ij}$ is the 2-form and $C_{ijk}$ is the 3-form. The 2- and 3-form appear as $B^i = \frac12 \epsilon^{ijk} B_{jk}$ and $C = \frac1{3!} \epsilon^{ijk} C_{ijk}$ where $\epsilon^{ijk} = \pm 1$ is the alternating symbol, i.e. a tensor \emph{density}.

\subsection{Including the Roman's mass} \label{A:SL5Massive}
As discussed in \cite{Ciceri:2016dmd,Cassani:2016ncu}, the Roman's mass of IIA SUGRA appears like a deformation of the differential structure of ExFT and EGG, similar to a gauging of lower-dimensional gauged SUGRAs. In particular, the generalised Lie derivative \eqref{eq:gen_diff} now takes the form
\begin{equation}
 \gL_{\xi} V^{ab} = \gL^{(0)}_{\xi} V^{ab} + \frac12 Z^{cd,[a} V^{b]e} \xi^{fg} \epsilon_{cdefg}  \,, \label{eq:DefGenLie}
\end{equation}
where $\gL^{(0)}$ is the undeformed generalised Lie derivative \eqref{eq:gen_diff} and $Z^{ab,c}$ satisfies $Z^{[ab,c]} = 0$ and encodes the deformation of the generalised Lie derivative. For the Roman's mass $m$, the only non-vanishing component of $Z^{ab,c}$ is
\begin{equation}
 Z^{45,4} = m \,.
\end{equation}

The deformation $Z^{ab,c}$ generates an $\SL{5}$ transformation and thus can easily be worked out for the generalised Lie derivative acting in another representation. In particular, to describe AdS$_7$ vacua, we require the massive generalised Lie derivative acting in the $\mbf{5}$ which is
\begin{equation}
 \gL_{\xi} {\cal B}^a = \gL_{\xi}^{(0)} {\cal B}^a - \frac14 Z^{bc,a} {\cal B}^d \xi^{ef} \epsilon_{bcdef} \,. \label{eq:DefGenLie5}
\end{equation}
The differential operators $d$ of \eqref{eq:7Dd} are also modified. Their deformations by $Z^{ab,c}$ can be determined by requiring them to be covariant under the deformed generalised Lie derivative \eqref{eq:DefGenLie}. In fact, the $d$ operator appearing in the differential conditions remains unmodified
\begin{equation}
 d {\cal C}_a = d^{(0)} {\cal C}_a \,, \label{eq:Defd}
\end{equation}
where $d^{(0)}$ is the unmodified $d: \Gamma\left({\cal R}_3\right) \longrightarrow \Gamma\left({\cal R}_2\right)$ given in \eqref{eq:7Dd}.

\section{$\SO{5,5}$ ExFT conventions and ExFT/IIB dictionary} \label{A:SO55}

\subsection{Embedding IIB into $\SO{5,5}$ ExFT} \label{A:IIBSO55}
To connect the $\SO{5,5}$ ExFT with IIB SUGRA we decompose $\SO{5,5} \longrightarrow \SL{4} \times \SL{2}_S \times \SL{2}_A$, where $\SL{2}_S$ corresponds to S-duality while $\SL{2}_A$ is an accidental symmetry in the decomposition relevant to six dimensions and which will be broken by the IIB solution to the section condition \cite{Blair:2013gqa,Hohm:2013vpa}. For our purposes, we will need the decomposition of the $\mbf{16}$ and $\mbf{10}$ representations of $\SO{5,5}$ which is
\begin{equation}
\begin{split}
\mbf{16} &\longrightarrow \left(\mbf{4},\mbf{2},\mbf{1}\right) \oplus \left(\obf{4},\mbf{1},\mbf{2}\right) \,, \\
\mbf{10} &\longrightarrow \left(\mbf{1},\mbf{2},\mbf{2}\right) \oplus \left(\mbf{1},\mbf{1},\mbf{6}\right) \,.
\end{split} \label{eq:SO55IIBDecomp}
\end{equation}

Thus, a generalised vector field becomes
\begin{equation}
 {\cal A}^M = \left( {\cal A}^{U,i},\, {\cal A}^\alpha{}_i \right) \,,
\end{equation}
where we use $i = 1, \ldots, 4$ for the $\SL{4}$ spatial indices, $\alpha = 1, 2$ as $\SL{2}_S$ indices and $U, V = +, -$ for the $\SL{2}_A$ indices. We identify these components with spacetime tensors as follows
\begin{equation}
 {\cal A}^{+,i} = V^i \,, \qquad {\cal A}^{-,i} = \frac1{3!} \epsilon^{ijkl}\, \omega_{(3)jkl} \,, \qquad {\cal A}^\alpha{}_i = \omega_{(1)}{}^\alpha{}_i \,,
\end{equation}
where $V$ is a spacetime tensor, $\omega_{(p)}$ are spacetime p-forms, $\alpha$ is as before a fundamental $\SL{2}_S$ index and $\epsilon^{ijkl}$ is the 4-dimensional alternating symbol, i.e. tensor \emph{density}.

Similarly, a tensor in the $\mbf{10}$ decomposes as ${\cal B}^I = \left( {\cal B}^{U,\alpha},\, {\cal B}^{ij} \right)$ which contain the spacetime tensors
\begin{equation}
 {\cal B}^{+,\alpha} = \omega_{(0)}{}^\alpha \,, \qquad {\cal B}^{-,\alpha} = \frac1{4!} \epsilon^{ijkl}\, \omega_{(4)ijkl} \,, \qquad {\cal B}^{ij} = \frac12 \epsilon^{ijkl}\, \omega_{(2)kl} \,,
\end{equation}
where $\omega_{(p)}$ are $p$-forms and $\alpha = 1, 2$ is an $\SL{2}_S$ index.

Furthermore, with these conventions the $\SO{5,5}$ invariant metric is given by
\begin{equation}
\eta_{IJ} = \begin{pmatrix}
\epsilon_{\alpha\beta} \epsilon_{UV} & 0 \\ 0 & \epsilon_{ijkl}
\end{pmatrix} \,,
\end{equation}
with inverse
\begin{equation}
\eta^{IJ} = \begin{pmatrix}
\epsilon^{\alpha\beta} \epsilon^{UV} & 0 \\ 0 & \epsilon^{ijkl}
\end{pmatrix} \,.
\end{equation}
We employ the following summation convention over the $\mbf{10}$ indices
\begin{equation}
{\cal B}_1{}^I {\cal B}_2{}^J \eta_{IJ} = {\cal B}_1{}^I {\cal B}_{2\,I} = {\cal B}_1{}^{U,\alpha} {\cal B}_{2\,U,\alpha} + \frac12 {\cal B}_1{}^{ij} {\cal B}_{2\,ij} \,. \label{eq:10Sum}
\end{equation}
The identity matrix in the $\mbf{10}$ takes the following form due to the summation convention \eqref{eq:10Sum}
\begin{equation}
\delta_{I}^J = \begin{pmatrix}
\delta_\alpha^\beta \delta_U^V & 0 \\ 0 & 2 \delta_{ij}^{kl}
\end{pmatrix} \,,
\end{equation}
where $\delta_{ij}^{kl} = \frac12 \left( \delta_{i}^{k} \delta_{j}^l - \delta_{i}^{l} \delta_{j}^k \right)$.

Finally, the $\left(\gamma_I\right)^{MN}$-matrices are given by
\begin{equation}
\begin{split}
\left(\gamma_{\alpha\,U}\right)^{V\,i\,\,\beta}{}_j &= \sqrt{2}\, \delta_\alpha^\beta\, \delta_U^V\, \delta^i_j \,, \\
\left(\gamma_{ij}\right)^{V\,k\,\,W\,l} &= 2 \sqrt{2} \epsilon^{VW} \delta_{ij}^{kl} \,, \\
\left(\gamma_{ij}\right)^\beta{}_k{}^\gamma{}_l &= - \sqrt{2} \epsilon_{ijkl} \epsilon^{\beta\gamma} \,,
\end{split}
\end{equation}
and the $\left(\gamma_I\right)_{MN}$-matrices are
\begin{equation}
\begin{split}
\left(\gamma_{\alpha\,U}\right)_{V\,i\,\,\beta}{}^j &= \sqrt{2} \epsilon_{\alpha\beta} \epsilon_{UV} \delta_i{}^j \,, \\
\left(\gamma_{ij}\right)_{V\,k\,\,W\,l} &= \sqrt{2} \epsilon_{VW} \epsilon_{ijkl} \,, \\
\left(\gamma_{ij}\right)_\beta{}^k{}_\gamma{}^l &= - 2 \sqrt{2} \delta_{ij}^{kl} \epsilon_{\beta\gamma} \,.
\end{split}
\end{equation}

With the above decomposition, the ``extended derivatives'' are given by
\begin{equation}
\partial_M = \left( \partial_{U,i},\, \partial_{\alpha}{}^i \right) \,
\end{equation}
with only $\partial_{+,i} \neq 0$. This corresponds to the IIB solution of the section condition \eqref{eq:section}
\begin{equation}
\left(\gamma_I\right)^{MN} \partial_M \otimes \partial_N = 0 \,,
\end{equation}
which we use.

\subsection{IIB parameterisation of the generalised metric} \label{A:IIBGM}
Here we give the IIB parameterisation of the $\SO{5,5}$ generalised metric in the $\mbf{16}$ and $\mbf{10}$ representations. The generalised metric in the $\mbf{16}$ is given by
\begin{equation}
\begin{split}
\mathcal{M}_{+i\,+j} &= e^{1/2} g_{ij} + e^{-3/2} \left( C_{(4)}^2\, g_{ij} + \frac{1}{4} C_{ik\,\alpha}\, \beta^{kr\,\alpha} C_{js\,\gamma} \beta^{st\,\gamma}\, g_{rt} \right) \\
& \quad -\frac{1}{2} e^{-3/2} C_{(4)} \left( g_{ir}\, C_{jk\,\alpha} \beta^{kr\,\alpha} + (i\leftrightarrow j)\right) + e^{1/2}\, C_{ik\,\alpha}\, C_{jl\,\gamma}\, g^{kl}\, H^{\alpha\gamma} \\
\mathcal{M}_{+i\,-j} &= e^{-3/2} \left( C_{(4)} \, g_{ij} - \frac{1}{2} C_{ik\,\alpha} \beta^{kl\,\alpha} g_{lj} \right) \,, \\
\mathcal{M}_{-i\,-j} &= e^{-3/2} g_{ij} \,, \\
\mathcal{M}_{+i\,\,\alpha}\,{}^j &= e^{-3/2} \left( C_{(4)}\, g_{ik}\, \beta^{jk}{}_\alpha - \frac{1}{2} C_{ik\,\gamma}\, \beta^{kl\,\gamma}\, g_{lm}\, \beta^{jm}{}_\alpha \right) - e^{1/2} C_{ik\,\gamma} g^{kj} H^{\gamma}{}_{\alpha} \,, \\
\mathcal{M}_{-i\,\,\alpha}\,{}^j &= e^{-3/2} g_{ik}\, \beta^{jk}{}_\alpha \,, \\
\mathcal{M}_\alpha{}^i{}_\beta{}^j &= e^{1/2} g^{ij} H_{\alpha\beta} + e^{-3/2} \beta^{ik}{}_\alpha\, \beta^{jl}{}_\beta\, g_{kl} \,.
\end{split} \label{eq:SO55GenMetricParam1}
\end{equation}
Here $g_{ij}$ is the internal 4-d Einstein-frame metric, $C_{(4)} = \frac1{4!} \epsilon^{ijkl} C_{ijkl}$ is the dual of the fully internal 4-form, $C_{ij\,\alpha}$ denotes the $\SL{2}$-dual of R-R 2-forms and $\beta^{ij}{}_\alpha = \frac12 \epsilon^{ijkl} C_{kl\,\alpha}$ is its dual. Throughout we dualise with $\epsilon^{ijkl} = \pm 1$, the four-dimensional alternating symbol, i.e. the tensor \emph{density}. $H_{\alpha\beta}$ is the $\SL{2}$ matrix parameterised by the axio-dilaton $\tau = e^{\psi} + i\, C_0$,
\begin{equation}
H_{\alpha\beta} = \frac{1}{\mathrm{Im}\,\tau}
\begin{pmatrix}
|\tau|^2 & \mathrm{Re}\,\tau \\ \mathrm{Re}\,\tau & 1
\end{pmatrix} \,. \label{eq:HParam}
\end{equation}
All our $\SL{2}_S$ indices are raised/lowered by the $\SL{2}$ invariant $\epsilon_{\alpha\beta} = \epsilon^{\alpha\beta} = \pm 1$ in a Northwest/Southeast convention. The $\epsilon_{\alpha\beta}$'s are normalised as
\begin{equation}
\epsilon_{\alpha\gamma} \epsilon^{\beta\gamma} = \delta_\alpha^\beta \,.
\end{equation}

The generalised metric in the $\mbf{10}$ is given by
\begin{equation}
 \begin{split}
  \gM_{+\alpha\,+\beta} &= \frac{1}{e}\left(e^2+C_{(4)}^2\right)H_{\alpha\beta} + \frac{1}{4e} \star\left( C_{(2)\alpha} \wedge C_{(2)\gamma} \right) \star\left(C_{(2)\beta} \wedge C_{(2)\delta} \right) H^{\gamma\delta} \\
  & \quad +\frac{C_{(4)}}{2e}\left(H_{\alpha_i\gamma}\star\left(C_{(2)\beta}\wedge C_{(2)\delta}\right)\epsilon^{\gamma\delta} + (\alpha\leftrightarrow\beta)\right) + \frac{e}{2}\, C_{(2)ij\,\alpha_i}\, C_{(2)kl\,\beta}\, g^{ik} g^{jl} \,, \\
  \gM_{+\alpha\,-\beta} &= \frac{C_{(4)}}{e}H_{\alpha\beta} + \frac{1}{2e}\star\left(C_{(2)\alpha}\wedge C_{(2)}{}^{\gamma}\right) H_{\gamma\beta}  \,, \\
  \gM_{-\alpha\,-\beta} &= \frac{1}{e} H_{\alpha\beta} \,, \\
  \gM_{+\alpha}{}^{ij} &= \frac{1}{e} \left( C_{(4)} H_{\alpha\beta} + \frac{1}{2} \star \left( C_{(2)\alpha} \wedge C_{(2)}{}^\gamma \right) H_{\gamma \beta} \right) \beta^{j_1j_2\,\beta} + e\, g^{ik}\,g^{jl}\, C_{(2)\alpha\,kl} \,, \\
  \gM_{\alpha-}{}^{ij} &= \frac{1}{e} H_{\alpha\beta}\, \beta^{ij\,\beta} \,, \\
  \gM^{ij \, kl} &= e \left( g^{ik} g^{jl} - g^{il} g^{jk} \right) + \frac{1}{e} \beta^{ij}{}_{\alpha}\, \beta^{kl}{}_{\beta}\, H^{\alpha\beta} \,.
 \end{split} \label{eq:SO55GenMetricParam2}
\end{equation}

\subsection{IIB parameterisation of the ExFT tensor hierarchy} \label{A:IIBtensor}
To complete the embedding of type IIB supergravity into exceptional field theory, one needs to embed the supergravity fields with legs along both the internal and external directions. These are encoded into the the ExFT tensor hierarchy fields $\mathcal{A}_\mu$, $\mathcal{B}_{\mu\nu}$, \ldots The map between supergravity and ExFT fields can be obtained by comparing how both transform under gauge transformations or by comparing their corresponding field strengths. We summarise the findings in the next section \ref{A:EFT-IIB-dictionary-TH} and give details of the derivations in sections \ref{A:SO55TH} - \ref{A:EFT-IIB-dictionaryDeriv}.

\subsubsection{Summary of IIB parameterisation} \label{A:EFT-IIB-dictionary-TH}
The 10-dimensional IIB metric is given by
\begin{equation}
 ds_{10}^2 = g_{ij} \KKD y^i \KKD y^j + g_{\mu\nu} dx^\mu dx^\nu \,,
\end{equation}
where $g_{ij}$ is the internal four-dimensional metric as computed from the generalised metric \eqref{eq:SO55GenMetricParam1}, \eqref{eq:SO55GenMetricParam2} and
\begin{equation}
 \KKD y^i= dy^i +(\imath_{A_\mu} dy^i) dx^\mu\,,
\end{equation} 
are the Kaluza-Klein covariantised derivatives of the internal coordinates with $A_\mu{}^i$ the Kaluza-Klein vector field. The ``external'' metric $g_{\mu\nu}$ is related to the ExFT metric ${\cal G}_{\mu\nu}$ by
\begin{equation}
 {\cal G}_{\mu\nu} = g_{\mu\nu} |g_{int}|^{-1/4} \,,
\end{equation}
where $|g_{int}|$ denotes the determinant of the internal metric $g_{ij}$.

For the remainder of this appendix, we will follow the conventions of \cite{Baguet:2015xha} and denote the 10-dimensional type IIB supergravity gauge fields by a hat, i.e. $\hat{C}_{(\hat{2})}{}^\alpha$ and $\hat{C}_{(\hat{4})}$, unlike in the main part of this paper. We will reserve the unhatted objects for later purposes in this appendix. Under the splitting of the 10 dimensions into six external and four internal directions, we write them as
\begin{equation}
\hat{C}_{(\hat{2})}{}^\alpha= \frac{1}{2}\bar C_{\mu\nu}{}^\alpha dx^\mu\wedge dx^\nu+\bar C_{\mu n}{}^\alpha dx^\mu\wedge \tilde{D}y^n+\frac{1}{2}\bar C_{mn}{}^\alpha \tilde{D}y^m\wedge \tilde{D}y^n\,,
\end{equation} 
and, analogously, for $\hat{C}_{(\hat{4})}$. The fields $\bar{C}_{\mu\nu}{}^\alpha$, $\bar{C}_{\mu n}{}^\alpha$, \ldots\, are the components of the KK-redefined form-fields $\bar{C}_{\mu\nu\,(0)}{}^\alpha$, $\bar{C}_{\mu\, (1)}{}^\alpha$, \ldots\, defined in \eqref{eq:field-redefinitions-bar}. The barred fields that are totally internal, i.e. $\bar{C}_{(2)}{}^\alpha$ and $\bar{C}_{(4)}$, are embedded into ExFT through the generalised metric \eqref{eq:SO55GenMetricParam1}, \eqref{eq:SO55GenMetricParam2}. The rest can be read off from the ExFT tensor hierarchy fields as (see \eqref{eq:dictionaries-gauge-fields}) 
\begin{equation}
\begin{split}
A_\mu&=(\mathcal{A}_\mu)_{(v)}\,,\\
\bar C_{\mu\,(1)}{}^\alpha&=(\mathcal{A}_\mu)_{(1)}{}^\alpha\,,\\
\bar C_{\mu\nu\,(0)}{}^\alpha&=\sqrt{2}\,(\mathcal{B}_{\mu\nu})_{(0)}{}^\alpha+\iota_{(\mathcal{A}_{[\mu})_{(v)}}(\mathcal{A}_{\nu]})_{(1)}{}^\alpha\\
\bar C_{\mu\,(3)}&=(\mathcal{A}_\mu)_{(3)}+\frac{1}{2}\epsilon_{\alpha\beta}\bar{C}_{\mu\,(1)}{}^\alpha\wedge\bar{C}_{(2)}{}^\beta\,,\\
\bar C_{\mu\nu\,(2)}&=-\sqrt{2}\,(\mathcal B_{\mu\nu})_{(2)}+\iota_{(\mathcal{A}_{[\mu})_{(v)}}(\mathcal{A}_{\nu]})_{(3)}+\frac{1}{2}\epsilon_{\alpha\beta}\bar{C}_{\mu\nu\,(0)}{}^\alpha\bar{C}_{(2)}{}^\beta\,,
\end{split} \label{eq:IIBTensorFieldsSummary}
\end{equation}
where $(\mathcal{A}_\mu)_{(v)}$, $(\mathcal{A}_\mu)_{(1)}{}^\alpha$, \ldots, $(\mathcal{B}_{\mu\nu})_{(0)}{}^\alpha$, \ldots\, are components of the tensor hierarchy fields $\mathcal{A}_\mu$ and $\mathcal{B}_{\mu\nu}$. The fields $\bar{C}_{\mu\nu\rho\,(1)}$ and $\bar{C}_{\mu\nu\rho\sigma\,(0)}$ involve further fields of the tensor hierarchy. However, they can also be determined (up to gauge transformations) from $\bar{C}_{(4)}$, $\bar{C}_{\mu\,(3)}$ and $\bar{C}_{\mu\nu\,(2)}$ through the self-duality condition of the 10-dimensional four-form. 

In addition to the dictionaries between tensor hierarchy and supergravity gauge fields, one can also embed the supergravity field strengths $\hat{F}_{(\hat{3})}{}^\alpha$ and $\hat{F}_{(\hat{5})}$  into the ExFT field strengths.  As in the case of gauge fields, we write the  10-dimensional field strengths as
\begin{equation}
\begin{split}
\hat{F}_{(\hat{3})}{}^\alpha&=\frac{1}{3!}\bar{F}_{\mu\nu\rho}{}^\alpha dx^\mu\wedge dx^\nu\wedge dx^\rho+\frac{1}{2}\bar{F}_{\mu\nu m}{}^\alpha dx^\mu\wedge dx^\nu\wedge \tilde{D}y^m\\
&\quad+\frac{1}{2}\bar{F}_{\mu m n}{}^\alpha dx^\mu\wedge\tilde{D}y^m\wedge \tilde{D}y^n+\frac{1}{3!}\bar{F}_{m n p}{}^\alpha \tilde{D}y^m\wedge\tilde{D}y^n\wedge \tilde{D}y^p\,,
\end{split}
\end{equation}
and analogous for $\hat{F}_{(\hat{5})}$. The barred $F$ fields are the components of the form-fields \eqref{eq:KK-fieldstrenghts-REDEF1}. Since the internal space is four-dimensional, the only 10-dimensional field strength with a totally internal part is $\hat{F}_{(\hat{3})}{}^\alpha$, given by
\begin{equation}
\bar{F}_{(3)}{}^\alpha=d\bar{C}_{(2)}{}^\alpha\,,
\end{equation} 
with $\bar{C}_{(2)}{}^\alpha$ the internal part of the 10-dimensional two-form, which is embedded into the ExFT through the generalised metric. Since all ExFT field strength have at least two external indices, the components of the 10-dimensional field strengths with one external leg can only be obtained directly from the gauge fields. These are (see \eqref{eq:field-strengths-redef-1} and \eqref{eq:field-strengths-redef-2})
\begin{equation}
\begin{split}
\bar F_{\mu\,(2)}{}^\alpha & =D_\mu^{KK}\bar C_{(2)}{}^\alpha - d\bar{C}_{\mu\,(1)}{}^\alpha\,, \\
\bar F_{\mu\,(4)}&=D_\mu^{KK}\bar C_{(4)}-d\bar C_{\mu\,(3)}-\frac{1}{2}\epsilon_{\alpha\beta}\bar C_{(2)}{}^\alpha\wedge \bar F_{\mu\,(2)}{}^\beta-\frac{1}{2}\epsilon_{\alpha\beta}\bar C_{\mu\,(1)}{}^\alpha\wedge \bar F_{(3)}{}^\beta\,.
\end{split}
\end{equation}
The rest of the components can be read off from the field strengths of the ExFT tensor hierarchy fields as (see \eqref{eq:Dictionary-field-strengths})
\begin{equation}
\begin{split}
F_{\mu\nu}&= (\mathcal{F}_{\mu\nu})_{(v)}\,,\\
\bar F_{\mu\nu\,(1)}{}^\alpha &=(\mathcal{F}_{\mu\nu})_{(1)}{}^\alpha+\iota_{F_{\mu\nu}}\bar{C}_{(2)}{}^\alpha\,,\\
\bar F_{\mu\nu\rho\,(0)}{}^\alpha&=\sqrt{2}\,(\mathcal{H}_{\mu\nu\rho})_{(0)}{}^\alpha\,,\\
\bar F_{\mu\nu\,(3)}&=(\mathcal{F}_{\mu\nu})_{(3)}+\epsilon_{\alpha\beta}\bar F_{\mu\nu\,(1)}{}^\alpha\wedge\bar{C}_{(2)}{}^\beta+\iota_{F_{\mu\nu}}\bar{C}_{(4)}+\frac{1}{2}\epsilon_{\alpha\beta}\bar{C}_{(2)}{}^\alpha\wedge\iota_{F_{\mu\nu}}\bar{C}_{(2)}{}^\beta\,,\\
\bar F_{\mu\nu\rho\,(2)}&=-\sqrt{2}\,(\mathcal{H}_{\mu\nu\rho})_{(2)}+\epsilon_{\alpha\beta}\bar{F}_{\mu\nu\rho\,(2)}{}^\alpha \bar{C}_{(2)}{}^\beta\,,
\end{split} \label{eq:IIBTensorFieldStrengthsSummary}
\end{equation}
where $F_{\mu\nu}$ is the field strength of the KK gauge field $A_\mu$ (see \eqref{eq:KK-fieldstrength-1}) and $(\mathcal{F}_{\mu\nu})_{(v)}$, \ldots,  $(\mathcal{H}_{\mu\nu\rho})_{(0)}{}^\alpha$, \ldots\, are the components of the ExFT field strengths $\mathcal{F}_{\mu\nu}$ and $\mathcal{H}_{\mu\nu\rho}$ defined in \eqref{eq:ExFTFieldStrength} and \eqref{eq:ExFTFieldStrengths}. As for the gauge fields, the components $\bar{F}_{\mu\nu\rho\sigma,(1)}$ and $\bar{F}_{\mu\nu\rho\sigma\delta,(0)}$ can be obtained from the 10-dimensional self-duality condition for $\hat{F}_{(\hat{5})}$.

\subsubsection{Tensor hierarchy of $\SO{5,5}$ Exceptional Field Theory} \label{A:SO55TH}
The tensor hierarchy of $\SO{5,5}$ ExFT containts the fields $\mathcal{A}_\mu $, $\mathcal{B}_{\mu\nu}$, $\mathcal{C}_{\mu\nu\rho}$, \ldots as listed in equation \eqref{eq:6DTH}. As discussed in section \ref{A:IIBSO55}, taking the type IIB solution to the section constraint \cite{Blair:2013gqa,Hohm:2013vpa} these decompose into
\begin{equation}
\begin{split}
\mathcal{A}_\mu &= A_{\mu ,(v)}+A_{\mu\,(1)}{}^\alpha+A_{\mu\,(3)}\,, \\
\mathcal{B}_{\mu\nu} &= B_{\mu\nu\,(0)}{}^\alpha+B_{\mu\nu\,(2)}+B_{\mu\nu\,(4)}{}^\alpha\,,\\
\mathcal{C}_{\mu\nu\rho}&= C_{\mu\nu\rho\,(1)}+\bar C_{\mu\nu\rho\,(1)}+C_{\mu\nu\rho\,(3)\,\alpha}\,.
\end{split}
\end{equation}
The gauge variations of $\mathcal{A}_\mu$ and $\mathcal{B}_{\mu\nu}$ are given by
\begin{equation}
\begin{split}
\delta\mathcal{A}_\mu &= \mathcal{D}_\mu \Lambda -d\,\Xi_\mu\,,\\
\delta\mathcal{B}_{\mu\nu}&=2\mathcal{D}_{[\mu}\Xi_{\nu]}+\Lambda\wedge\mathcal{F}_{\mu\nu}-\mathcal{A}_{[\mu}\wedge\delta \mathcal{A}_{\nu]}-d\,\Theta_{\mu\nu}\,,
\end{split}
\end{equation}
where $\Lambda\in\textbf{16}$, $\Xi_\mu\in\textbf{10}$ and $\Theta_{\mu\nu}\in\overline{\textbf{16}}$ are generalised gauge parameters, the derivative $\mathcal{D_\mu}$ is defined as
\begin{equation}
\mathcal{D}_\mu=\partial_\mu-\mathcal{L}_{\mathcal{A}_\mu}\,,
\end{equation}
and $\mathcal{F}_{\mu\nu}$ is the field strength of $\mathcal{A}_\mu$ defined in \eqref{eq:ExFTFieldStrength}.

\subsubsection*{Gauge variations and field strength of $\mathcal{A}_\mu$}
In the type IIB solution of the section constraint, the variation $\delta\mathcal{A}_\mu$ decomposes as
\begin{equation}
\begin{split} \label{eq:EFT-variations-1}
\delta A_{\mu\,(v)}&= D_\mu^{KK}\Lambda_{(v)}\,, \\
\delta A_{\mu\,(1)}{}^\alpha&= D_\mu^{KK}\Lambda_{(1)}{}^\alpha+L_{\Lambda_{(v)}} A_{\mu\,{(1)}}{}^\alpha-\,\,d\,\tilde\Xi_{\mu\,(0)}{}^\alpha\,, \\
\delta A_{\mu\,(3)}&=D_\mu^{KK}\Lambda_{(3)}+L_{\Lambda_{(v)}}A_{\mu\,{(3)}}-\epsilon_{\alpha\beta}dA_{\mu\,(1)}{}^\alpha\wedge \Lambda_{(1)}{}^\beta+  \,\,d\,\tilde\Xi_{\mu\,(2)}\,,
\end{split}
\end{equation}
where now $L$ is the usual Lie derivatives, the derivative $D_\mu^{KK}$ is defined as 
\begin{equation}
D_\mu^{KK}=\partial_\mu-L_{A_{\mu\,(v)}}\,,
\end{equation}
and the fields $\tilde{\Xi}_{\mu\,(0)}{}^\alpha$ and $\tilde{\Xi}_{\mu\,(2)}$ are 
\begin{equation}
\begin{split} \label{eq:xi-tilde-redefinition}
\tilde\Xi_{\mu\,(0)}{}^\alpha &= \sqrt{2}\,\Xi_{\mu\,(0)}{}^\alpha+\iota_{\Lambda_{(v)}}A_{\mu\,(1)}\,, \\
\tilde\Xi_{\mu\,(2)} &= \sqrt{2}\,\Xi_{\mu\,(2)}-\iota_{\Lambda_{(v)}}A_{\mu\,(3)}\,,
\end{split}
\end{equation}
with  $\Xi_{\mu\,(0)}{}^\alpha$ and $\Xi_{\mu\,(2)}$ being the zero- and two- form components of the gauge parameter $\Xi_\mu$. The field strength $\mathcal{F}_{\mu\nu}$, defined in \eqref{eq:ExFTFieldStrength}, decomposes as
\begin{equation}
\begin{split} \label{eq:EFT-fieldstrengths-1}
(\mathcal{F}_{\mu\nu})_{(v)} &= 2\partial_{[\mu}A_{\nu]\,(v)}-[A_{\mu\,(v)},A_{\nu\,(v)}]\,, \\
(\mathcal{F}_{\mu\nu\,(1)})_{(1)}{}^\alpha &= 2D^{KK}_{[\mu}A_{\nu]\,(1)}{}^\alpha+ d\tilde{B}_{\mu\nu\,(0)}{}^\alpha\,,\\
(\mathcal{F}_{\mu\nu})_{(3)} &= 2D^{KK}_{[\mu}A_{\nu]\,(3)}-d\tilde{B}_{\mu\nu\,(2)}-\epsilon_{\alpha\beta}A_{[\mu\,(1)}{}^\alpha\wedge dA_{\nu]\,(1)}{}^\beta\,,
\end{split}
\end{equation}
where, analogously to \eqref{eq:xi-tilde-redefinition},  the fields $\tilde{B}_{\mu\nu\,(0)}{}^\alpha$ and $\tilde{B}_{\mu\nu\,(2)}$ are defined as
\begin{equation}
\begin{split}
\tilde{B}_{\mu\nu\,(0)}{}^\alpha &= \sqrt{2}B_{\mu\nu\,(0)}{}^\alpha+\iota_{A_{[\mu\,(v)}}A_{\nu]\,(1)}{}^\alpha\,, \\
\tilde{B}_{\mu\nu\,(2)} &= \sqrt{2}B_{\mu\nu\,(2)}-\iota_{A_{[\mu\,(v)}}A_{\nu]\,(3)}\,,
\end{split}
\end{equation}
with $B_{\mu\nu\,(0)}{}^\alpha$ and $B_{\mu\nu\,(2)}$ components of $\mathcal{B}_{\mu\nu}$. 

\subsubsection*{Gauge variations and field strength of $\mathcal{B}_{\mu\nu}$}
The fields $\tilde{B}_{\mu\nu\,(0)}{}^\alpha$ and $\tilde{B}_{\mu\nu\,(2)}$ transform under gauge variations as
\begin{equation}
\begin{split} \label{eq:EFTvariations-2}
\delta\tilde{B}_{\mu\nu\,(0)}{}^\alpha &= \sqrt{2}\, (\delta\mathcal{B}_{\mu\nu})_{(0)}{}^\alpha+\iota_{\delta A_{[\mu\,(v)}}A_{\nu]\,(1)}{}^\alpha+\iota_{A_{[\mu\,(v)}}\delta A_{\nu]\,(1)}{}^\alpha \\
&= 2D^{KK}_{[\mu}\tilde{\Xi}_{\nu]\,(0)}{}^\alpha+L_{\Lambda_{(v)}}\tilde{B}_{\mu\nu\,(0)}{}^\alpha+\iota_{(\mathcal{F}_{\mu\nu})_{(v)}}\Lambda_{(1)}{}^\alpha\,, \\
\delta\tilde{B}_{\mu\nu\,(2)} &= \sqrt{2}\,(\delta\mathcal{B}_{\mu\nu})_{(0)}-\iota_{\delta A_{[\mu\,(v)}}A_{\nu]\,(3)}-\iota_{A_{[\mu\,(v)}}\delta A_{\nu]\,(3)} \\ 
&=2 D^{KK}_{[\mu}\left(\tilde{\Xi}_{\nu]\,(2)}
-\frac{1}{2}\epsilon_{\alpha\beta}A_{\nu]\,(1)}{}^\alpha\wedge\Lambda_{(1)}^\beta\right)
+\epsilon_{\alpha\beta}dA_{[\mu\,(1)}{}^\alpha\tilde{\Xi}_{\nu]\,(0)}{}^\beta
+L_{\Lambda_{(v)}}\tilde{B}_{\mu\nu\,(2)} \\ 
& \quad -\frac{1}{2}\epsilon_{\alpha\beta}\Lambda_{(1)}{}^\alpha\wedge d\tilde{B}_{\mu\nu\,(0)}{}^\beta
-\iota_{(\mathcal{F}_{\mu\nu})_{(v)}}\Lambda_{(3)}
-\frac{1}{2}\epsilon_{\alpha\beta}(\mathcal{F}_{\mu\nu})_{(1)}{}^\alpha\wedge\Lambda_{(1)}{}^\beta+d\tilde{\Theta}_{\mu\nu\,(1)}\,,
\end{split}
\end{equation}
where the field $\tilde{\Theta}_{\mu\nu\,(1)}$ is a redefinition of the one-form part of $\Theta_{\mu\nu}$.  Finally, the field strengths of the fields $\tilde{B}_{\mu\nu\,(0)}{}^\alpha$ and $\tilde{B}_{\mu\nu\,(2)}$ can be obtained from the field strength $\mathcal{H}_{\mu\nu\rho}$ (without tilde), defined in \eqref{eq:ExFTFieldStrengths}, as
\begin{equation}
\begin{split} \label{eq:EFT-fieldstrengths-H}
\tilde{H}_{\mu\nu\rho\,(0)}{}^\alpha &\equiv \sqrt{2}\,(\mathcal{H}_{\mu\nu\rho})_{(0)}{}^\alpha=3 D^{KK}_{[\mu}\tilde{B}_{\nu\rho]\,(0)}{}^\alpha-3\,\iota_{(\mathcal{F}_{[\mu\nu})_{(v)}} A_{\rho]\,(1)}{}^\alpha \,,\\
\tilde{H}_{\mu\nu\rho\,(2)} &\equiv \sqrt{2}\,(\mathcal{H}_{\mu\nu\rho})_{(2)} \\
&= 3D^{KK}_{[\mu}\tilde{B}_{\nu\rho]\,(2)} + 3\,\iota_{(\mathcal{F}_{[\mu\nu})_{(v)}} A_{\rho]\,(3)} + 3\epsilon_{\alpha\beta}A_{[\mu\,(1)}{}^\alpha\wedge D_\nu^{KK}A_{\rho]\,(1)}{}^\beta \\
& \quad + 3\epsilon_{\alpha\beta}\,dA_{[\mu\,(1)}{}^\alpha\tilde{B}_{\nu\rho]\,(0)}{}^\beta+d\tilde{C}_{\mu\nu\rho\,(1)}\,,
\end{split}
\end{equation}
where $\tilde{C}_{\mu\nu\rho\,(1)}$ is again some redefinition of the one-form part of $\tilde{C}_{\mu\nu\rho}$.

\subsubsection{KK decompositon of type IIB supergravity}
The bosonic field content of type IIB supergravity is given by a metric field $\hat{g}$, an axio-dilaton field $H$, a $SL(2)$-doublet of two-form fields $\hat{C}_{(\hat{2})}{}^\alpha$ and a four-form field $\hat{C}_{(\hat{4})}$. The sub-index $(\hat{p})$ indicates that the object is a $p$-form from the 10-dimensional point of view. The gauge variations of $\hat{C}_{(\hat{2})}{}^\alpha$ and a $\hat{C}_{(\hat{4})}$ are
\begin{equation}
\begin{split} \label{eq:gauge-transformations-10d}
\delta\hat C_{\hat\mu\hat\nu}{}^\alpha &= d\hat\lambda_{(\hat 1)}{}^\alpha\,, \\
\delta\hat C_{\hat\mu\hat\nu\hat\rho\hat\sigma} &= d\hat\lambda_{(\hat 3)}+\frac{1}{2}\epsilon_{\alpha\beta}\hat\lambda_{(\hat 1)}{}^\alpha\wedge\hat F_{(\hat 3)}{}^\beta\,,
\end{split}
\end{equation}
and their field strengths
\begin{equation}
\begin{split}
\hat{F}_{(\hat{3})}{}^\alpha &= d \hat{C}_{(\hat{2})}{}^\alpha\,, \\
\hat{F}_{(\hat{5})} &= d\hat{C}_{(\hat{4})}-\frac{1}{2}\epsilon_{\alpha\beta}\hat{C}_{(\hat{2})}{}^\alpha\wedge d\hat{C}_{(\hat{2})}{}^\beta \,.
\end{split}
\end{equation}

Next we split the 10-dimensional space into a six-dimensional external and a four-dimensional internal spaces. Throughout the rest of this section, we will use the field redefinitions of \cite{Baguet:2015xha}. Our conventions for the coordinates are: $x^{\hat{\mu}}$ are the ten dimensional coordinates,  $x^\mu$ are the external ones and $y^n$ the internal, with $\hat\mu=1,...,10$, $\mu=1,...,6$ and $n=1,...,4$. The two-forms fields $\hat{C}_{(\hat 2)}{}^\alpha$ decomposes under this splitting as
\begin{equation}
\begin{split} \label{eq:total-C2}
\hat{C}_{(\hat{2})}{}^\alpha &= \frac{1}{2}\hat{C}_{\hat\mu\hat\nu}{}^\alpha dx^{\hat{\mu}}\wedge dx^{\hat\nu} \\
&= \frac{1}{2}\hat C_{\mu\nu}{}^\alpha dx^\mu\wedge dx^\nu+\hat C_{\mu n}{}^\alpha dx^\mu\wedge dy^n+\frac{1}{2}\hat C_{mn}{}^\alpha dy^m\wedge dy^n \\
&\equiv \frac{1}{2}\hat C_{\mu\nu\,(0)}{}^\alpha dx^\mu\wedge dx^\nu +dx^\mu\wedge\hat C_{\mu\,(1)}{}^\alpha+\hat C_{(2)}{}^\alpha \,,
\end{split}
\end{equation}
where now the subscript $(p)$ indicates that the object is a $p$-form from the point of view of the internal space. The four-form $\hat{C}_{(\hat{4})}$ decomposes in an analogous way. Next, in a standard Kaluza-Klein manner, we redefine these form-fields by projecting the 10-dimensional curved indices into six-dimensional ones using the projector $P_\mu{}^{\hat\mu}=e_\mu{}^ae_a{}^{\hat\mu}$, where $a$ are the external flat indices and $e_{\hat\mu}{}^{\hat{a}}$ is the 10-dimensional metric vielbein  in a frame where it is upper-triangular. We obtain
\begin{equation}
\begin{split} \label{eq:field-redefinitions-bar}
\bar C_{(2)}{}^\alpha &= \hat C_{(2)}{}^\alpha\,, \\
\bar C_{\mu\,(1)}{}^\alpha &= \hat C_{\mu\,(1)}{}^\alpha-\iota_{A_\mu}\hat C_{(2)}{}^\alpha\,, \\
\bar C_{\mu\nu\,(0)}{}^ \alpha &= \hat  C_{\mu\nu\,(0)}{}^\alpha + 2 \iota_{A_{[\mu}}\hat C_{\nu]\,(1)}{}^\alpha-\iota_{A_\mu}\iota_{A_\nu}\hat C_{(2)}{}^\alpha\,, \\
\bar C_{(4)} &= \hat C_{(4)}\,, \\
\bar C_{\mu\,(3)} &=\hat C_{\mu\,(3)}-\iota_{A_\mu}\hat C_{(4)}\,, \\
\bar C_{\mu\nu\,(2)} &= \hat  C_{\mu\nu\,(2)} + 2 \iota_{A_{[\mu}}\hat C_{\nu]\,(3)}-\iota_{A_\mu}\iota_{A_\nu}\hat C_{(4)}\,, \\
&\;\; \vdots
\end{split}
\end{equation} 
where $A_\mu$ is the KK gauge field, with field strength
\begin{equation}\label{eq:KK-fieldstrength-1}
F_{\mu\nu}=2\partial_{[\mu}A_{\nu]}-[A_\mu,A_\nu]\,.
\end{equation}
For computational purposes it is worth noticing that the 10-dimensional two-forms \eqref{eq:total-C2} can now be written as
\begin{equation}\label{eq:field-redefinitions-bar-2}
\hat{C}_{(\hat{2})}{}^\alpha= \frac{1}{2}\bar C_{\mu\nu}{}^\alpha dx^\mu\wedge dx^\nu+\bar C_{\mu n}{}^\alpha dx^\mu\wedge \tilde{D}y^n+\frac{1}{2}\bar C_{mn}{}^\alpha \tilde{D}y^m\wedge \tilde{D}y^n\,,
\end{equation} 
with 
\begin{equation}
\tilde{D}y^n=dy^n+(\iota_{A_\mu}dy^n)\, dx^\mu\,,
\end{equation} 
and analogously for any other 10-dimensional form. Furthemore, because of the Chern-Simons term in the five-form field strength, the four-form needs some extra redefinitions (see for instance \cite{Baguet:2015xha,Ciceri:2014wya}), namely
\begin{equation}
\begin{split}
C_{(4)}&=\bar C_{(4)}\,,\qquad
C_{\mu\,(3)}=\bar C_{\mu\,(3)}-\frac{1}{2}\epsilon_{\alpha\beta}\bar{C}_{\mu\,(1)}{}^\alpha\wedge\bar{C}_{(2)}{}^\beta\,,\\ C_{\mu\nu\,(2)}&=\bar C_{\mu\nu\,(2)}-\frac{1}{2}\epsilon_{\alpha\beta}\bar C_{\mu\nu\,(0)}{}^\alpha\bar{C}_{(2)}{}^\beta\,,
\end{split}
\end{equation}
For completeness, we also define
\begin{equation}
C_{(2)}{}^\alpha =\bar C_{(2)}{}^\alpha\,,\qquad C_{\mu\,(1)}{}^\alpha =\bar C_{\mu\,(1)}{}^\alpha\,,\qquad C_{\mu\nu\,(0)}{}^\alpha =\bar C_{\mu\nu\,(0)}{}^\alpha\,.
\end{equation}
Analogous definitions apply also for the field strengths. In particular, now,
\begin{equation}
\begin{split}\label{eq:KK-fieldstrenghts-REDEF1}
\bar{F}_{(3)}{}^\alpha&=\hat{F}_{(3)}{}^\alpha\,,\\
\bar{F}_{\mu\,(2)}{}^\alpha&=\hat{F}_{\mu\,(2)}{}^\alpha-\iota_{A_\mu}\hat{F}_{(3)}{}^\alpha\,,\\
& \;\;\vdots\\
\bar{F}_{\mu\,(4)}&=\hat{F}_{\mu\,(4)}\,,\\
\bar{F}_{\mu\nu\,(3)}&=\hat{F}_{\mu\nu\,(3)}+2\iota_{A_{[\mu}}\hat{F}_{\nu]\,(4)}\,,\\
&\;\;\vdots
\end{split}
\end{equation}
where we recall that there is no internal five-form because the internal space is four-dimensional. Furthermore, 
\begin{equation}
\begin{split} \label{eq:KK-fieldstrenghts-REDEF2}
F_{(3)}{}^\alpha&=\bar{F}_{(3)}{}^\alpha\,,\qquad
F_{\mu\,(2)}{}^\alpha=\bar{F}_{\mu\,(2)}{}^\alpha\,,\qquad
F_{\mu\nu\,(1)}{}^\alpha=\bar{F}_{\mu\nu\,(1)}{}^\alpha-\iota_{F_{\mu\nu}}C_{(2)}{}^\alpha\,,\\[7pt]
F_{\mu\nu\rho\,(0)}{}^\alpha&=\bar{F}_{\mu\nu\rho\,(0)}{}^\alpha\,,\qquad
F_{\mu\,(4)}=\bar{F}_{\mu\,(4)}\,,\\[5pt]
F_{\mu\nu\,(3)}&=\bar{F}_{\mu\nu\,(3)}-\epsilon_{\alpha\beta}F_{\mu\nu\,(1)}{}^\alpha\wedge C_{(2)}{}^\beta-\iota_{F_{\mu\nu}}C_{(4)}+\frac{1}{2}\epsilon_{\alpha\beta}C_{(2)}{}^\alpha\wedge\iota_{F_{\mu\nu}}C_{(2)}{}^\beta\,,\\[5pt]
F_{\mu\nu\rho\,(2)}&=\bar{F}_{\mu\nu\rho\,(2)}-\epsilon_{\alpha\beta}F_{\mu\nu\rho\,(0)}{}^\alpha C_{(2)}{}^\beta\,,
\end{split}
\end{equation}
where $F_{\mu\nu}$ is  the KK field strenght \eqref{eq:KK-fieldstrength-1}.

\subsubsection*{Gauge variations and field strength of $\hat C_{(\hat 2)}{}^\alpha$}
Following the above redefinitions, the fields coming from the decomposition of $\hat C_{(\hat 2)}{}^\alpha$ transform under the gauge transformations \eqref{eq:gauge-transformations-10d} as
\begin{equation}
\begin{split}
\delta C_{(2)}{}^\alpha&=d\lambda_{(1)}{}^\alpha\,,\\[5pt]
\delta C_{\mu(1)}{}^\alpha&=D_\mu^{KK}\lambda_{(1)}{}^\alpha-d\lambda_{\mu\,(0)}{}^\alpha\,,\\[5pt]
\delta C_{\mu\nu\,(0)}{}^\alpha&=2D_{[\mu}^{KK}\lambda_{\nu](0)}{}^\alpha+\iota_{F_{\mu\nu}}\lambda_{(1)}{}^\alpha\,,
\end{split}
\end{equation}
where $F_{\mu\nu}$ is the KK field strength \eqref{eq:KK-fieldstrength-1} and the derivative $D_\mu^{KK}$ is, as above, defined as
\begin{equation}
D_\mu^{KK}=\partial_\mu-L_{A_\mu}\,.
\end{equation}
Analogous to the gauge fields, the $\lambda$-parameters are defined as
\begin{equation}
\begin{split}
\lambda_{(1)}{}^\alpha&=\bar\lambda_{(1)}{}^\alpha=\hat\lambda_{(1)}{}^\alpha\,,\\
\lambda_{\mu\,(0)}{}^\alpha&=\bar\lambda_{\mu\,(0)}{}^\alpha=\hat\lambda_{\mu\,(0)}{}^\alpha-\iota_{A_\mu}\hat\lambda_{(1)}{}^\alpha\,.
\end{split}
\end{equation}
After redefinitions, the field strengths coming from  $\hat F_{(\hat 3)}{}^\alpha$ become 
\begin{equation}
\begin{split}\label{eq:field-strengths-redef-1}
F_{(3)}{}^\alpha&=dC_{(2)}{}^\alpha\,,\\
F_{\mu\,(2)}{}^\alpha&=D_{\mu}^{KK}C_{(2)}{}^\alpha-dC_{\mu\,(1)}{}^\alpha\,,\\
F_{\mu\nu\,(1)}{}^\alpha&=2 D_{[\mu}^{KK}C_{\nu]\,(1)}{}^\alpha+dC_{\mu\nu\,(0)}{}^\alpha\,,\\
F_{\mu\nu\rho\,(0)}{}^\alpha&=3D_{[\mu}^{KK}C_{\nu\rho]\,(0)}{}^\alpha-3\iota_{F_{[\mu\nu}}C_{\rho]\,(1)}{}^\alpha\,.
\end{split}
\end{equation}

\subsubsection*{Gauge variations and field strength of $\hat C_{(\hat 4)}$}
The redefined fields coming from $\hat C_{(\hat 4)}$ transform under gauge transformations as
\begin{equation}
 \begin{split}
  \delta C_{(4)} &= d\lambda_{(3)} + \frac{1}{2} \epsilon_{\alpha\beta} \lambda_{(1)}{}^\alpha \wedge F_{(3)}{}^\beta\,,\\
  \delta C_{\mu\,(3)} &= D_\mu^{KK} \lambda_{(3)} - d\lambda_{\mu\,(2)} + \epsilon_{\alpha\beta} \lambda_{(1)}{}^\alpha \wedge dC_{\mu\,(1)}{}^\beta\,,\\
  \delta C_{\mu\nu\,(2)} &= 2 D_{[\mu}^{KK} \left(\lambda_{\nu]\,(2)} + \frac{1}{2} \epsilon_{\alpha\beta} \lambda_{(1)}{}^\alpha \wedge C_{\nu]\,(1)}{}^\beta\right) + d\lambda_{\mu\nu\,(1)} + \iota_{F_{\mu\nu}} \lambda_{(3)} \\
  &\quad + \frac{1}{2} \epsilon_{\alpha\beta} (\lambda_{(1)}{}^\alpha \wedge F_{\mu\nu\,(1)}{}^\beta - 2\lambda_{[\mu\,(0)}{}^\alpha dC_{\nu]\,(1)}{}^\beta + dC_{\mu\nu\,(0)}{}^\alpha \wedge \lambda_{(1)}{}^\beta) \,,
\end{split}
\end{equation}
where the new $\lambda$-fields are defined, analogous to the gauge fields, as
\begin{equation}
\begin{split}
\lambda_{(3)}&=\bar\lambda_{(3)}-\frac{1}{2}\epsilon_{\alpha\beta}\bar\lambda_{(1)}{}^\alpha\wedge C_{(2)}{}^\beta\,,\\
\lambda_{\mu\,(2)}&=\bar\lambda_{\mu\,(2)}-\frac{1}{2}\epsilon_{\alpha\beta}\Bigl(\bar\lambda_{(1)}{}^\alpha\wedge C_{\mu\,(1)}{}^\beta+\bar\lambda_{\mu\,(0)} C_{(2)}{}^\beta\Bigr)\,,\\
\lambda_{\mu\nu\,(1)}&=\bar\lambda_{\mu\nu\,(1)}-\frac{1}{2}\epsilon_{\alpha\beta}\bar\lambda_{(1)}{}^\alpha C_{\mu\nu\,(0)}{}^\beta\,,
\end{split}
\end{equation}
together with
\begin{equation}
\begin{split}
\bar\lambda_{(3)}&=\hat\lambda_{(3)}\,,\\
\bar\lambda_{\mu\,(2)}&=\hat\lambda_{\mu\,(2)}-\iota_{A_\mu}\hat\lambda_{(3)}\,,\\
\bar\lambda_{\mu\nu\,(1)}&=\hat\lambda_{\mu\nu\,(1)}+2\iota_{A_{[\mu}}\hat\lambda_{\nu]\,(2)}-\iota_{A_\mu}\iota_{A_\nu}\hat\lambda_{(3)}\,.
\end{split}
\end{equation}
After redefinitions, the field strengths coming from $\hat{F}_{(\hat{5})}$ become
\begin{equation}
\begin{split}\label{eq:field-strengths-redef-2}
F_{\mu\,(4)}&=D_\mu^{KK}C_{(4)}-dC_{\mu\,(3)}-\frac{1}{2}\epsilon_{\alpha\beta}C_{(2)}{}^\alpha\wedge F_{\mu\,(2)}{}^\beta+\frac{1}{2}\epsilon_{\alpha\beta}C_{(2)}{}^\alpha\wedge dC_{\mu\,(1)}{}^\beta\,,\\
F_{\mu\nu\,(3)}&=2D_{[\mu}^{KK}C_{\nu]\,(3)}+dC_{\mu\nu\,(2)}-\epsilon_{\alpha\beta}C_{[\mu\,(1)}{}^\alpha\wedge dC_{\nu]\,(1)}{}^\beta\,,\\
F_{\mu\nu\rho\,(2)}&=3D_{[\mu}^{KK}C_{\nu\rho]\,(2)}-dC_{\mu\nu\rho\,(1)}-3\iota_{F_{[\mu\nu}}C_{\rho]\,(3)}-3\epsilon_{\alpha\beta}dC_{[\mu\nu\,(0)}{}^\alpha\wedge C_{\rho]\,(1)}{}^\beta\\
&\quad-3\epsilon_{\alpha\beta}C_{[\mu\,(1)}{}^\alpha\wedge D_\nu^{KK}C_{\rho]\,(1)}{}^\beta\,. 
\end{split}
\end{equation}

\subsubsection*{Summary of variations}
Combining the results above together with diffeomorphism variations along a vector $\chi$ in the internal space one finally obtains
\begin{equation}
\begin{split}\label{eq:KKvariations-final-1}
\delta C_{(2)}{}^\alpha&=d\lambda_{(1)}{}^\alpha+L_\chi C_{(2)}{}^\alpha\,,\\[5pt]
\delta C_{\mu(1)}{}^\alpha&=D_\mu^{KK}\lambda_{(1)}{}^\alpha-d\lambda_{\mu\,(0)}{}^\alpha+L_{\chi}C_{\mu(1)}{}^\alpha\,,\\[5pt]
\delta C_{\mu\nu\,(0)}{}^\alpha&=2D_{[\mu}^{KK}\lambda_{\nu](0)}{}^\alpha+\iota_{F_{\mu\nu}}\lambda_{(1)}{}^\alpha+L_\chi C_{\mu\nu\,(0)}{}^\alpha\,,
\end{split}
\end{equation}
and
\begin{equation}
 \begin{split}\label{eq:KKvariations-final-2}
  \delta C_{(4)} &= d\lambda_{(3)} + \frac{1}{2} \epsilon_{\alpha\beta} \lambda_{(1)}{}^\alpha \wedge F_{(3)}{}^\beta + L_\chi C_{(4)}\,, \\
  \delta C_{\mu\,(3)} &= D_\mu^{KK} \lambda_{(3)} - d\lambda_{\mu\,(2)} + \epsilon_{\alpha\beta} \lambda_{(1)}{}^\alpha \wedge dC_{\mu\,(1)}{}^\beta + L_\chi C_{\mu\,(3)}\,,\\
  \delta C_{\mu\nu\,(2)} &= 2 D_{[\mu}^{KK} \left(\lambda_{\nu]\,(2)} + \frac{1}{2} \epsilon_{\alpha\beta} \lambda_{(1)}{}^\alpha \wedge C_{\nu]\,(1)}{}^\beta\right)
  + d\lambda_{\mu\nu\,(1)} + \iota_{F_{\mu\nu}}\lambda_{(3)}\\
  & \quad + \frac{1}{2} \epsilon_{\alpha\beta} (\lambda_{(1)}{}^\alpha \wedge F_{\mu\nu\,(1)}{}^\beta - 2\lambda_{[\mu\,(0)}{}^\alpha dC_{\nu]\,(1)}{}^\beta + dC_{\mu\nu\,(0)}{}^\alpha \wedge \lambda_{(1)}{}^\beta) + L_\chi C_{\mu\nu\,(2)}\,,
\end{split}
\end{equation}
The KK gauge field $A_\mu$ transforms as
\begin{equation}
\delta A_\mu=D_\mu^{KK}\chi\,.
\end{equation}

\subsubsection{Dictionaries $\SO{5,5}$ ExFT - IIB supergravity}\label{A:EFT-IIB-dictionaryDeriv}
By comparing \eqref{eq:EFT-variations-1} and \eqref{eq:EFTvariations-2} with \eqref{eq:KKvariations-final-1} and \eqref{eq:KKvariations-final-2} we can identify
\begin{equation}
\begin{split}\label{eq:dictionaries-gauge-fields}
A_\mu&=(\mathcal{A}_\mu)_{(v)}\,,\qquad C_{\mu\,(1)}{}^\alpha=(\mathcal{A}_\mu)_{(1)}{}^\alpha\,,\qquad C_{\mu\nu\,(0)}{}^\alpha=\tilde B_{\mu\nu\,(0)}{}^\alpha\\
C_{\mu\,(3)}&=(\mathcal{A}_\mu)_{(3)}\,,\qquad C_{\mu\nu\,(2)}=-\tilde B_{\mu\nu\,(2)}\,,
\end{split}
\end{equation}
and analogously for the gauge parameters
\begin{equation}
\begin{split}
\chi&=\Lambda_{(v)}\,,\qquad\lambda_{(1)}{}^\alpha=\Lambda_{(1)}{}^\alpha\,,\qquad\lambda_{\mu\,(0)}{}^\alpha=\tilde\Xi_{\mu\,(0)}{}^\alpha\\
\lambda_{(3)}&=\Lambda_{(3)}\,,\qquad \lambda_{\mu\,(2)}=-\tilde\Xi_{\mu\,(2)}\,.
\end{split}
\end{equation}
We can also establish dictionaries between field strenghts. Comparing \eqref{eq:EFT-fieldstrengths-1}, \eqref{eq:EFT-fieldstrengths-H},\eqref{eq:field-strengths-redef-1} and \eqref{eq:field-strengths-redef-2} we obtain
\begin{equation}
\begin{split}\label{eq:Dictionary-field-strengths}
F_{\mu\nu}&= (\mathcal{F}_{\mu\nu})_{(v)}\,,
\qquad F_{\mu\nu\,(1)}{}^\alpha =(\mathcal{F}_{\mu\nu})_{(1)}{}^\alpha\,,
\qquad F_{\mu\nu\rho\,(0)}{}^\alpha=\tilde{H}_{\mu\nu\rho\,(0)}{}^\alpha\\
F_{\mu\nu\,(3)}&=(\mathcal{F}_{\mu\nu})_{(3)}\,,
\qquad F_{\mu\nu\rho\,(2)}=-\tilde{H}_{\mu\nu\rho\,(2)}\,.
\end{split}
\end{equation}

\section{$S^2$ conventions} \label{A:S2}
We describe the $S^2$ by three functions $y_u$, $u = 1, \ldots, 3$ satisfying
\begin{equation}
y_u y^u = 1 \,,
\end{equation}
where we raise/lower $u, v = 1, \ldots, 3$ indices with $\delta_{uv}$. In terms of these functions, the round metric on $S^2$ and its volume form are given by
\begin{equation}
ds_{S^2}^2 = dy_u dy^u \,, \qquad vol_{S^2} = \frac12 \epsilon_{uvw} y^u\, dy^v \wedge dy^w \,. \label{eq:MetricVolS2}
\end{equation}
The Killing vectors of the round $S^2$ are given by
\begin{equation}
v_u^i = g^{ij} \epsilon_{uvw} y^v \partial_j y^w \,,
\end{equation}
where $i, j = 1, 2$ denote a local coordinate basis and $g^{ij}$ is the inverse metric of the round $S^2$. Alternatively, the Killing vectors can be defined as in \cite{Lee:2014mla}.

We also make repeated use of the 1-forms that are Hodge dual to $dy_u$ with respect to the round metric \eqref{eq:MetricVolS2}
\begin{equation}
\theta_u = \star dy_u = \epsilon_{uvw} y^v dy^w \,.
\end{equation}
These form a ``dual span'' of the $T^*(S^2)$ to the Killing vectors, i.e.
\begin{equation}
\imath_{v_u} \theta_v = \delta_{uv} - y_u\, y_v \,.
\end{equation}
Note that the 1-forms $dy_u$, $\theta_u$ and Killing vectors $v_u$ satisfy
\begin{equation}
y_u dy^u = y_u \theta^u = y_u v^u = 0 \,.
\end{equation}

All the objects we introduced above transform naturally under the $\SU{2}_R$ symmetry generated by the Killing vector fields.
\begin{equation}
 \begin{split}
  L_{v_u} v_v &= - \epsilon_{uvw}\, v^w \,, \\
  L_{v_u} y_v &= - \epsilon_{uvw}\, y^w \,, \\
  L_{v_u} dy_v &= - \epsilon_{uvw} \, dy^w \,, \\
  L_{v_u} \theta_v &= - \epsilon_{uvw} \theta^w \,.
 \end{split}
\end{equation}

\section{Dictionary between AdS$_6$ conventions} \label{A:AdS6Match}
Upon imposing the Cauchy-Riemann equations \eqref{eq:CR} and identifying the holomorphic functions as in \eqref{eq:HolMatch}, we find the following match between our objects and those of \cite{DHoker:2016ujz}.
\begin{equation}
\begin{split}
r &= \frac18 {\cal G} \,, \qquad
|dk| = \frac14 \underline{\kappa}^2 \,, \qquad
\Delta = \frac{3 \underline{\kappa}^4 {\cal G}}{128} {\cal \tilde{D}} \,,
\end{split}
\end{equation}
where, as in \cite{Hong:2018amk},
\begin{equation}
 {\cal \tilde{D}} = 1 + \frac{2\,|\partial {\cal G}|^2}{\underline{\kappa}^2\,{\cal G}} \,.
\end{equation}
Here, to differentiate our $\kappa$ from the objects denoted by the same symbols in \cite{DHoker:2016ujz} we denoted theirs by an underline: $\underline{\kappa}$.

Our $\SL{2}$ doublet of 2-forms, $C_{(2)}{}^\alpha$, are related to the complex 2-form, ${\cal C}_{(2)}$, of \cite{DHoker:2016ujz} by
\begin{equation}
 {\cal C}_{(2)} = - C_{(2)}{}^1 + i\, C_{(2)}{}^2 \,.
\end{equation}
Similarly, our axio-dilaton, $H^{\alpha\beta}$ is mapped to the complex scalar ${\cal B}$ of \cite{DHoker:2016ujz} via
\begin{equation}
 {\cal B} = - 2\, \frac{1 + (H_{12})^2 + (H_{22})^2}{1 + (H_{12} + i\, H_{22})^2} \,.
\end{equation}

We can similarly match our minimal consistent truncation with that found in \cite{Hong:2018amk}. To differentiate between our scalar field $X$, gauge fields $A^A$, $A^4$, two-form fields $B_{(2)}$ and those of \cite{Hong:2018amk}, we will denote the objects of \cite{Hong:2018amk} by an underline, i.e. $\underline{X}$, $\underline{A}^A$, $\underline{A}^4$, $\underline{B}$. We use the same notation for the field strengths, i.e. our objects are $\tilde{F}_{(2)}{}^A$, $\tilde{F}_{(2)}{}^4$ and $\tilde{F}_{(3)}$ and those of \cite{Hong:2018amk} are $\tilde{\underline{F}}_{(2)}{}^A$, $\tilde{\underline{F}}_{(2)}{}^4$ and $\tilde{\underline{F}}_{(3)}$. The map is now given by
\begin{equation}
 \begin{split}
  X &= \underline{X}^{-1} \,, \qquad A^A = \frac{\sqrt{2}}{3} \underline{A}^A \,, \qquad  A^4 = \underline{A}^4 \,, \qquad B = \underline{B} \,, \\
  \tilde{F}_{(2)}{}^A &= \frac{\sqrt{2}}{3} \underline{\tilde{F}}_{(2)}{}^A \,, \qquad \tilde{F}_{(2)}{}^4 = \underline{\tilde{F}}_{(2)}{}^4 \,, \qquad \tilde{F}_{(3)} = \underline{\tilde{F}}_{(3)} \,,
 \end{split}
\end{equation}
and our function $\bar{\Delta}$ is related to ${\cal D}$ of \cite{Hong:2018amk} by
\begin{equation}
 \bar{\Delta} = \frac{3\underline{\kappa}^4{\cal G}}{128} \underline{X}^{-4} {\cal D} \,.
\end{equation}

\bibliographystyle{JHEP}
\bibliography{NewBib}

\end{document}